\pdfoutput=1

\documentclass[11pt,twoside,a4paper,cmspaper,final,collab]{cms-tdr}
\def\svnVersion{294884}\def\cmsCernNo{2013-037}\def\cmsCernDate{\today}\def\cmsMessage{Published in the Journal of Instrumentation as \href{http://dx.doi.org/10.1088/1748-0221/10/06/P06005}{\doi{10.1088/1748-0221/10/06/P06005.}}}

\begin{document}\cmsNoteHeader{EGM-13-001}

\hyphenation{had-ron-i-za-tion}
\hyphenation{cal-or-i-me-ter}
\hyphenation{de-vices}
\RCS$Revision: 294732 $
\RCS$HeadURL: svn+ssh://svn.cern.ch/reps/tdr2/papers/EGM-13-001/trunk/EGM-13-001.tex $
\RCS$Id: EGM-13-001.tex 294732 2015-06-30 13:42:54Z baffi $
\newlength\cmsFigWidth
\ifthenelse{\boolean{cms@external}}{\setlength\cmsFigWidth{0.85\columnwidth}}{\setlength\cmsFigWidth{0.4\textwidth}}
\ifthenelse{\boolean{cms@external}}{\providecommand{\cmsLeft}{top}}{\providecommand{\cmsLeft}{left}}
\ifthenelse{\boolean{cms@external}}{\providecommand{\cmsRight}{bottom}}{\providecommand{\cmsRight}{right}}

\newcommand{\brem}{bremsstrahlung }
\newcommand{\bremns}{bremsstrahlung}
\newcommand{\Xz}{\ensuremath{\mathrm{X}_0}\xspace}
\newcommand{\sPlot}{sPlot\xspace}
\newcommand{\ETxy}[2]{\ifx#2\relax\ensuremath{E_\text{T, #1}}\else\ensuremath{E_\text{T, #1}^\text{#2}}\fi\xspace}

\cmsNoteHeader{EGM-13-001}\title{Performance of electron reconstruction and selection with the CMS detector in proton-proton collisions at $\sqrt{s}$ = 8\TeV}

\date{\today}

\abstract{
The performance and strategies used in electron reconstruction and selection at CMS are presented based on data corresponding to an integrated luminosity of 19.7\fbinv, collected in proton-proton collisions at $\sqrt{s} = 8\TeV$ at the CERN LHC. The paper focuses on prompt isolated electrons with transverse momenta ranging from about 5 to a few 100\GeV. A detailed description is given of the algorithms used to cluster energy in the electromagnetic calorimeter and to reconstruct electron trajectories in the tracker. The electron momentum is estimated by combining the energy measurement in the calorimeter with the momentum measurement in the tracker. Benchmark selection criteria are presented, and their performances assessed using \Z, $\Upsilon$, and \JPsi decays into $\Pep$+$\Pem$ pairs. The spectra of the observables relevant to electron reconstruction and selection as well as their global efficiencies are well reproduced by Monte Carlo simulations. The momentum scale is calibrated with an uncertainty smaller than 0.3\%. The momentum resolution for electrons produced in \Z boson decays ranges from 1.7 to 4.5\%, depending on electron pseudorapidity and energy loss through bremsstrahlung in the detector material.
}

\hypersetup{%
pdfauthor={CMS Collaboration},%
pdftitle={Performance of electron reconstruction and selection with the CMS detector in proton-proton collisions at sqrt(s)=8 TeV},%
pdfsubject={CMS},%
pdfkeywords={CMS, physics, electron detector}}

\maketitle
\tableofcontents

\section{Introduction}
\label{sec:intro}

Electron reconstruction and selection is of great importance in many analyses performed using data from the CMS detector, such as standard model precision measurements, searches and measurements in the Higgs sector, and searches for processes beyond the standard model.
These scientific analyses require excellent electron reconstruction and selection
efficiencies together with small misidentification probability
over a large phase space,
 excellent momentum resolution, and small systematic uncertainties.
A high level of performance has been achieved in steps, evolving from
the initial algorithms for electron reconstruction developed in the
context of online selection~\cite{Sphicas:2002gg}.
The basic principles of offline electron reconstruction, outlined in the CMS Physics Technical Design Report~\cite{Bayatian:2006zz,Baffioni:2006cd},
rely on a combination of the energy measured in the electromagnetic calorimeter (ECAL) and the momentum measured in the tracking detector (tracker), to optimize the performance over a wide range of transverse momentum (\pt).
Throughout the paper, ``energy'' and ``momentum'' refer, respectively, to the energy of the electromagnetic shower initiated by the electron in the ECAL and to the track momentum measurement in the tracker, while the term ``electron momentum'' is used to refer to the combined information.
The energy calibration and resolution in the ECAL were discussed in Ref.~\cite{Chatrchyan:2013dga}, and general issues in track reconstruction in Ref.~\cite{trkpaper}.
Preliminary results on electron reconstruction and selection were also
given in Refs.~\cite{CMS:2010bta,CMS:2011hqb,CMS:2013hoa}.
One of the main challenges for precise reconstruction of electrons in CMS is the tracker material, which causes significant bremsstrahlung along the electron trajectory. In addition,
this \brem spreads over a large volume due to the CMS magnetic field.
Dedicated techniques have been developed to account for this effect \cite{Baffioni:2006cd}. These procedures have been optimized using simulation, and commissioned with data taken since 2009.

\begin{figure}[htb]
\centering
\includegraphics[width=0.95\textwidth]{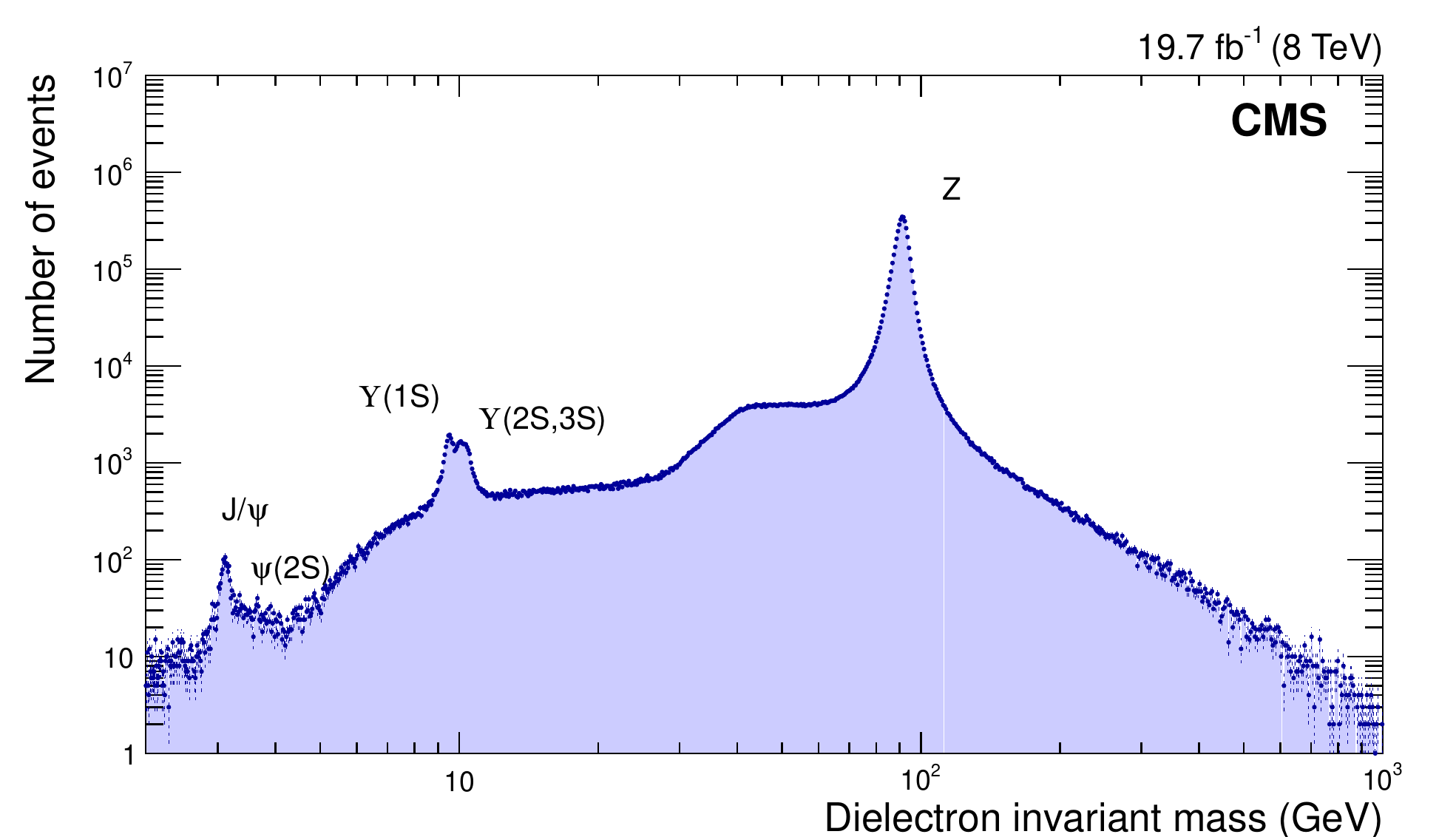}
\caption{\label{Scale_meeall}
Two-electron invariant mass spectrum for data collected with
dielectron triggers. Electron momenta are obtained by combining information from the tracker and the ECAL.}

\end{figure}

This paper describes the reconstruction and selection algorithms for isolated primary electrons, and their
performance in terms of momentum calibration, resolution, and measured efficiencies. The results are based on data collected in proton-proton collisions at $\sqrt{s}=8$\TeV at the CERN LHC that correspond to an integrated luminosity of 19.7\fbinv.
Figure~\ref{Scale_meeall} shows the two-electron invariant mass spectrum from data collected with dielectron triggers.
The step near 40\GeV is due to the thresholds used in the triggers.
The \JPsi, $\Pgy$, $\PgUa$, the overlapping $\PgUb$ and $\PgUc$ mesons, and the \Z boson resonances can be seen,
and are  used to assess the performance of the electron momentum calibration and resolution, and to measure the reconstruction and selection efficiencies.

A crucial and challenging process used as a benchmark in the paper is the decay of the Higgs boson into four leptons
through on-shell Z boson and virtual Z boson (Z*) intermediate states~\cite{legacyHzz}. In the case of a decay into four electrons or two muons and two electrons, one electron
can have a very small \pt that requires good performance
down to $\pt\approx5\GeV$.
At the other extreme, electrons with \pt above a few 100\GeV are often used to search
for high-mass resonances~\cite{zprime} and other new processes
beyond the standard model.

The paper is organized as follows. Sections~\ref{sec:detector} and \ref{sec:samples} briefly describe the CMS detector, the online selections, the data, and
Monte Carlo (MC) simulations used in this analysis.
The electron reconstruction algorithms,
 together with the performance of the electron-momentum calibration and resolution, are detailed in Section~\ref{sec:reco}.
The different steps in electron
selection, namely the identification and the isolation techniques, are described in Section~\ref{sec:selection}.
Measurements of reconstruction and selection efficiencies and misidentification probabilities are presented in Section~\ref{sec:TandP}, and
results are summarized in Section~\ref{sec:conclusion}.

\section{CMS detector}
\label{sec:detector}

The central feature of the CMS apparatus is a superconducting solenoid of 6\unit{m} internal diameter, providing a magnetic field of 3.8\unit{T}. The field volume contains a silicon pixel and strip tracker, a lead tungstate crystal ECAL, and a brass and scintillator hadron calorimeter (HCAL), each one composed of a barrel and two endcap sections. Muons are measured in gas ionization detectors embedded in the steel flux return yoke outside of the solenoid. Extensive forward calorimetry complements the coverage provided by the barrel and endcap detectors.
A more detailed description of the CMS detector together with a definition of the coordinate system and relevant kinematic variables can be found in Ref.~\cite{Chatrchyan:2008aa}. In this section,
the origin of the coordinate system is at the geometrical centre of the detector, however, in all later sections, unless otherwise specified, the origin is defined to be the reconstructed interaction point (collision vertex).

The tracker and the ECAL, being the main detectors involved in the reconstruction and identification of electrons, are described in greater detail in the following paragraphs. The HCAL, which is used at different steps of electron reconstruction and selection, is also described below.

The CMS tracker is a cylindric detector 5.5\unit{m} long and 2.5\unit{m} in diameter, equipped
with silicon that provides a total surface of 200\unit{m$^2$} for an active detection region of $\abs{\eta} \leq 2.5$ (the acceptance). The inner part is based on silicon pixels and the
outer part on silicon strip detectors. The pixel tracker (66 million channels) consists of 3 central layers
covering a radial distance $r$ from 4.4\unit{cm} up to 10.2\unit{cm},
complemented by two forward endcap disks
covering  $6 \leq r \leq 15$\unit{cm} on each side.
With this geometry, a deposition of hits in at least 3 layers or disks per track for almost the entire acceptance is ensured.
The strip detector (9.3 million channels) consists of 10 central layers,
complemented by 12 disks in each endcap.
The central layers cover radial distances $r\leq108\unit{cm}$
and $\abs{z}\leq 109$\unit{cm}.
The disks cover up to $\abs{z} \leq 280$\unit{cm} and $r \leq 113$\unit{cm}.
Since the tracker extends to $\abs{\eta}=2.5$, precise
detection of electrons is only possible up to this pseudorapidity,
despite the larger coverage of the ECAL. In this paper the acceptance
of electrons is restricted to $\abs{\eta}\leq2.5$, corresponding to the
region where electron tracks can be reconstructed in the tracker.

\begin{figure}[htb]
\centering
\includegraphics[width=0.50\textwidth]{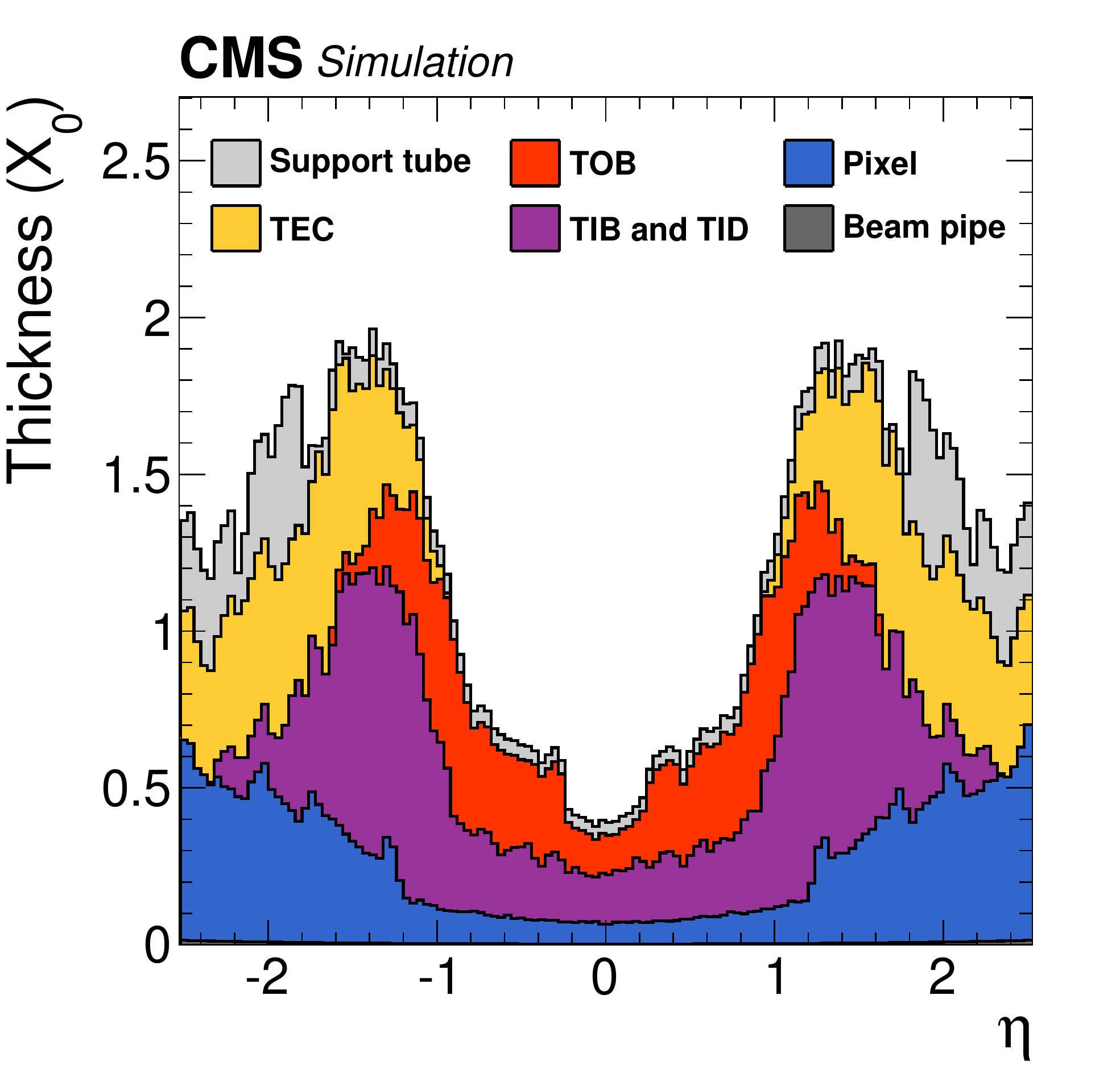}
\caption{\label{fig:Material}
Total thickness of tracker material traversed by a particle produced at the centre of the detector expressed in units of $\Xz$, as a function of particle pseudorapidity $\eta$ in the $\abs{\eta}\leq2.5$ acceptance region. The contribution to the total material of each of the subsystems that comprise the CMS tracker is given separately for the pixel tracker, strip tracker consisting of the tracker endcap (TEC), the tracker outer barrel (TOB), the tracker inner barrel (TIB), and the tracker inner disks (TID),
together with contributions from the support tube that surrounds the tracker, and from the beam pipe, which is visible as a thin line at the bottom of the figure~\cite{trkpaper}.
}

\end{figure}

A consequence of the presence of the silicon tracker is a significant amount of material in front of the ECAL, mainly due to the mechanical structure, the services, and the cooling system.
Figure~\ref{fig:Material} shows the thickness of the tracker as a function of $\eta$ in the $\abs{\eta}\leq2.5$ acceptance region,
presented in terms of radiation lengths $\Xz$~\cite{trkpaper}.
It rises from $\approx$0.4\unit{\Xz} near $\abs{\eta}\approx0$, to $\approx$2.0\unit{\Xz} near
$\abs{\eta}\approx1.4$, and decreases to $\approx$1.4\unit{\Xz} near $\abs{\eta} \approx 2.5$.
This material, traversed by electrons before reaching the ECAL, induces a potential loss of
electron energy via bremsstrahlung.
The emitted photons can also convert to $\Pep\Pem$ pairs, and the produced electrons and positrons can radiate photons
through \bremns, leading to
the early development of an electromagnetic shower in the tracker.

The ECAL is a homogeneous and hermetic calorimeter made of PbWO$_4$ scintillating
crystals. It is composed of a central barrel covering the pseudorapidity
region $\abs{\eta}\leq 1.479$ with the
internal surface located at $r = 129$\unit{cm}, and complemented by two endcaps
covering $1.479 \leq \abs{\eta} \leq 3.0$ that are located at $z =\pm 315.4$\unit{cm}.
The large density (8.28\unit{g/cm$^{3}$}), the small radiation length (0.89\unit{cm}), and the small
Moli\`ere radius (2.3\unit{cm}) of the PbWO$_4$ crystals result in a compact calorimeter with excellent separation of close clusters.
A preshower detector consisting of two planes of silicon sensors interleaved with a total of 3\unit{$\Xz$} of lead is located in front of the endcaps, and covers $1.653 \leq \abs{\eta} \leq 2.6$.

The ECAL barrel is made of 61\,200 trapezoidal crystals with front-face transverse sections of 22 $\times$ 22$\unit{mm}^2$, giving a granularity of 0.0174 in $\eta$ and 0.0174\unit{rad} in $\phi$, and a length of 230\unit{mm} (25.8\unit{$\Xz$}). The crystals are installed using a quasi-projective geometry, with each one tilted by an angle of 3$^{\circ}$
relative to the projective axis that passes through the centre of CMS,
to minimize electron and photon passage through uninstrumented regions.
 The crystals are organized in 36 supermodules, 18 on each side of $\eta = 0$. Each supermodule contains 1\,700 crystals, covers 20 degrees in $\phi$, and is made of four modules along $\eta$.
 This structure has a few thin uninstrumented regions between the modules at $\abs{\eta}$ = 0, 0.435, 0.783, 1.131, and 1.479 for the end of the barrel and the transition to the endcaps, and at every 20$^\circ$ between supermodules in $\phi$.

The ECAL endcaps consist of a total of 14\,648 trapezoidal crystals with front-face transverse sections of $28.62 \times 28.62$\unit{mm$^2$}, and lengths of 220\unit{mm} (24.7\unit{$\Xz$}).
The crystals are grouped in  5$\times$5 arrays.
Each endcap is separated into two half-disks.
The crystals are installed within a quasi-projective geometry, with their
main axes pointing 1\,300\unit{mm} in $z$ beyond the
centre of CMS (-1\,300\unit{mm} for the endcap at $z>0$), resulting in tilts of 2 to 8$^{\circ}$ relative to the projective axis that passes through the centre of CMS.

The HCAL is a sampling calorimeter, with brass as the passive material, and plastic scintillator tiles serving as active material, providing coverage for $\abs{\eta} < 2.9$. The calorimeter cells are grouped in projective towers of granularity
0.087 in $\eta$ and 0.087\unit{rad} in $\phi$
in the barrel, and
0.17 in $\eta$ and 0.17\unit{rad} in $\phi$ in the endcaps, the exact granularity depending on $\abs{\eta}$. A more forward steel and quartz-fiber hadron calorimeter extends the coverage up to $\abs{\eta} < 5.2$.

\section{Data and simulation}
\label{sec:samples}

The data sample corresponds to an integrated luminosity of
19.7\fbinv~\cite{CMS-PAS-LUM-13-001}, collected at $\sqrt{s}=8\TeV$.
The results take advantage of the final calibration and
alignment conditions of the CMS detector,
obtained using the procedures described in Refs.~\cite{Chatrchyan:2013dga,Chatrchyan:2014wfa}.

The first level (L1) of the CMS trigger system, composed of specially designed hardware processors, uses information from the calorimeters
and muon detectors to select events of interest in 3.6\mus. The high-level trigger
(HLT) processor farm decreases the event rate from about 100\unit{kHz} (L1 rate) to about 400\unit{Hz} for data
storage~\cite{Chatrchyan:2008aa}.

The electron and photon candidates at L1 are based on ECAL trigger towers defined by arrays of $5\times 5$
crystals in the barrel and similar but more complex arrays of crystals in the endcaps. The central trigger tower with largest transverse energy $\ET= E\sin(\theta)$, together with its next-highest adjacent \ET tower form a L1 candidate. Requirements are set on the energy distribution among the central and neighbouring towers, on the amount of energy in the HCAL downstream the central tower, and on the \ET of the electron candidate.
The HLT electron candidates are constructed through associations of energy in ECAL crystals grouped into clusters (as discussed in Section~\ref{sec:recosupercluster}) around the corresponding L1 electron candidate and a reconstructed track
with direction compatible with the location of ECAL clusters. Their selection relies on
 identification and isolation criteria, together with minimal thresholds on \ET. The identification criteria are based on the
transverse profile of the cluster of energy in the ECAL, the amount of energy in the HCAL downstream the ECAL cluster, and the degree of association
between the track and the ECAL cluster. The isolation criterion makes use of the energies that surround the HLT electron candidate in the tracker, in the ECAL,
and in the HCAL.

The electron triggers, corresponding to the first selection step of most analyses using electrons, require the presence of at least one, two or three electron
candidates at L1 and HLT. Table \ref{tab:L1HLT} shows the lowest unprescaled L1 and HLT $\ET$ thresholds.

\begin{table}[htb]
\centering
\topcaption{\label{tab:L1HLT} Lowest, unprescaled $\ET$ threshold values in \GeV used for the L1 and HLT single-, double- and triple-electron triggers.}
\begin{tabular}{cccc}

\hline
 & Single & Double & Triple \\
\hline
L1  & 20 & 13, 7 & 12, 7, 5\\

HLT & 27 & 17, 8 & 15, 8, 5\\
\hline
\end{tabular}

\end{table}

The performance of electron reconstruction and selection is checked with events
selected by the double-electron triggers. These are mainly used to collect electrons from
Z boson decays, but also from low-mass resonances, usually at a smaller rate.
To study efficiencies, two additional dedicated double-electron triggers are introduced to maximize the number
of $\Z\to \Pep\Pem$ events collected
without biasing the efficiency of one of the electrons.
Both triggers require a tightly selected HLT electron candidate, and either a second looser HLT electron
or a cluster in the ECAL, that together have an invariant mass above 50\GeV.
Finally, studies of background distributions and misidentification probabilities are performed using events with $\Z\to \Pep\Pem$
or $\Z\to \mu^+\mu^-$ decays that contain a single additional jet misidentified as an electron,
the latter also using triggers with two relatively high-\pt muons.

Several simulated samples are exploited to optimize reconstruction and selection algorithms, to
evaluate efficiencies, and to compute systematic uncertainties.
The reconstruction algorithms are tuned mostly on simulated events with two back-to-back electrons with uniform distributions in $\eta$ and \pt,
with $1<\pt<100\GeV$.
Simulated Drell--Yan (DY) events, corresponding to generic quark + antiquark $\to \Z/\gamma^* \to \Pep\Pem$ production, are used to study various reconstruction and selection efficiencies.
 Results from the \MADGRAPH 5.1~\cite{madgraph} and \POWHEG~\cite{Nason:2004rx,Frixione:2007vw,Alioli:2010xd} generators
 are compared to evaluate systematic uncertainties.
These programs are
interfaced to \PYTHIA 6.426~\cite{pythia} for showering of partons and for jet fragmentation.
The \PYTHIA tune Z2*~\cite{FWD-11-003} is used to generate the underlying event.

Pileup signals caused by additional proton-proton interactions in the same
time frame of the event of interest are added to the simulation.
 There are on average approximately 15 reconstructed interaction vertices for each
recorded interaction,
corresponding to about 21 concurrent interactions per beam crossing.

The generated events are processed through a full \GEANTfour-based~\cite{Agostinelli:2002hh,GEANT}
detector simulation and reconstructed with the same algorithms as used for the data.
A realistic description of the detector conditions (tracker alignment, ECAL calibration and
alignment, electronic noise) is implemented in the simulation.
In addition, for some specific tasks requiring a more precise understanding of the detector,
a run-dependent version of the simulation is used to match the evolution of the
detector response with time observed in data.
This run-dependent simulation includes the evolution of the
 transparency of the crystals and of the noise in the ECAL, and
accounts in each event
for the effect of energy deposition from interactions in a significantly increased time window relative to the one containing the event of interest.

\section{Electron reconstruction}
\label{sec:reco}
Electrons are reconstructed by associating a track reconstructed in the silicon detector with a cluster of energy in the ECAL.
A mixture of a stand-alone approach~\cite{Baffioni:2006cd} and the complementary global ``particle-flow'' (PF) algorithm~\cite{CMS:2009nxa,CMS:2010byl} is used to maximize the performance.

This section specifies the algorithms used for clustering the energy deposited in the ECAL, building the electron track, and associating the two inputs to estimate the electron properties. Most of these algorithms have been optimized using simulation, and adjusted during data taking periods. A large part of the section is dedicated to the estimation of electron momentum, the chain of momentum calibration, and the performance of the momentum scale and resolution.

\subsection{Clustering of electron energy in the ECAL}
\label{sec:recosupercluster}

The electron energy usually spreads out over several crystals of the ECAL.
This spread can be quite small
 when electrons lose little energy via
bremsstrahlung before reaching ECAL. For example, electrons of 120\GeV in a test beam
that impinge directly on the centre of a crystal deposit about 97\% of the energy
in a 5$\times$5 crystal array~\cite{Adzic:2007mi}.
For an electron produced within CMS, the effect induced by radiation of photons can be large: on average, 33\% of the electron energy is radiated before it reaches the ECAL where the intervening material is minimal ($\eta \approx 0$), and about 86\% of its energy is radiated where the intervening material is the largest ($\abs{\eta} \approx 1.4$).

To measure the initial energy of the electron accurately, it is essential to collect the energy of the radiated
photons that mainly spreads along the $\phi$ direction because of the bending of the electron trajectory in the magnetic field. The spread in the $\eta$ direction is usually negligible, except for very low \pt ($\pt \lesssim  5\GeV$).
Two clustering algorithms, the ``hybrid'' algorithm in the barrel, and the ``multi-5$\times$5'' in the endcaps, are used for this purpose and are described in the following paragraphs.
For the clustering step, the $\eta$ and $\phi$ directions  and \ET are defined relative to the centre of CMS.

The hybrid algorithm exploits the geometry of the ECAL barrel (EB)
and properties of the shower shape, collecting the energy in a small
window in $\eta$ and an extended window in $\phi$~\cite{Bayatian:2006zz}.
The starting point is a seed crystal, defined as the one containing most of the energy deposited in any considered
region, that has a minimum \ET of $\ETxy{seed}{}> \ETxy{seed}{min}$. Arrays of $5\times1$ crystals in  $\eta\times\phi$
are added around the seed
crystal, in a range of $N_{\text{steps}}$ crystals in both directions of $\phi$, if
their energies exceed a minimum threshold of $E_{\text{array}}^{\text{min}}$.
The contiguous arrays are
grouped into clusters, with each distinct cluster required to have a seed
array with energy greater than a threshold of $E_{\text{seed-array}}^{\text{min}}$ in order to be collected in the final global cluster, called the
supercluster (SC).
These threshold values are summarized in Table~\ref{tab:hybridmulti_th}.
They were originally tuned to provide best ECAL-energy resolution for electrons with $\pt \approx 15\GeV$,
 but eventually minor adjustments were made to provide the current performance over a wider range of \pt values.

The multi-5$\times$5 algorithm is used in the ECAL endcaps (EE), where crystals are not arranged in an $\eta \times \phi$ geometry. It starts with the seed crystals, the ones with local maximal energy  relative to their four direct neighbours, which must fulfill an \ET requirement of $\ETxy{seed}{} > \ETxy{EEseed}{min}$. Around these seeds and beginning with the largest \ET, the energy is collected in clusters of 5$\times$5 crystals, that can partly overlap.
These clusters are then grouped into an SC if their total \ET satisfies $\ETxy{cluster}{} > \ETxy{cluster}{min}$, within a range in $\eta$ of $\pm \eta^{\text{range}}$, and a range in $\phi$ of $\pm \phi^{\text{range}}$ around each seed crystal.
 These threshold values are summarized in Table~\ref{tab:hybridmulti_th}.
 The
energy-weighted positions of all clusters belonging to an SC are then extrapolated
to the planes of the preshower, with the most energetic cluster used as reference point.
The maximum distance in $\phi$ between the clusters and their reference point are used to define the preshower clustering
range along $\phi$, which is then extended by $\pm0.15$\unit{rad}. The range along $\eta$ is set to 0.15 in both directions.
The preshower energies within these ranges around the reference point are then added to the SC energy.

\begin{table}[htb]
\centering
\topcaption{\label{tab:hybridmulti_th} Threshold values of parameters used in the hybrid superclustering algorithm in the barrel, and in the multi-5$\times$5 superclustering algorithm in the endcaps.}
\begin{tabular}{ll@{}p{1em}@{}ll}
\hline
\multicolumn{2}{c}{Barrel} & &  \multicolumn{2}{c}{Endcaps}\\

Parameter& Value&& Parameter & Value\\
\hline
$\ETxy{seed}{min}$ & 1\GeV&&$\ETxy{EEseed}{min}$ & 0.18\GeV\\
$E_\text{seed-array}^\text{min}$ & 0.35\GeV&&$\ETxy{cluster}{min}$ & 1\GeV \\
$E_\text{array}^\text{min}$ & 0.1\GeV&&$\eta^\text{range}$ & 0.07 \\
$N_\text{steps}$ & 17 ($\approx$0.3\unit{rad})&&$\phi^\text{range}$ & 0.3\unit{rad}\\
\hline
\end{tabular}
\end{table}

The SC energy corresponds to the sum of the energies of all its clusters.
The SC position is calculated as the energy-weighted mean of the cluster positions.
Because of the non-projective geometry of the crystals and the lateral shower shape, a simple energy-weighted mean of the crystal positions biases the estimated position  of each cluster towards the core of the shower. A better position estimate is obtained by taking a weighted mean, calculated using the logarithm of the crystal energy, and applying a correction based on the depth of the shower~\cite{Bayatian:2006zz}.

\begin{figure}[htb]
\centering
\includegraphics[width=0.49\textwidth]{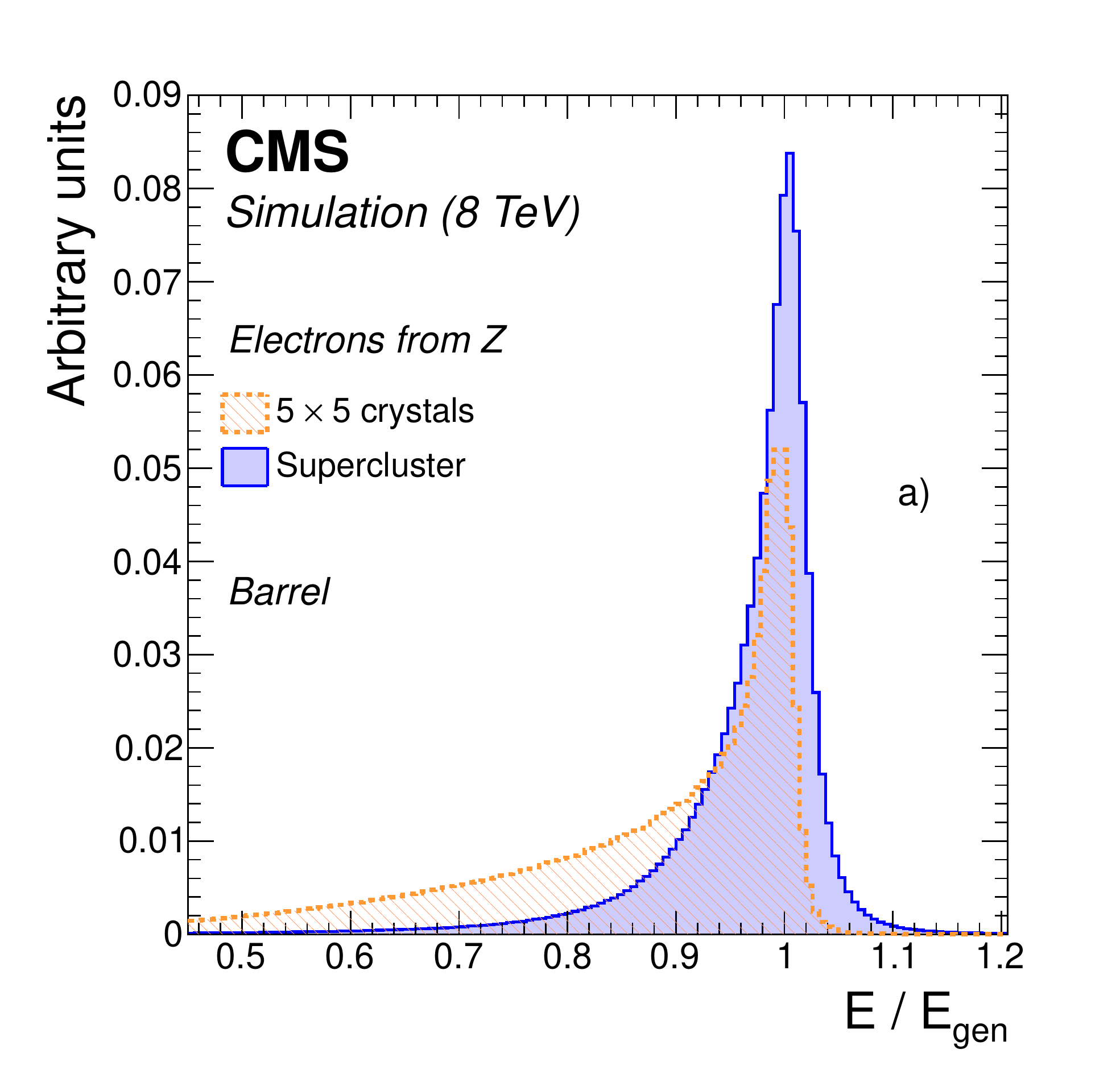}
\includegraphics[width=0.49\textwidth]{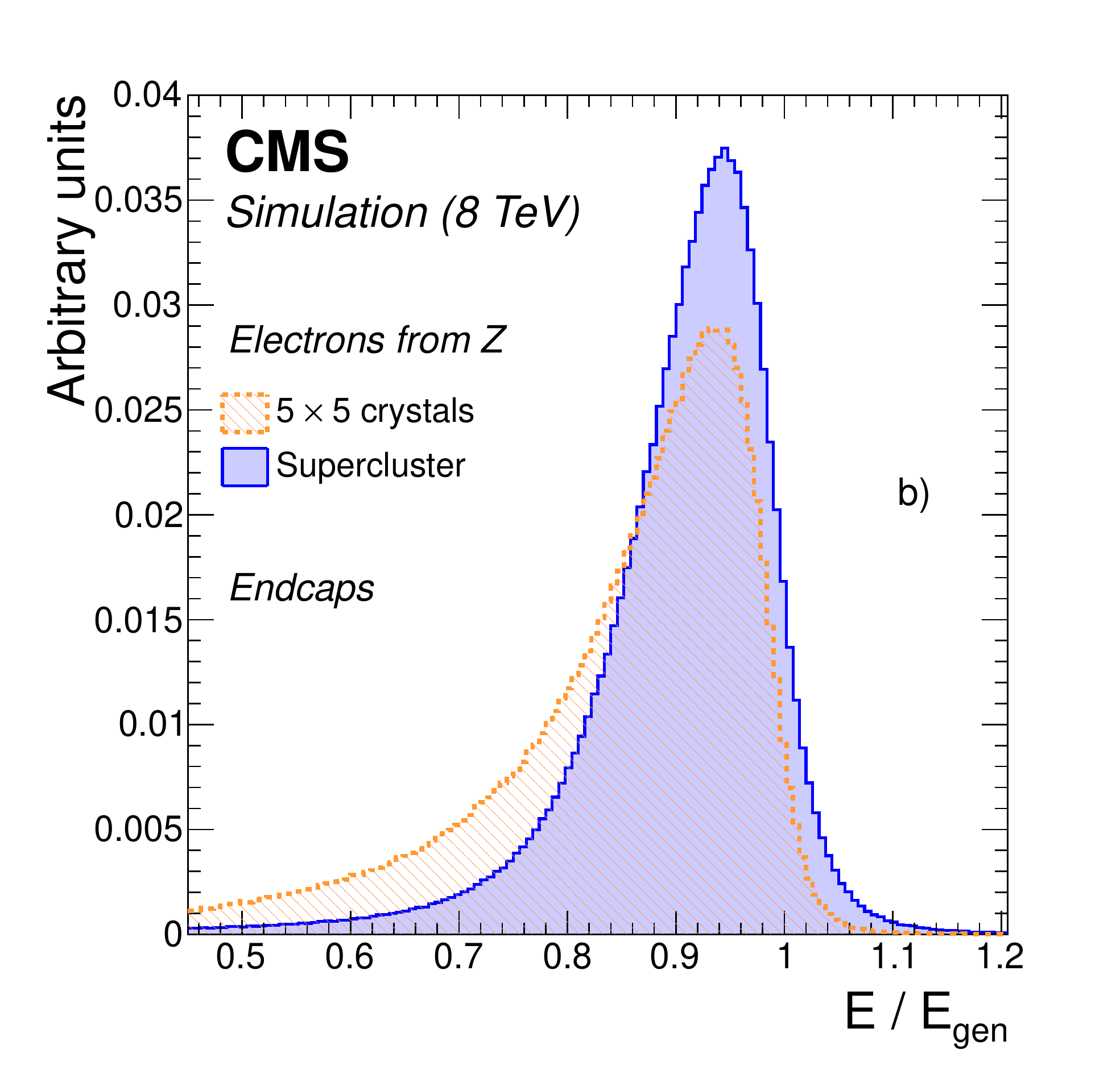}
\caption{\label{fig:SC} Comparison of the distributions of the ratio of reconstructed over generated energy for simulated electrons from \Z boson decays in a) the barrel, and b) the endcaps, for energies reconstructed using superclustering (solid histogram) and a matrix of 5$\times$5 crystals (dashed histogram). No energy correction is applied to any of the distributions.}

\end{figure}

Figure~\ref{fig:SC} illustrates the effect of superclustering on the recovery of energy from simulated $\Z \to \Pep\Pem$ events, comparing the energy reconstructed within the SC to the one reconstructed using a simple matrix of 5$\times$5 crystals around the most energetic crystal in a) the barrel and b) the endcaps. The tails at small values of the reconstructed energy $E$ over the generated one ($E_{\text{gen}}$) are seen to be significantly reduced through the superclustering.

In addition, as part of the PF-reconstruction algorithm, another clustering algorithm is introduced that aims at reconstructing the particle showers individually. The PF clusters are reconstructed by aggregating around a seed all contiguous crystals with energies of two standard deviations ($\sigma$) above the electronic noise
observed at the beginning of the data-taking run,
with $E_{\text{seed}}> 230\MeV$ in the barrel,
and $E_{\text{seed}}> 600\MeV$ or $\ETxy{seed}{}> 150\MeV$
in the endcaps.
An important difference relative to the stand-alone approach is that it is possible to share the energy of one crystal
among two or more clusters. Such clusters are used in different steps of electron reconstruction, and are
hereafter referred to as PF clusters.

\subsection{Electron track reconstruction}

Electron tracks can be reconstructed in the full tracker using the standard Kalman filter (KF) track reconstruction procedure used for all charged particles~\cite{trkpaper}. However, the large radiative losses for electrons in the tracker material compromise this procedure and lead in general to a reduced hit-collection efficiency (hits are lost when the change in curvature is large because of \bremns), as well as to a poor estimation of track parameters. For these reasons, a dedicated tracking procedure is used for electrons. As this procedure can be very time consuming, it has to be initiated
from seeds that are likely to correspond to initial electron trajectories. The key point for reconstruction is to collect the hits efficiently, while preserving an optimal estimation of track parameters over the large range of energy fractions lost through \bremns.

\subsubsection{Seeding}
\label{sec:seeding}
The first step in electron track reconstruction, also called seeding, consists of finding and selecting the two or three first hits in the tracker from which the track can be initiated. The seeding is of
primary importance since its performance greatly affects the reconstruction efficiency.
Two complementary algorithms are used and their results combined. The ECAL-based seeding starts from the SC energy and position, used to estimate the electron trajectory in the first layers of the tracker, and selects electron seeds from all the reconstructed seeds.
 The tracker-based seeding relies on tracks that are reconstructed using the general algorithm for charged particles, extrapolated towards the ECAL and matched to an SC.
These algorithms were first commissioned with data taken in 2010, using electrons from $\PW$ boson decays. The distributions in data were found to agree with expectations, even at low \pt, and tuning of the parameters obtained from simulation has been left essentially unchanged.

In the ECAL-based seeding, the SC energy and position are used to extrapolate the electron trajectory towards the collision vertex, relying on the fact that the energy-weighted average position of the clusters is on the helix corresponding to the initial electron energy, propagated through the magnetic field without emission of radiation.
The back propagation of the helix parameters through the magnetic field from the SC is performed for both positive and negative charge hypotheses. The intersections of helices with the innermost layers or disks predict the seeding hits. The SC are selected to limit the number of misidentified seeds using an \ET requirement of $\ET^{\mathrm{SC}} > 4\GeV$, together with a hadronic veto selection  of $H/E_{\mathrm{SC}} < 0.15$, with $E_{\mathrm{SC}}$ being the energy of the SC, and $H$ the sum of the HCAL tower energies within a cone of
$\Delta R = \sqrt{\smash[b]{(\Delta \eta) ^2 + (\Delta \phi) ^2}} = 0.15$
around the electron direction. This procedure reduces computing time.

On the other hand, tracker seeds are formed by combining pairs or triplets of hits with the vertices obtained from pixel tracks.
Combinations of first and second hits from tracker seeds are located in the barrel pixel layers (BPix), the forward pixel disks (FPix), and in the TEC to improve the coverage in the forward region. Only a subset of the seeds leads eventually to tracks.

For each SC, a seed selection is performed by comparing hits of each tracker seed and
 the SC-predicted hits within windows in $\phi$ and z (or in transverse distance $r$ in the forward regions where hits are only in the disks). The windows for the first and second hits are optimized using simulation to maximize the efficiency, while reducing the number of misidentified candidates to a level that can be handled within the CPU time available for electron track reconstruction. The overall efficiency of the ECAL-based seeding is $\approx$92\% for simulated electrons from \Z boson decay.

The windows for the first hit are wide,
and adapted to the uncertainty in the measurement of $\phi_{\mathrm{SC}}$, and the spread of the beam spot in $z$ ($\sigma_z$, changing with beam conditions, and typically about 5\unit{cm} in 2012).
The first $\phi$ window is chosen to depend on $\ET^{\mathrm{SC}}$, to reduce
the misidentified candidates, and asymmetrical, to take into account the uncertainty on the collected energy of the SC. When the first hit of a tracker seed is matched, the information is used to refine the parameters of the helix, and to search for a second-hit compatibility with more restricted windows. A seed is selected if its first two hits are matched with the predictions from the SC.

Tables~\ref{tab:ecalseed_firstwindow} and~\ref{tab:ecalseed_secondwindow} give the values of the first and second window acceptance parameters. For electrons with $5<\ET^{\mathrm{SC}}<35\GeV$, the first window size in $\phi$ ($\delta \phi$) is a function of $1/\ET^{\mathrm{SC}}$. The point given at 10\GeV represents the median of the dependence on $\ET^{\mathrm{SC}}$. 

\begin{table}[htb]
\centering
\topcaption{\label{tab:ecalseed_firstwindow}Values of the $\delta z$, $\delta r$ and $\delta \phi$ parameters used for the first window of seed selection, for three ranges of $\ET^{\mathrm{SC}}$, with $\sigma_z$ being the
standard deviation of the beam spot along the $z$ axis.
For electron candidates with negative charge, the same $\delta \phi$ window is used, but with opposite signs.}
\begin{tabular}{cccc}
\hline
 \multirow{2}{*}{$\ET^{\mathrm{SC}} (\GeVns{})$} &  $\delta z$ & $\delta r $& $\delta \phi$ (rad)  \\
             & (BPix) & (FPix or TEC) & (positive charge) \\
\hline

$\le$5 & $\pm 5 \sigma_z$ & $\pm 5 \sigma_z$ &$[-0.075;0.155]$ \\
 10 & $\pm 5 \sigma_z$& $\pm 5 \sigma_z$ &$[-0.046;0.096]$ \\
$\ge$35 & $\pm 5 \sigma_z$& $\pm 5 \sigma_z$ &$[-0.026;0.054]$ \\
\hline
\end{tabular}

\end{table}

\begin{table}[htb]
\centering
\topcaption{\label{tab:ecalseed_secondwindow}Values of the $\delta z$, $\delta r$ and $\delta \phi$ parameters used in different regions of the tracker for the second window of seed selection.}
\begin{tabular}{ccccc}
\hline
$\delta z$ (cm) & $\delta r$ (cm)& $\delta r$ (cm)&$\delta \phi$ (rad)  &  $\delta \phi$ (rad)\\
(BPix) & (FPix) &  (TEC) & (BPix) & (FPix or TEC)\\
\hline
$\pm 0.09$ & $\pm 0.15$ & $\pm 0.2$ &$\pm 0.004$ & $\pm 0.006$\\
\hline
\end{tabular}

\end{table}

\begin{figure}[htbp]
\centering
\includegraphics[width=0.49\textwidth]{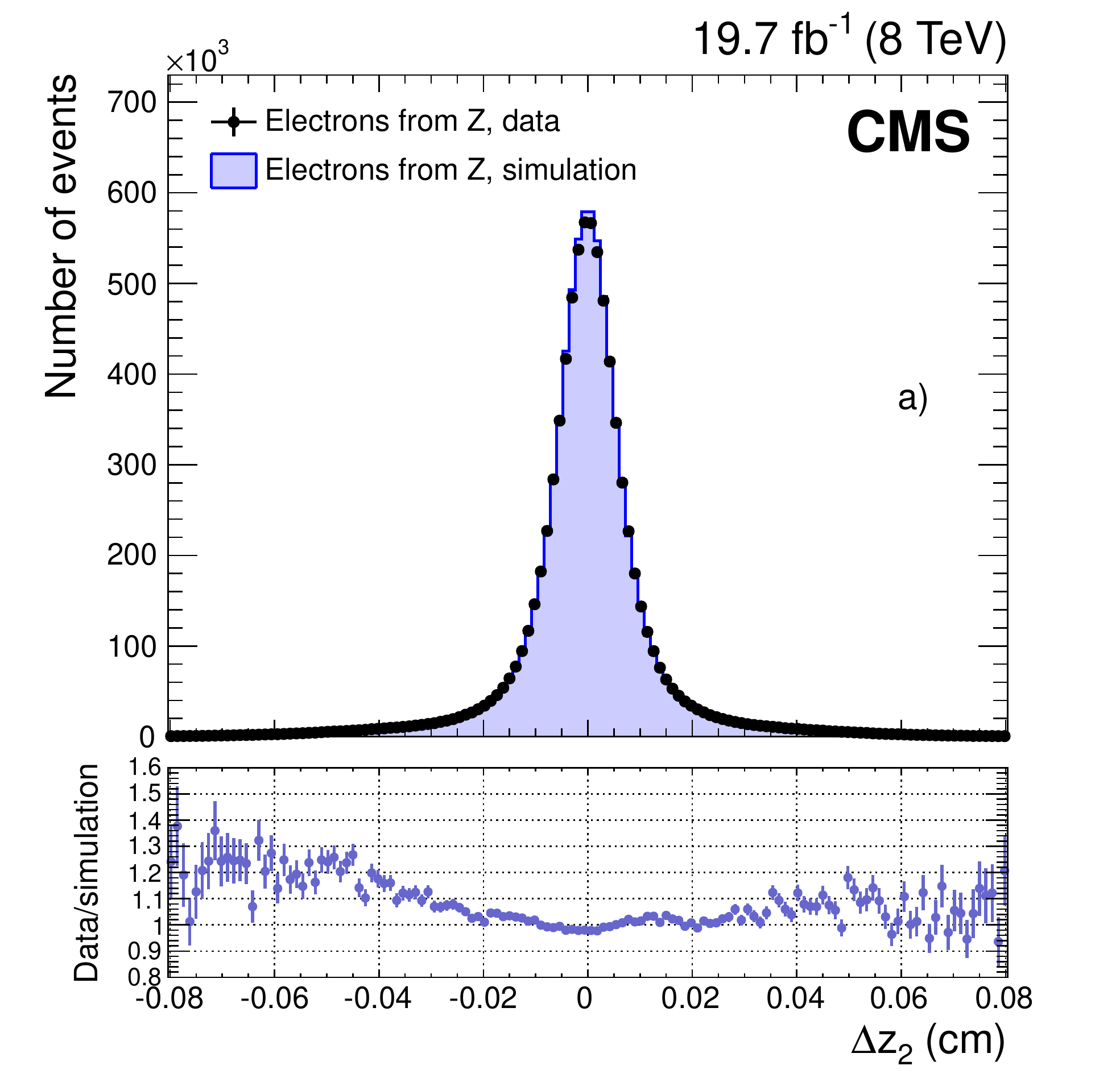}
\includegraphics[width=0.49\textwidth]{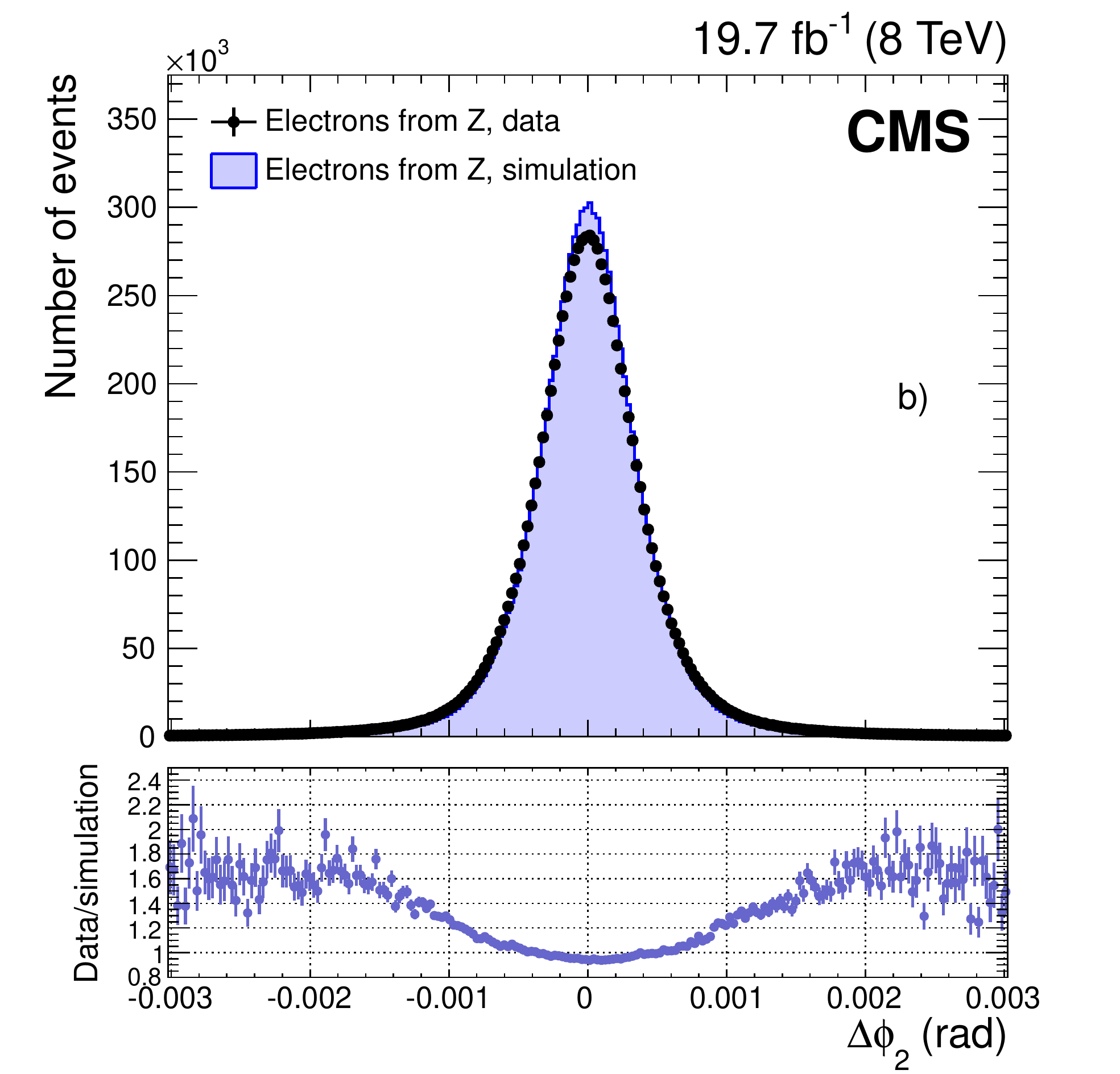}
\caption{\label{fig:seeding}  Distributions of the difference between predicted and measured
values of the $z_2$ and $\phi_2$ variables for
hits in the second window of the ECAL-based seeding, for electrons from $\Z \to \Pep\Pem$ decays in data (dots) and simulation (histograms): a) $\Delta z_2$ (barrel pixel), and b) $\Delta \phi_2$ (all tracker subdetectors). The data-to-simulation ratios are shown below the main panels.}

\end{figure}

Figure~\ref{fig:seeding} a) and b) show respectively the differences $\Delta z_2$ and $\Delta \phi_2$ between the measured and predicted
positions in $z$ (in the barrel pixels, BPix), and in $\phi$ (in all the tracker subdetectors), for the second window of each electron track seed,
in $\Z \to \Pep\Pem$ events in data and in simulation.
The distributions in data are slightly wider than in simulation, with the effect more pronounced in $\Delta \phi_2$, which is related directly to the difference in energy resolution between data and simulation.

Tracker-based seeding is developed as part of the PF-reconstruction algorithm, and complements the seeding efficiency, especially for low-\pt or nonisolated electrons, as well as for electrons in the barrel-endcap transition region.

The algorithm starts with tracks reconstructed with the KF algorithm.
The electron trajectory can be reconstructed accurately using the KF approach when \brem is negligible. In this case, the KF algorithm collects hits up to the ECAL, the KF track is well matched to the closest PF cluster, and its momentum is measured with good precision.
As a first step of the seeding algorithm, each KF track, with direction compatible with the position of the closest PF cluster that fulfills the matching-momentum criterion of $r_{\text{th}}< E/p <3$, has its seed selected for electron track reconstruction.
 The cutoff $r_{\text{th}}$ is set to 0.65 for electrons with
$2<\pt<6\GeV$, and to 0.75 for electrons with $\pt \geq  6\GeV$.

For tracks that fail the above condition, indicating potential presence of significant
bremsstrah\-lung, a second selection is attempted. As the KF algorithm cannot follow the change of curvature of the electron trajectory because of the bremss\-trahlung, it either stops collecting hits, or keeps collecting them, but with a bad quality identified through a large value of the $\chi^2_{\mathrm{KF}}$. The KF tracks with a small number of hits or a large $\chi^2_{\mathrm{KF}}$ are therefore refitted using a dedicated Gaussian sum filter (GSF)~\cite{Adam:paper},
as described in Section~\ref{tracking}.

The number of hits and the quality of the KF
track $\chi^2_{\mathrm{KF}}$, the quality of the GSF track $\chi^2_{\mathrm{GSF}}$, and
the geometrical and energy matching of the ECAL and tracker information are used in a multivariate (MVA) analysis~\cite{TMVA}
to select the tracker seed as an electron seed.

The electron seeds found using the two algorithms are combined, and the overall efficiency of the seeding is predicted $>$95\% for simulated electrons from \Z boson decay.

\subsubsection{Tracking}
\label{tracking}
\label{sec:fbrem}

\begin{figure}[htbp]
\centering
\includegraphics[width=0.49\textwidth]{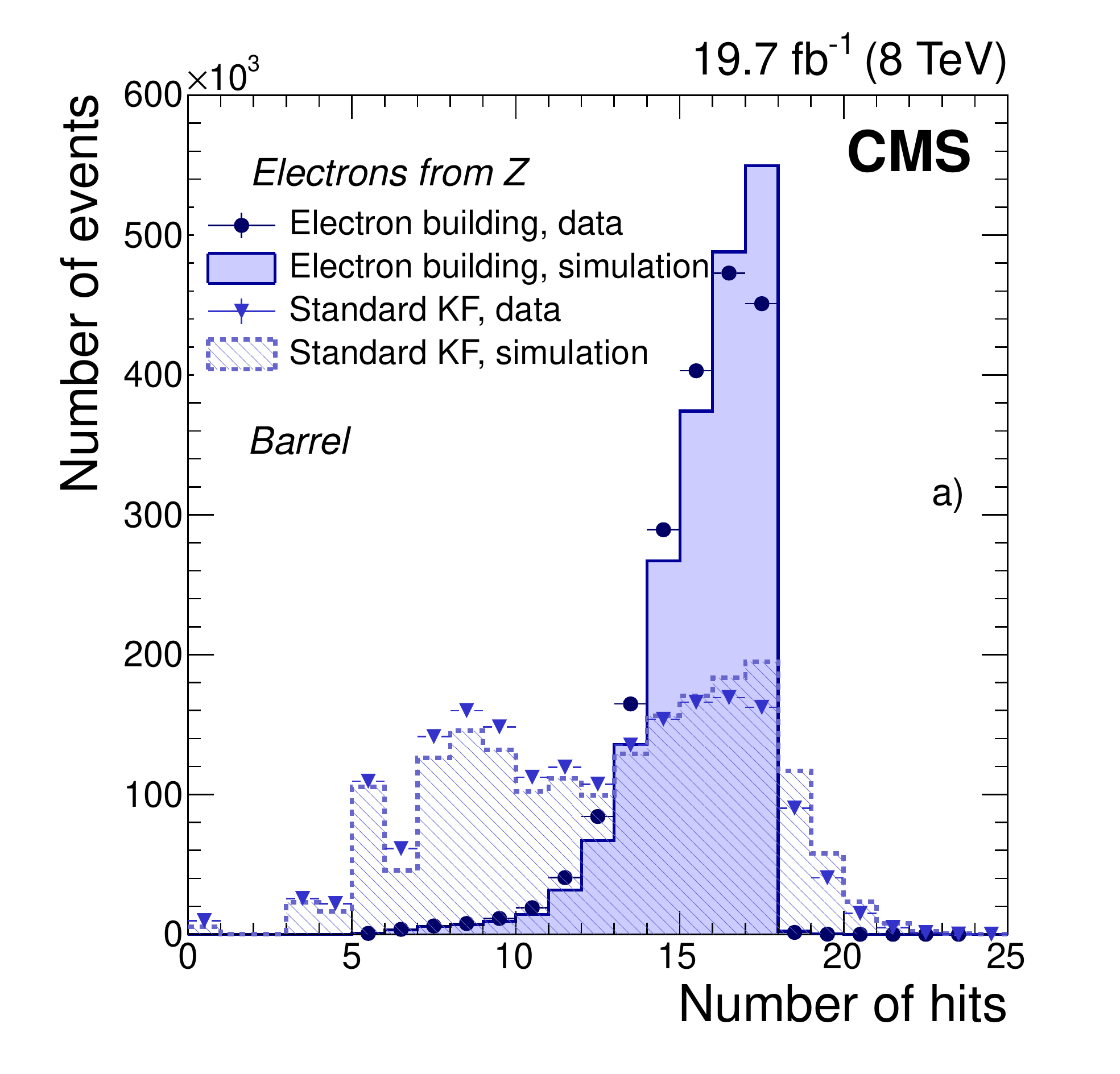}
\includegraphics[width=0.49\textwidth]{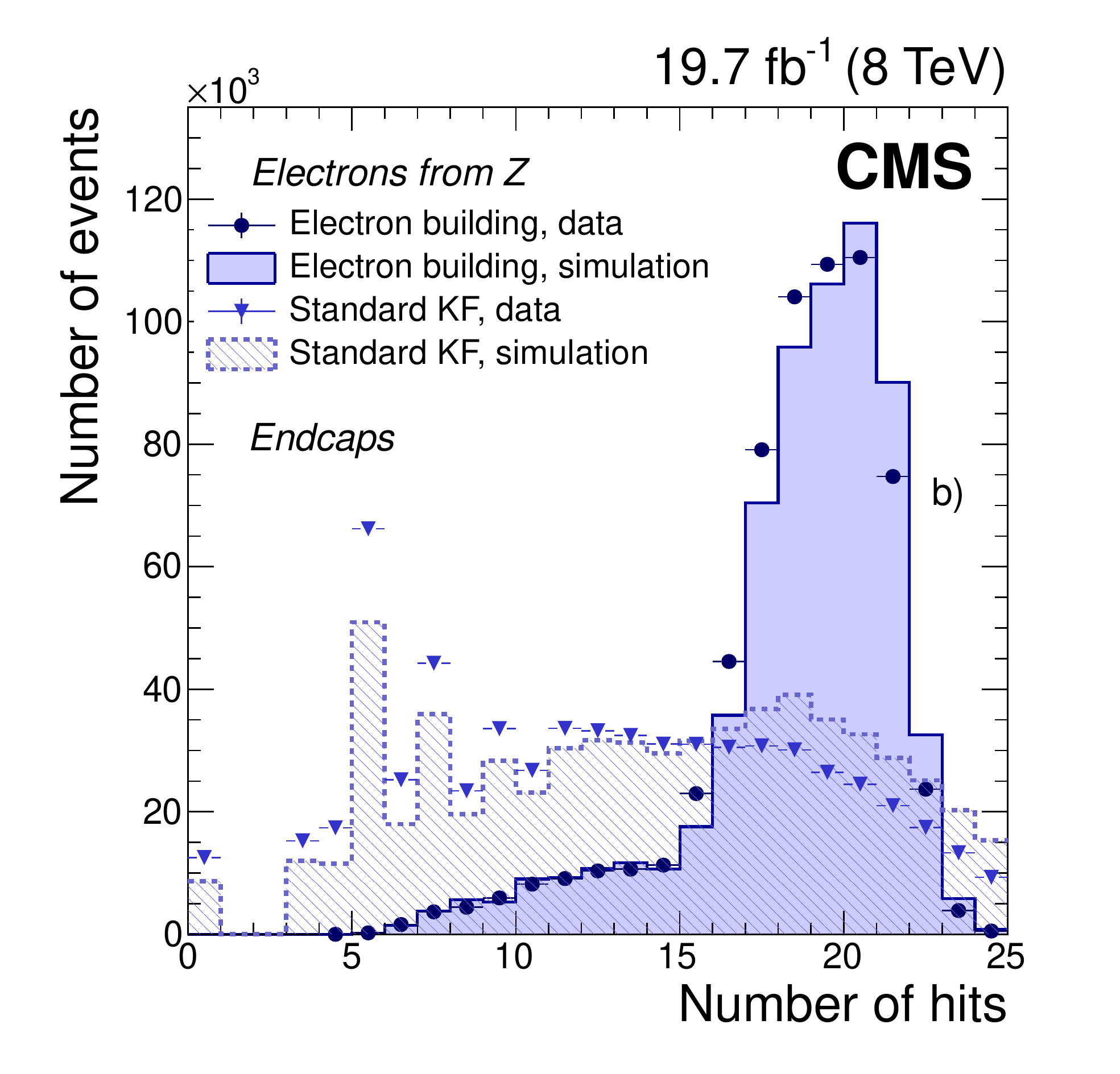}
\caption{\label{fig:nhits} Comparison of the number of hits collected with the dedicated electron building and KF procedures in data (symbols) and in simulation (histograms), for electrons obtained using a $\Z \to \Pep\Pem$ selection, a) in the barrel, and b) in the endcaps.}

\end{figure}

The selected electron seeds are used to initiate electron-track building, which is followed by track fitting. The track building is based on the combinatorial KF method, which for each electron seed proceeds iteratively from the track parameters provided in each layer, including one-by-one the information from each successive layer~\cite{trkpaper}. The electron energy loss is modelled through a Bethe--Heitler function. To follow the electron trajectory in case of \brem and to maintain good efficiency, the compatibility between the predicted and the found hits in each layer is chosen not to be too restrictive.
When several hits are found compatible with those predicted in a layer, then several trajectory candidates are created and developed, with a limit of five candidate trajectories for each layer of the tracker.
At most, one missing hit is allowed for an accepted trajectory candidate, and, to avoid including hits from converted bremsstrahlung photons in the reconstruction of primary electron tracks, an increased $\chi^2$ penalty is applied to trajectory candidates with one missing hit.
Figure \ref{fig:nhits} shows the number of hits collected using this procedure for electrons from a \Z boson sample in data and in simulation, compared with the KF procedure used for all the other charged particles in the barrel and in the endcaps.
The \Z boson selections in data and in simulation require both decay electrons to satisfy $\pt>20\GeV$,
several criteria pertaining to isolation and to rejection of converted photons,
and a condition of $\abs{m_{\Pep\Pem} - m_{\Z}} < 7.5\GeV$ on their invariant mass.
The structure in the figure reflects the geometry of the tracker. This comparison shows that shorter electrons tracks are obtained using the standard KF than using the dedicated electron building. The number of hits for the KF procedure is set to zero when there is no KF track associated with the electron. While the general behaviour is well reproduced, disagreement is observed between data and simulation  due to an imperfect description of the active tracker sensors in the simulation.

\begin{figure}[htbp]
\centering
\includegraphics[width=0.49\textwidth]{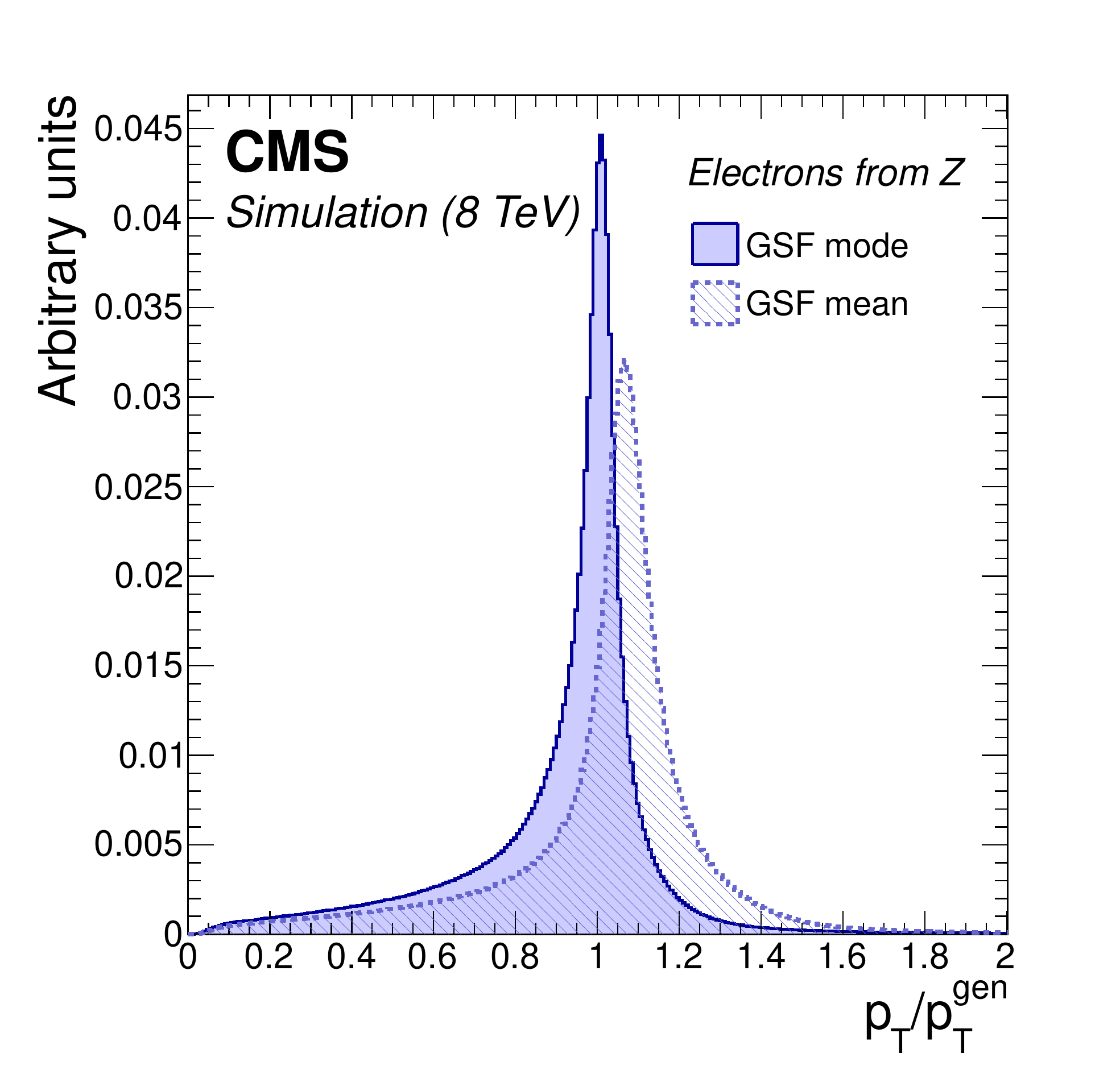}
\caption{\label{fig:pmode} Distribution of the ratio of reconstructed over generated electron \pt in simulated $\Z \to \Pep\Pem$ events, reconstructed through the most probable value of the GSF track components (solid histogram), and its weighted mean (dashed histogram). }

\end{figure}

Once the hits are collected,
a GSF fit is performed to estimate the track parameters. The energy loss in each layer is approximated by a mixture of Gaussian distributions.
A weight is attributed to each Gaussian distribution that describes the associated probability.
Two estimates of track properties are usually exploited at each measurement point that correspond either to the weighted mean of all the components, or to their most probable value (mode).
The former provides an unbiased average, while the latter peaks at the generated value and has a smaller standard deviation for the core
of the distribution~\cite{Baffioni:2006cd}. This is shown in Fig.~\ref{fig:pmode}, where the ratio $\pt / \pt^{\text{gen}}$
is compared for the two estimates, for simulated electrons from \Z boson decays. For these reasons, the mode
estimate is chosen to characterize all the parameters of electron tracks.

\begin{figure}[htb]
\centering
\includegraphics[width=0.49\textwidth]{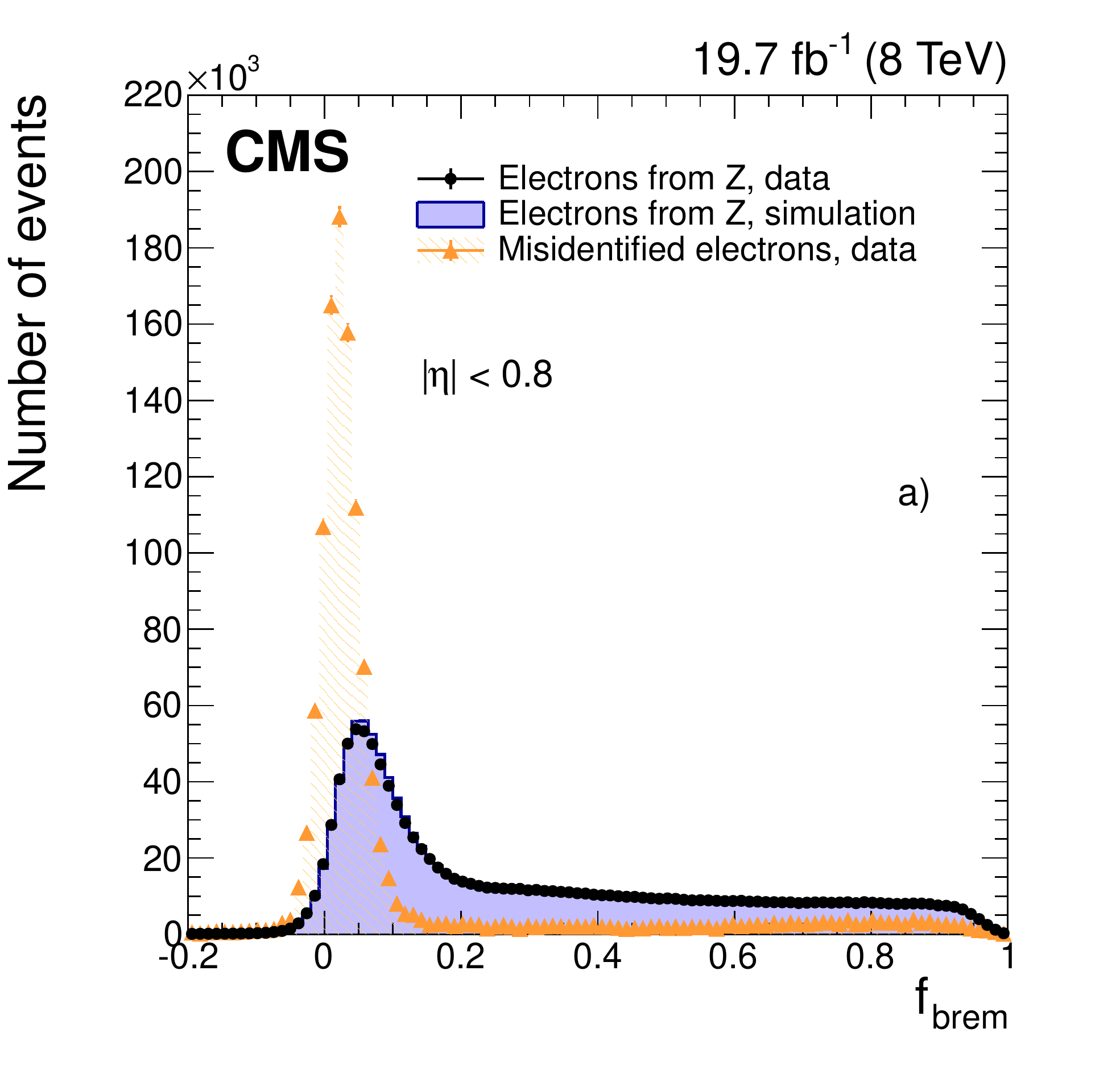}
\includegraphics[width=0.49\textwidth]{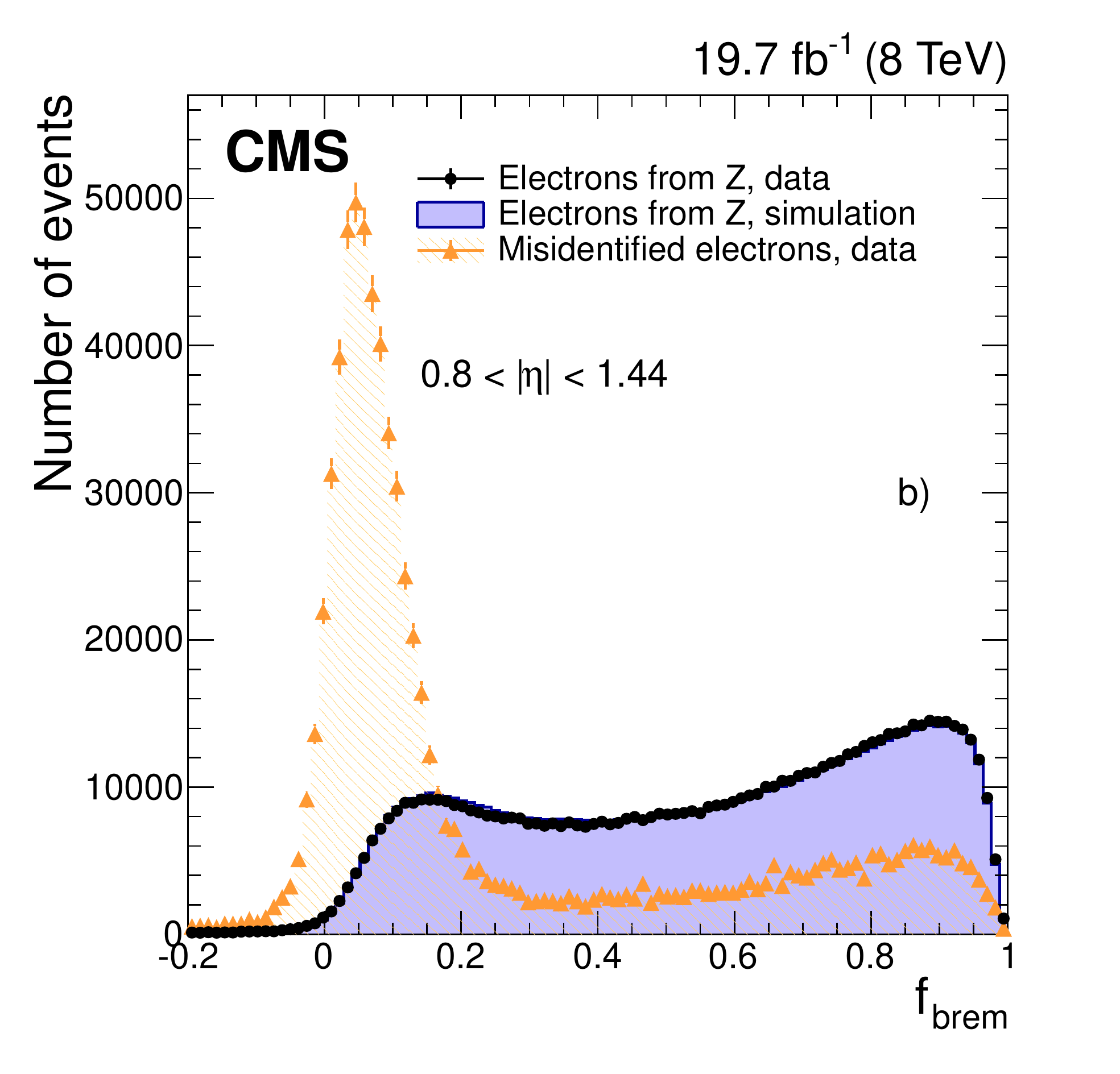}\\
\includegraphics[width=0.49\textwidth]{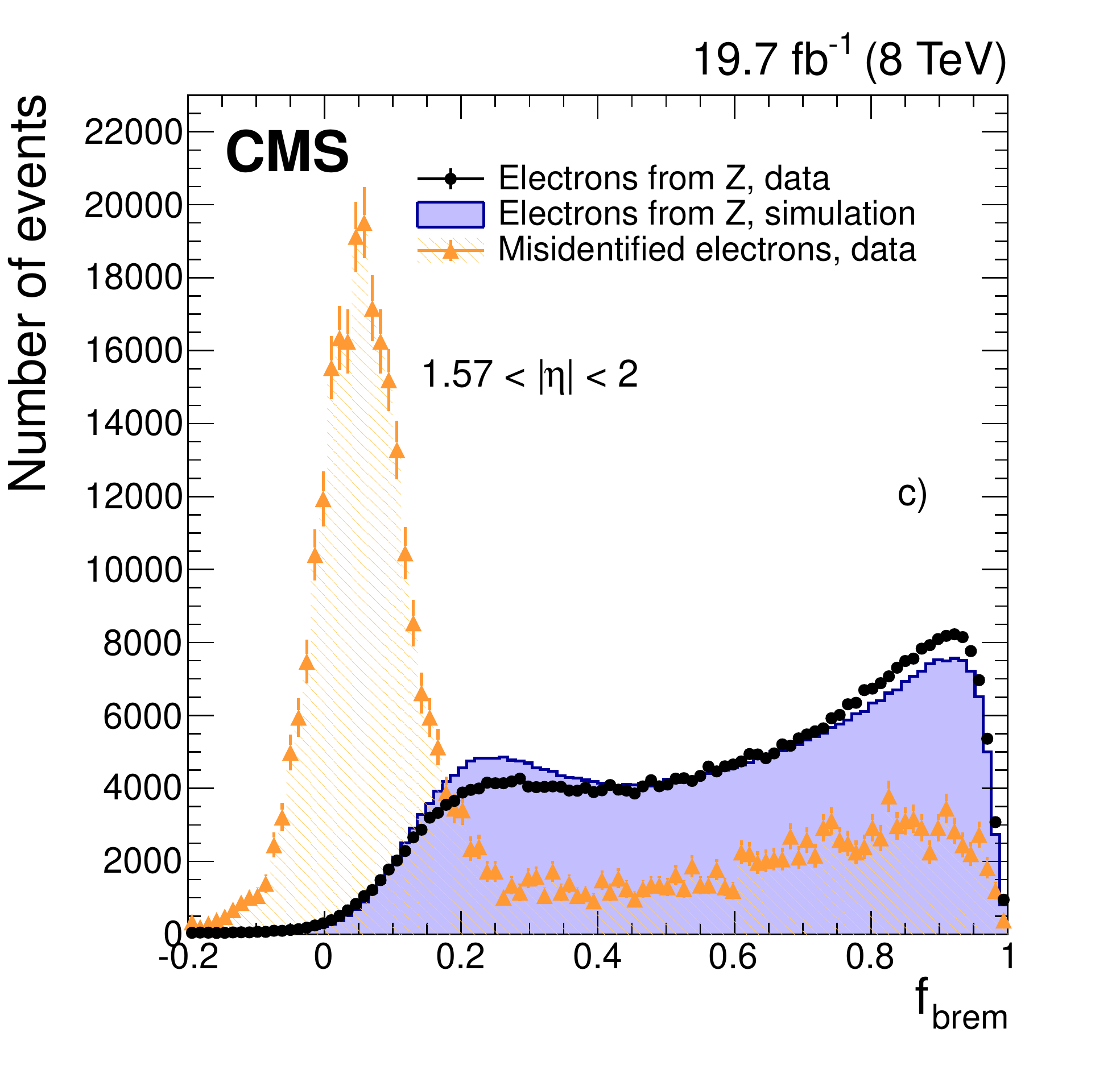}
\includegraphics[width=0.49\textwidth]{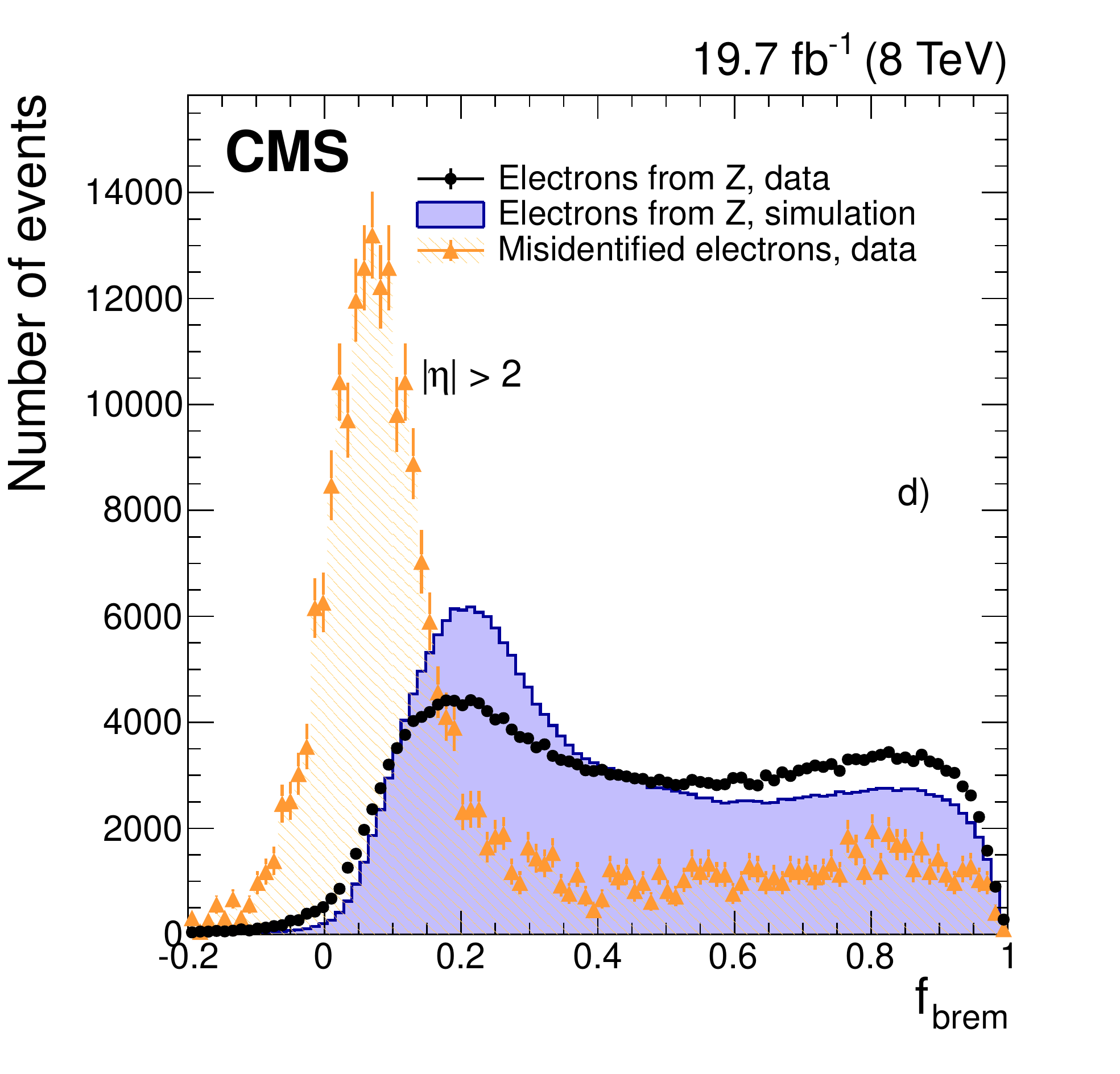}
\caption{\label{fig:fbrem}
Distribution of $f_{\text{brem}}$ for electrons from $\Z \to \Pep\Pem$ data (dots) and simulated (solid histograms) events, and from background-enriched events in data (triangles),
in a) the central barrel $\abs{\eta} < 0.8$, b) outer barrel $0.8 < \abs{\eta} < 1.44$, c) endcaps $1.57 < \abs{\eta} < 2$, and d) endcaps $\abs{\eta} > 2$.
The distributions are normalized to the area of the $\Z \to \Pep\Pem$ data distributions.}

\end{figure}

This procedure of track building and fitting provides electron tracks that can be followed up to the ECAL, and thereby extract track parameters at the surface of the ECAL. The fraction of energy lost through \brem is estimated using the momentum at the point of closest approach to the beam spot ($p_{\text{in}}$), and the momentum extrapolated to the surface of the ECAL from the track at the exit of the tracker ($p_{\text{out}}$), and is defined as $f_\text{brem} = [p_\text{in} - p_\text{out}]/ p_\text{in}$. This variable is used to estimate the
electron momentum, and it enters into the identification procedure. In Fig.~\ref{fig:fbrem}, this observable is shown for $\Z \to \Pep\Pem$ data and simulated events, as well as for misidentified electron candidates from jets in data enriched in Z+jets, in four regions of the ECAL barrel and endcaps. Each distribution is normalized to the area of the $\Z \to \Pep\Pem$ data.
As mentioned above, the \Z boson selections in data and in simulation require both decay electrons to satisfy $\pt>20\GeV$,
as well as several isolation
and photon conversion rejection criteria, and a condition  of $\abs{m_{\Pep\Pem} - m_{\Z}} < 7.5\GeV$ on their invariant mass.
The sample of misidentified electrons is obtained by selecting nonisolated electron candidates with $\pt>20\GeV$, in events selected
with a pair of identified leptons (electrons or muons) with invariant mass compatible with that of the \Z boson, and an imbalance
in transverse momentum smaller than 25\GeV.
When a \brem photon is emitted prior to the first three hits in the tracker, leading to an underestimation of $p_{\text{in}}$, or when the amount of radiated energy is very low, the $p_{\text{out}}$ and $p_{\text{in}}$ have similar values, and $p_{\text{out}}$ can be measured to be greater than $p_{\text{in}}$, leading thereby to negative values of $f_{\text{brem}}$.
In the central barrel region, the amount of intervening material is small, and the \brem fraction peaks at low values, contrary to the outer region, where the amount of material is large and leads to a sizable population of electrons emitting high fractions of their energies through \bremns. For the background, chiefly composed of hadron tracks misidentified as electrons, the \brem fraction generally peaks at very small values. The increased contribution of background at high values of \brem fraction that can be observed in Figs.~\ref{fig:fbrem}b), c), and d), is ascribed to residual early photon conversions and nuclear interactions within the tracker material.

The disagreement observed between data and simulation in the endcap region is attributed to
an imperfect modelling of the material in
simulation.
In fact, the $f_{\text{brem}}$ variable is a perfect tool for accessing the intervening
material, and a direct comparison of the mean value of $f_{\text{brem}}$ in data and in simulation in
narrow bins of $\eta$ indicates that the description of the material in certain regions is imperfect. For example,
a localized region near $\abs{\eta} \approx 0.5$ where there are complicated connections of the TOB to its wheels, and beyond $\abs{\eta} \approx 0.8$ where there is a region of inactive material, do not have the material properly represented in the simulation~\cite{photonPaper}.
The observed difference between data and simulation, relevant for updating the
simulated geometry in future analyses, is taken into account in the analysis of 8 TeV data, through specific
corrections applied to the electron momentum scale, resolution, and identification and reconstruction efficiencies extracted from $\Z\to \Pep\Pem$ events, as discussed in Sections~\ref{sec:scale_perf} and~\ref{sec:TandP}.

\subsection{Electron particle-flow clustering}
The PF clustering of electrons is driven by GSF tracks, and is independent of the way they are seeded. For each GSF track, several PF clusters, corresponding to the electron at the ECAL surface and the \brem photons emitted along its trajectory, are grouped together. The PF cluster corresponding to the electron at the ECAL surface is the one matched to the track at the exit of the tracker.
Since most of the material is concentrated in the layers of the tracker, for each layer a straight line is extrapolated to the ECAL, tangent to the electron track, and each matching PF cluster is added to the electron PF cluster. Most of the \brem photons are recovered in this way, but some
converted photons can be missed. For these photons, a specific procedure selects displaced KF tracks through a dedicated MVA algorithm, and kinematically associates them with the PF clusters.
In addition, for ECAL-seeded isolated electrons, any PF clusters  matched geometrically with the hybrid or multi-5$\times$5 SC are also added to the PF electron cluster.

\subsection{Association between track and cluster}
\label{sec:association}
The electron candidates are constructed from the association of a GSF track and a cluster in the ECAL. For ECAL-seeded electrons, the ECAL cluster associated with the track is simply the one reconstructed through the hybrid or the multi-5$\times$5 algorithm that led to the seed. For electrons seeded only through the tracker-based approach, the association is made with the electron PF cluster.

The track-cluster association criterion, just like the seeding selection, is designed to preserve highest efficiency and reduced misidentification probability, and it is therefore not very restrictive along the direction of the track curvature affected by \bremns. For ECAL-seeded electrons, this requires a  geometrical matching between the GSF track and the SC, such as:
\begin{itemize}
\item $\abs{\Delta \eta} = \abs{\eta_{\mathrm{SC}} - \eta_{\text{in}}^{\text{extrap}}} < 0.02$, with $\eta_{\mathrm{SC}}$ being the SC energy-weighted position in $\eta$, and $\eta_{\text{in}}^{\text{extrap}}$ the track $\eta$ extrapolated from the innermost track position and direction to the position of closest approach to the SC,
\item $\abs{\Delta \phi} = \abs{\phi_{\mathrm{SC}} - \phi_{\text{in}}^{\text{extrap}}} < 0.15$, with analogous definitions for $\phi$.
\end {itemize}
For tracker-seeded electrons, a global identification variable is defined using an MVA
technique that combines information on track observables (kinematics, quality, and KF track), the electron PF cluster observables (shape and pattern), and the association between the two (geometric and kinematic observables).
For electrons seeded only through the tracker-based approach, a weak selection is applied on this global identification variable. For electrons seeded through both approaches, a logical OR is applied on the two selections.

 The overall efficiency is $\approx$93\% for electrons from \Z decay, and the reconstruction efficiency measured in data is compared to simulation in Section~\ref{recoEff}.

\subsection{Resolving ambiguity}
Bremsstrahlung photons can convert into $\Pep\Pem$ pairs within the tracker and be reconstructed as electron candidates.  This is particularly important for $\abs{\eta}> 2$, where electron seeds can be used from layers of the tracker endcap that are located far from the interaction vertex and away from the bulk of the material. In such topologies, a single electron seed can often lead to several reconstructed tracks, especially when a \brem photon carries a significant fraction of the initial electron energy, so that the hits corresponding to the converted photon are located close to the expected position of the initial track. This creates ambiguities in electron candidates, when two nearby GSF tracks share the same SC.

To resolve this problem,
the following criteria are used, based on the small probability of a \brem photon to convert in the tracker material just after its point of emission.
The number of missing inner hits is obtained from the intersections between the track trajectory and the active inner layers.
\begin{itemize}
\item When two GSF tracks have a different number of missing inner hits, the one with the smallest number is retained.
\item When the number of missing inner hits is the same, and both candidates have an ECAL-based seed, the one with $E_{\mathrm{SC}}/p$ closest to unity is chosen, where $p$ is the track momentum evaluated at the interaction vertex.
\item The same criterion is also applied when both candidates have the same number of missing inner hits and just tracker-based seeds.
\item When the number of missing inner hits is the same, but only one candidate is just tracker-seeded, the track with an ECAL-based seed is chosen, because
the tracks from tracker-based seeds have a higher chance to be contaminated by track segments from conversions.
\end{itemize}

\subsection{Relative ECAL to tracker alignment with electrons}

Electrons are also used to probe subtle detector effects such as the ECAL alignment relative to the tracker. The tracker was first aligned using cosmic rays before the start of LHC operations, and constantly refined using proton-proton collisions, reaching an accuracy $<10$\unit{$\mu$m}~\cite{Chatrchyan:2014wfa}.
The relative alignment of the tracker to the ECAL for 2012 data is obtained using electrons from \Z boson decays. Tight identification and isolation criteria are applied to both electrons with $\ET>30\GeV$, and the dielectron invariant mass is required to be $\abs{m_{\Pep\Pem} - m_{\Z}} < 7.5\GeV$, to ensure a high signal purity of 97\%, needed for the alignment procedure. In addition, to disentangle \brem effects from position reconstruction, only electrons with little \brem and best energy measurement are considered. The distances $\Delta\eta$ and  $\Delta\phi$,
defined in Section~\ref{sec:association},
are compared between data and simulation, the ECAL being aligned with the tracker in the simulation. The position of each supermodule in the barrel and each half-disk in the endcaps is measured relative to the tracker by minimizing the differences between data and simulation as a function of the alignment coefficients.
Residual misalignments lower than $2\times10^{-3}$\unit{rad} in $\Delta\phi$ and $2\times10^{-3}$ units in $\Delta\eta$, are obtained using this procedure, which is compatible with expectations from simulation.

\subsection{Charge estimation}
The measurement of the electron charge is affected by \brem  followed by photon conversions. In particular, when the \brem photons convert upstream in the detector, they lead to very complex hit patterns, and the contributions from conversions can be wrongly included in the fitting of the electron track.

 A natural choice for a charge estimate is the sign of the GSF track curvature, which unfortunately can be altered by the misidentification probability in presence of conversions, especially for $\abs{\eta} > 2$, where it can reach about 10\% for reconstructed electrons from \Z boson decay without further selection.
This is improved by combining two other charge estimates, one that is based on the associated KF track matched to a GSF track when at least one hit is shared in the innermost region, and the second one that is evaluated using the SC position, and defined as the sign of the difference in $\phi$ between the vector joining the beam spot to the SC position and the vector joining the beam spot and the first hit of the electron GSF track.

The electron charge is defined by the sign shared by at least two of the three estimates, and is referred to as the ``majority method''. The misidentification probability of this algorithm is predicted by simulation to be 1.5\% for reconstructed electrons from \Z boson decays without further selection, offering thereby a global improvement on the charge-misidentification probability of about a factor 2 relative to the charge given by the GSF track curvature alone. It also reduces the misidentification probability at very large $\abs{\eta}$, where it is predicted to be $<$7\% for such electrons. Higher purity can be obtained by requiring all three measurements to agree, termed the ``selective method''. This yields a misidentification probability of $<$0.2\% in the central part of the barrel, $<$0.5\% in the outer part of the barrel, and $<$1.0\% in the endcaps, which can be achieved at the price of an efficiency loss that depends on \pt, but is typically $\approx$7\% for electrons from \Z boson decays. The selective algorithm is used mainly in analyses where the charge estimate is crucial, for example in the study of charge asymmetry in inclusive W boson production~\cite{Chatrchyan:2012xt}, or in searches for supersymmetry using same-charge dileptons~\cite{Chatrchyan:2013fea}.

The charge misidentification probability decreases strongly when the identification selections become more restrictive, mainly
because of the suppression of photon conversions.
Table~\ref{tab:chmisid} gives the measurement in data and simulation of the charge misidentification probability that can be achieved for a tight selection of electrons (corresponding to the HLT criteria) from $\Z\to\Pep\Pem$ decays in the barrel and in the endcaps, for the majority and the selective methods.
These values are estimated by comparing the number of same-charge and opposite-charge dielectron pairs
that are extracted from a fit to the dielectron invariant mass.
The misidentification probability is significantly reduced relative to the one at the reconstruction level.
A good agreement is found between data and simulation in both ECAL regions and for both charge-estimation methods.

\begin{table}[htb]
\centering
\topcaption{\label{tab:chmisid} Charge misidentification probability for a tight selection of electrons from $\Z\to\Pep\Pem$ decays in the barrel and in the endcaps, for the majority and for the selective methods used to estimate electron charge. Only statistical uncertainties are shown in the table.}
\begin{tabular}{ccc@{}p{1em}@{}cc}
\hline
& \multicolumn{2}{c}{Barrel}  &&  \multicolumn{2}{c}{Endcaps}\\
\cline{2-3}\cline{5-6}
Method&  Simulation & Data &&  Simulation & Data \\
\hline
majority&0.13 $\pm$ 0.01\% & 0.14 $\pm$ 0.01\% &&1.4 $\pm$ 0.2\% & 1.6 $\pm$ 0.2\%\\
selective &0.017 $\pm$ 0.002\% & 0.020 $\pm$ 0.002\%&&0.21 $\pm$ 0.02\% & 0.23 $\pm$ 0.02\% \\
\hline
\end{tabular}
\end{table}

\subsection{Estimation of electron momentum }
The electron momentum is estimated using a combination of the tracker and ECAL measurements. As for all electron observables, it is particularly sensitive to the pattern of brems\-strahlung photons and their conversions. To achieve the best possible measurement of electron momentum, electrons are classified according to their brems\-strahlung pattern, using observables sensitive to the emission and conversion of photons along the electron trajectory. The SC energy is corrected and calibrated, then the combination between the tracker and ECAL measurements is performed.

\subsubsection{Classification}
 For most of the electrons, the bremsstrah\-lung fraction in the tracker $f_{\text{brem}}$, defined in Section~\ref{sec:fbrem}, is
complemented by the \brem fraction in the ECAL, defined as $f_{\text{brem}}^{\mathrm{ECAL}} = [E_{\mathrm{SC}}^{\mathrm{PF}} - E_{\text{ele}}^{\mathrm{PF}}]/ E_\mathrm{SC}^\mathrm{PF}$,
where $E_{\mathrm{SC}}^{\mathrm{PF}}$ and $E_{\text{ele}}^{\mathrm{PF}}$ are the SC energy and the electron-cluster energy measured with the PF algorithm, that correspond respectively to the initial and final electron energies. The number of clusters in the SC is also used in the classification process.

Electrons are classified in the following categories:
\begin{itemize}
\item ``Golden'' electrons are those with little \brem and consequently provide the most accurate estimation of momentum. They are defined by an SC with a single cluster and $f_{\text{brem}}<0.5$.
\item ``Big-brem'' electrons have a large amount of \brem radiated in a single step, either very early or very late along the electron trajectory. They are defined by an SC with a single cluster and $f_{\text{brem}}>0.5$.
\item ``Showering'' electrons have a large amount of \brem radiated all along the electron trajectory, and are defined by an SC containing several clusters.
\end{itemize}
In addition, two special electron categories are defined. One is termed ``crack'' electrons, defined as electrons with the SC seed crystal adjacent to an $\eta$ boundary between the modules of the ECAL barrel, or between the ECAL barrel and endcaps, or at the high $\abs{\eta}$ edge of the endcaps. The second category, called ``bad track'', requires a calorimetric \brem fraction that is significantly larger than the track \brem fraction ($f_{\text{brem}}^{\mathrm{ECAL}} - f_{\text{brem}} > 0.15$), which identifies electrons with a poorly fitted track in the innermost part of the trajectory.

\begin{figure}[htb]
\centering
\includegraphics[width=0.49\textwidth]{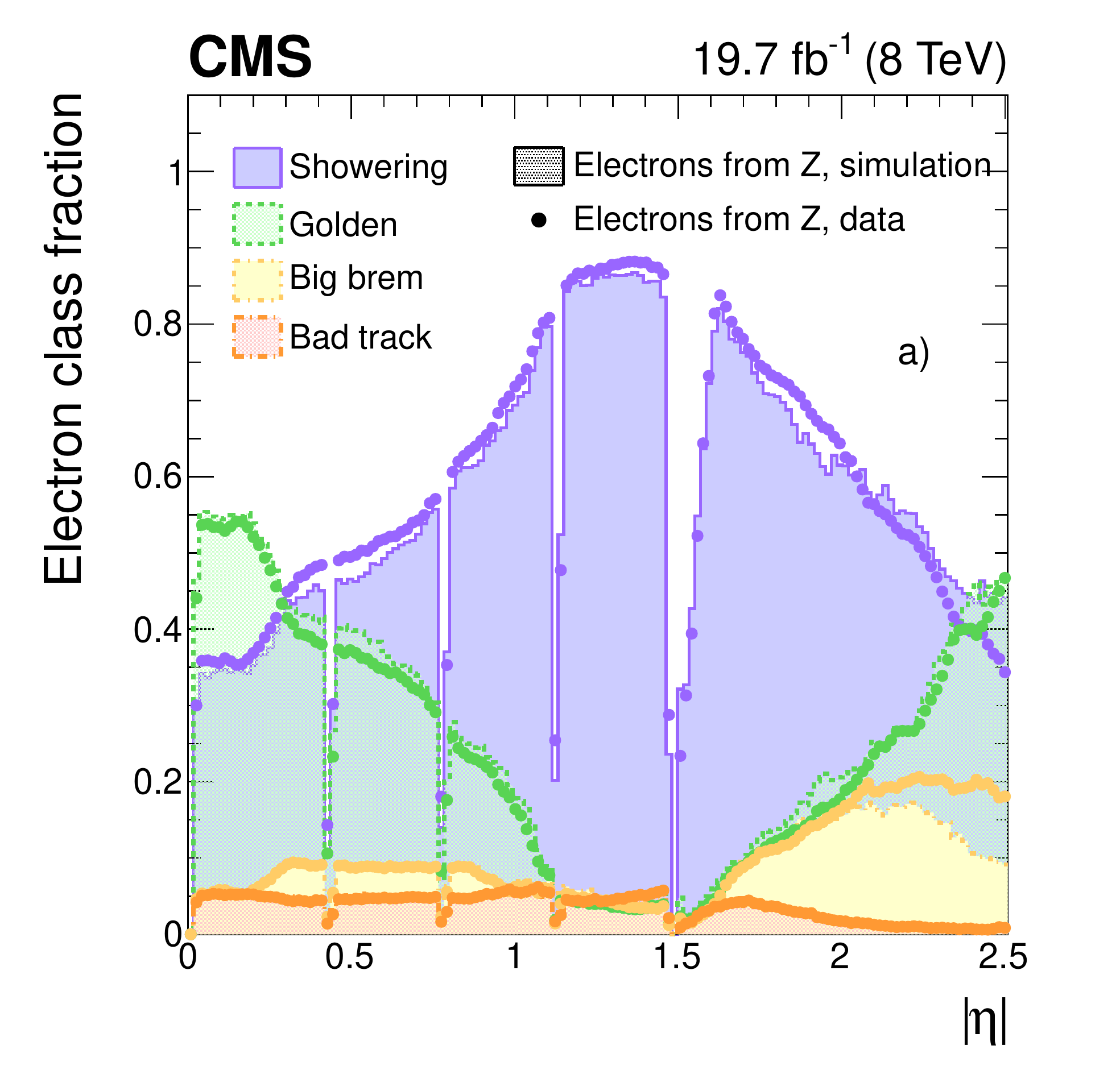}
\includegraphics[width=0.49\textwidth]{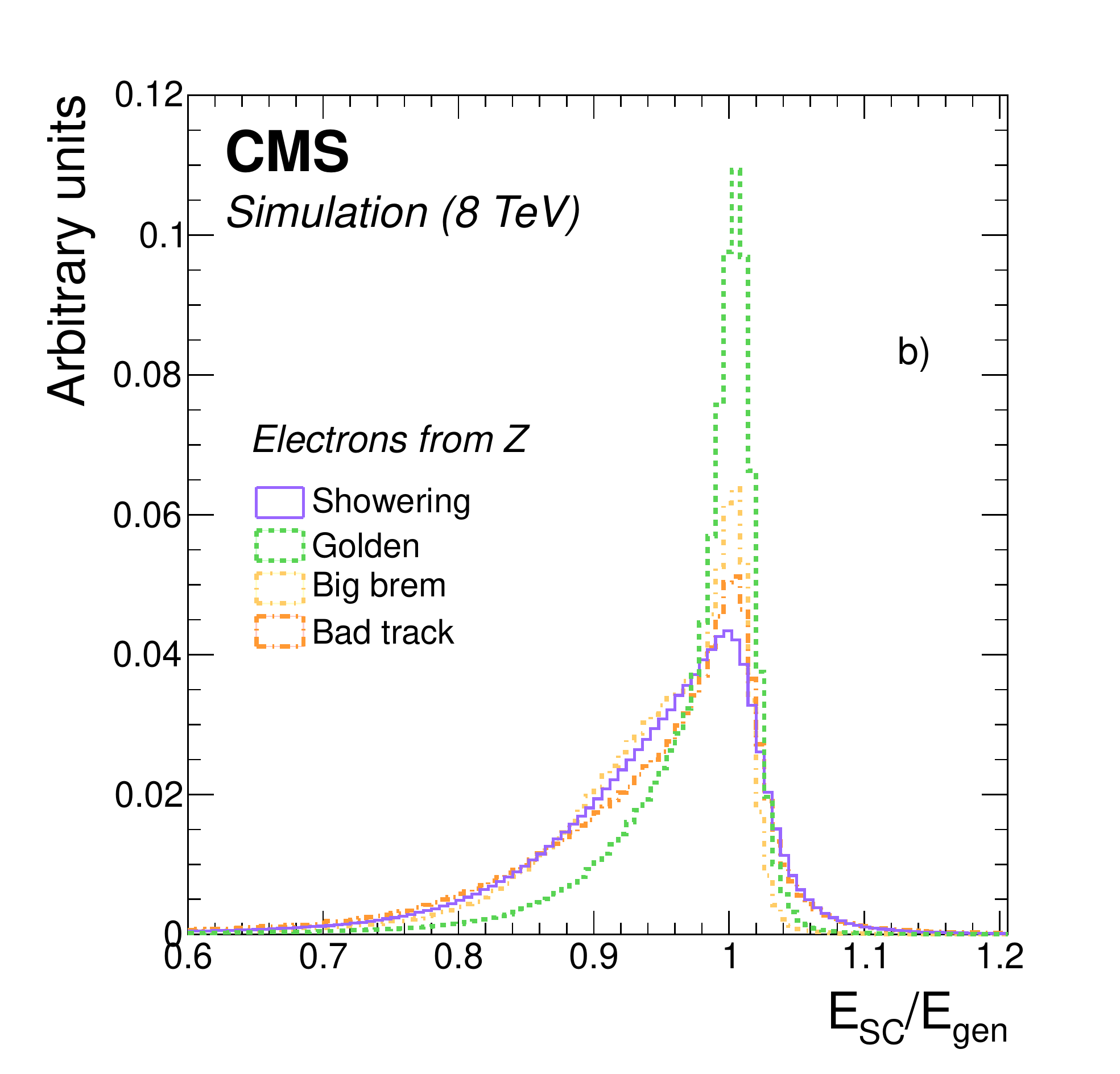}
\caption{\label{fig:classes}a) Fraction of population in different classes of electrons from \Z boson decays as a function of $\abs{\eta}$, for data (dots) and simulated (histograms) events, and b) distribution of $E_{\mathrm{SC}}/E_{\text{gen}}$ for the different classes of simulated electrons. Crack electrons are not shown in either plot.}

\end{figure}

Figure~\ref{fig:classes}~a) shows the fraction of the electron population in the above classes, as a function of $\abs{\eta}$ (defined relative to the centre of CMS), for data and simulated electrons from \Z boson decays. Crack electrons are not shown in the plot, but complement the proportion to unity. The distributions for the golden and showering classes reflect the $\eta$ distribution of the intervening material. Data and simulation agree well, except for the regions of $\eta$ with known mismodelling of material, and for $\abs\eta > 2$, where the number of clusters is overestimated in the simulation. The integrated proportions of electrons in the different classes for data and simulation are, respectively,  57.4\% and 56.8\% for showering, 25.5\% and 26.3\% for  golden, 8.4\% and 8.0\% for big-brem, 4.1\% and 4.1\% for bad track, and 4.6\% and 4.7\% for crack electrons. Figure~\ref{fig:classes}~b) shows the distributions in the ratio of reconstructed SC energy to the generated energy ($E_{\text{gen}}$) for the different classes.
The SC performs differently for each class, and
provides an energy estimate of limited quality for electrons with sizeable bremsstrahlung.
An improved energy estimate is achieved with additional corrections, as discussed in the following section.

\subsubsection{ECAL supercluster energy}

\paragraph*{Energy in individual crystals}
\label{par:calibSingle}
Several procedures are used to calibrate the energy response of individual crystals
before the clustering step~\cite{Chatrchyan:2013dga}.
The amplitude in each crystal is reconstructed using a linear combination of the
40~MHz sampling of the pulse shape. This amplitude is then converted into an energy value using factors measured separately for the ECAL barrel, endcaps, and the preshower detector.
The changes in the crystal response induced by radiation are corrected through
the ECAL laser-monitoring system~\cite{Anfreville:2007zz,Zhang:2005ip}, and the correction factors are checked using the reconstructed
dielectron invariant mass in $\Z \to \Pep\Pem$ events, and through the ratio of the ECAL energy and the track momentum
($E_{\mathrm{SC}}/p$) in $\PW \to \Pe\cPgn$ events. The inter-calibration factors between crystals are
obtained with data using different methods, e.g. the $\phi$ symmetry of the energy in minimum-bias
events  for a given $\eta$, the reconstructed invariant mass of
$\pi^0 \to \cPgg\cPgg$, $\eta \to \cPgg\cPgg$, and $\Z \to \Pep\Pem$ events,
and the $E_{\mathrm{SC}}/p$ ratio
of electrons in $\PW \to \Pe \cPgn$ events.

\paragraph*{Supercluster energy correction}
\label{sec:scEneCorr}
The SC energy is obtained by summing the individual energies in all the crystals of an SC, and the preshower energies of electrons in the endcaps. At this stage, the main effects impacting the estimation of SC energy are related to energy containment:
\begin{itemize}
\item energy leakage in $\phi$ or $\eta$ out of the SC,
\item energy leakage into the gaps between crystals, modules, supermodules, and the transition region between barrel and endcaps,
\item energy leakage into the HCAL downstream the ECAL,
\item energy loss in interactions in the material before the ECAL, and
\item additional energy from pileup interactions.
\end{itemize}

An MVA regression technique~\cite{Regression} is used to obtain the SC corrections that are needed
to account for these effects.
Simulated electrons with a uniform spectrum in $\eta$ and $\pt$ between 5 and 300\GeV
are used to train the regression algorithm, separately for electrons in the barrel and in the endcaps. The regression target is the ratio $E_{\text{gen}} / E_{\mathrm{SC}}$.
The first input observables are the SC energy to be corrected, and the SC position in $\eta$ and $\phi$, which are related to the intervening material.
The energy leakage out of the SC is assessed through the SC shape observables and its number of clusters, together with their individual respective positions, energies, and shape observables. The energy leakage in the gaps between modules, supermodules and in the transition region between the barrel and endcaps is explored through the position of the seed crystal of the SC.
The position of the seed cluster relative to the seed crystal is used together with the shower-shape observables to account for energy leakage between the crystals. The ratio $H/E_{\mathrm{SC}}$ (defined in Section~\ref{sec:seeding}) is used to estimate the energy leakage into the HCAL.
The effects of pileup interactions are assessed through the number of reconstructed interaction vertices and the average energy density $\rho$ in the event (defined as the median of the energy density distribution for particles
within the area of any jet in the event, reconstructed using the $k_{\rm{T}}$-clustering
algorithm~\cite{Catani:1993hr,Ellis:1993tq} with distance parameter of 0.6, $\pt^{\text{jet}} > 3\GeV$ and within
$\abs{\eta}<2.5$).

\begin{figure}[h!tbp]
\centering
\includegraphics[width=0.49\textwidth]{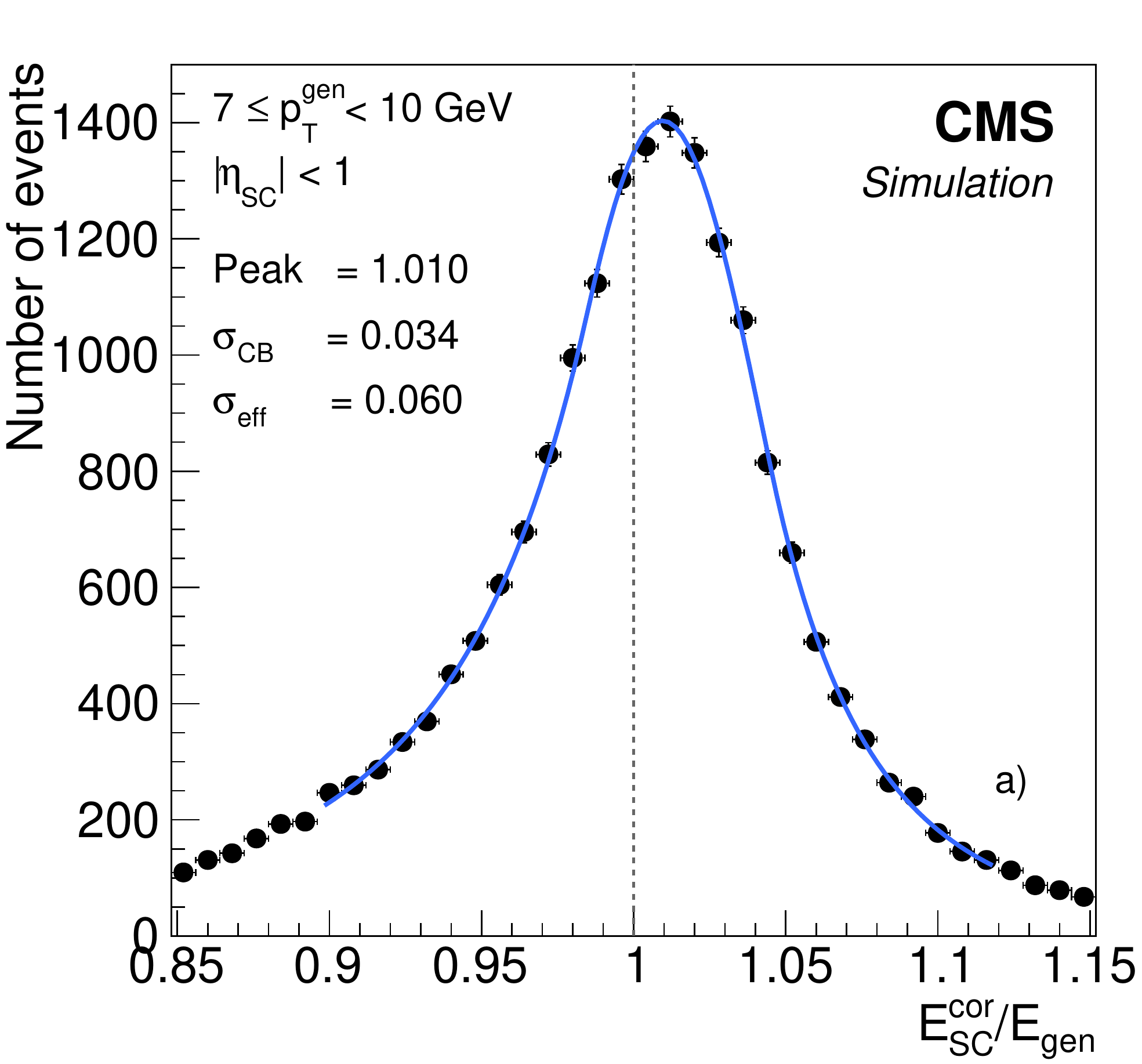}
\includegraphics[width=0.49\textwidth]{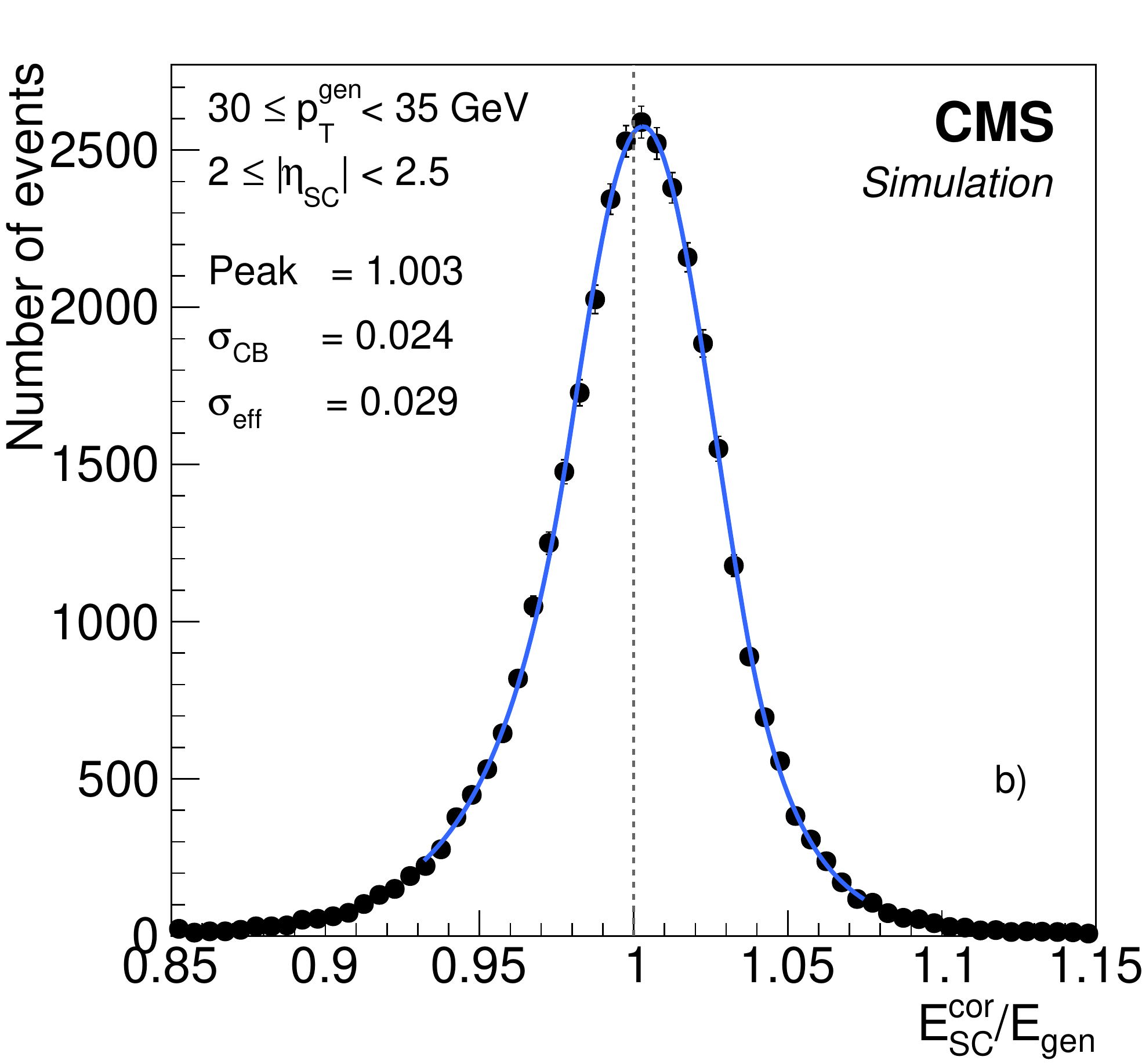}
\caption{\label{fig:regression_fit} Example distributions of the ratio of corrected over generated
supercluster energies ($E_{\mathrm{SC}}^{\text{cor}} / E_{\text{gen}}$) and their (double Crystal Ball)
fits, in two regions of $\eta$ and \pt after implementing the regression corrections: for electrons a) with $7 \leq \pt^{\text{gen}} < 10\GeV$ and  $\abs{\eta_{\mathrm{SC}}} < 1$, and b) with $30 \leq \pt^{\text{gen}} < 35\GeV$ and  $2 \leq \abs{\eta_{\mathrm{SC}}} < 2.5$, $\eta_{\mathrm{SC}}$ being defined relative to the centre of CMS. Electrons are generated with uniform distributions in $\eta$ and \pt. }
\end{figure}

Figure~\ref{fig:regression_fit} shows the distribution in the ratio of the corrected SC energy over the generated energy $E_{\mathrm{SC}}^{\text{cor}} / E_{\text{gen}}$, obtained through the regression
for two categories of simulated electrons: low-\pt electrons ($7\leq\pt<10\GeV$) in the central part of the barrel, and medium-\pt electrons ($30\leq\pt<35\GeV$) in the forward part of the endcaps.
The distributions are fitted with a ``double'' Crystal Ball function~\cite{CrystalBall}.
The Crystal Ball function is defined as:
\begin{equation}
f_{\mathrm{CB}}(x;\alpha,n,m_{\mathrm{CB}},\sigma_{\mathrm{CB}}) = N  \left\{
\begin{aligned}
 A \left[ B - \frac{x-m_{\mathrm{CB}}}{\sigma_{\mathrm{CB}}} \right]^{- n }, \quad\text{for} \quad \frac{x-m_{\mathrm{CB}}}{\sigma_{\mathrm{CB}}} \leq -\alpha \\
\exp \left( - \frac{(x-m_{\mathrm{CB}})^2}{2\sigma_{\mathrm{CB}}^2} \right), \quad\text{for}\quad\frac{x-m_{\mathrm{CB}}}{\sigma_{\mathrm{CB}}} > -\alpha
\end{aligned}
\right.
\label{eq:dCB}
\end{equation}
where $A$ and $B$ are functions of $\alpha$ and $n$, and $N$ is a normalization factor.
This function is intended to capture both the Gaussian core of the distribution
(described by $\sigma_{\mathrm{CB}}$) and non-Gaussian tails (described by
the parameters $n$ and $\alpha$).
The double Crystal Ball function is a modified Crystal Ball with the $\sigma_{\mathrm{CB}}$, $n$, and $\alpha$ parameters distinct for $x$ values below and above the peak position at $m_{\mathrm{CB}}$.

The peak position and the standard deviation of the Gaussian core of the distributions are estimated
through the fitted values of $m_{\mathrm{CB}}$ and $\sigma_{\mathrm{CB}}$, respectively.
The ``effective'' standard deviation $\sigma_{\text{eff}}$, defined as half of the smallest interval around the peak
position containing 68.3\% of the electrons, is used to assess the resolution, while taking
into account possible non-Gaussian tails.
A bias of at most 1\% affects the peak position, which reflects the
asymmetric nature of the $E_{\text{gen}} / E_{\mathrm{SC}}$ distribution.

\begin{figure}[h!btp]
\centering
\includegraphics[width=0.49\textwidth]{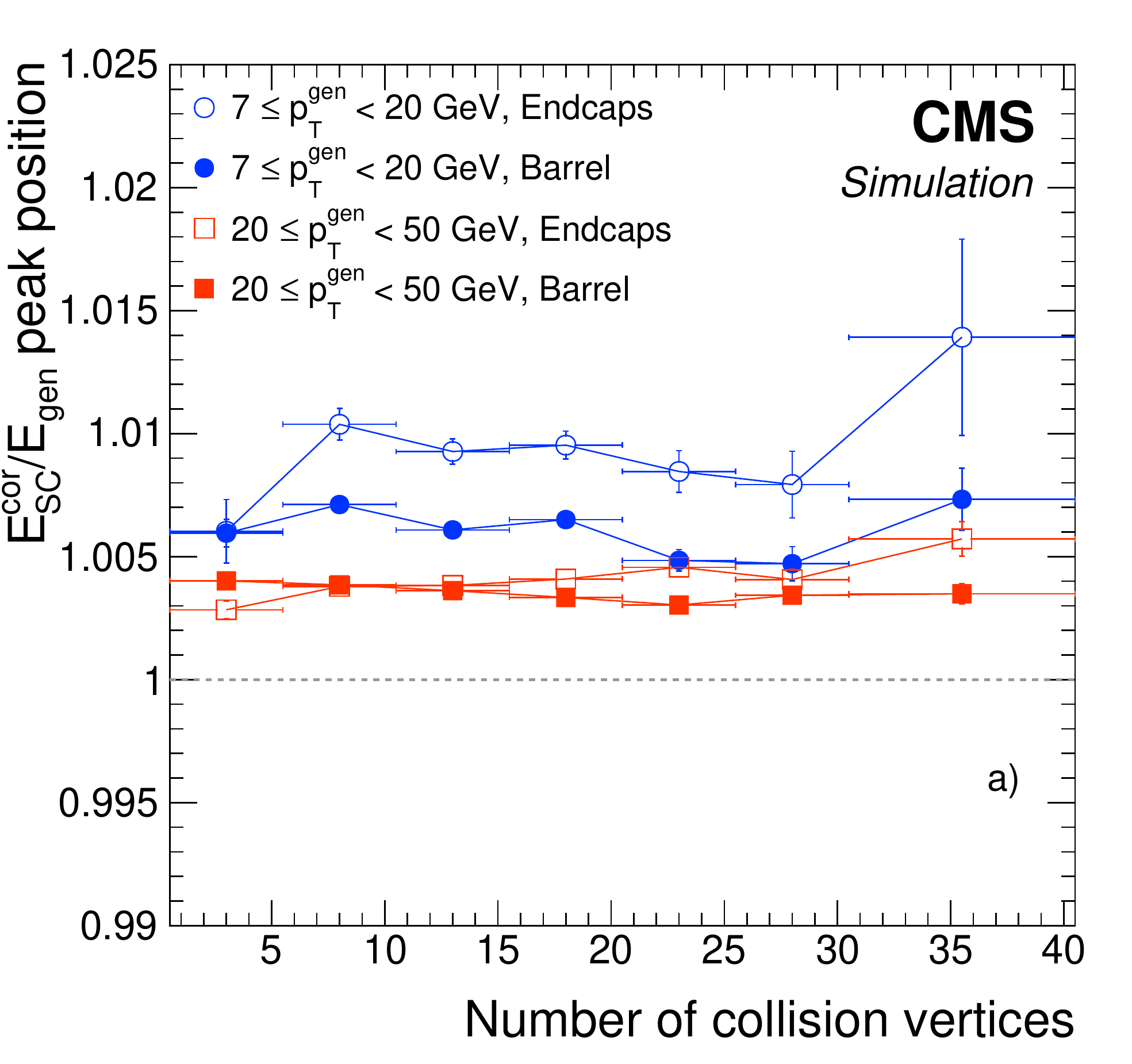}
\includegraphics[width=0.49\textwidth]{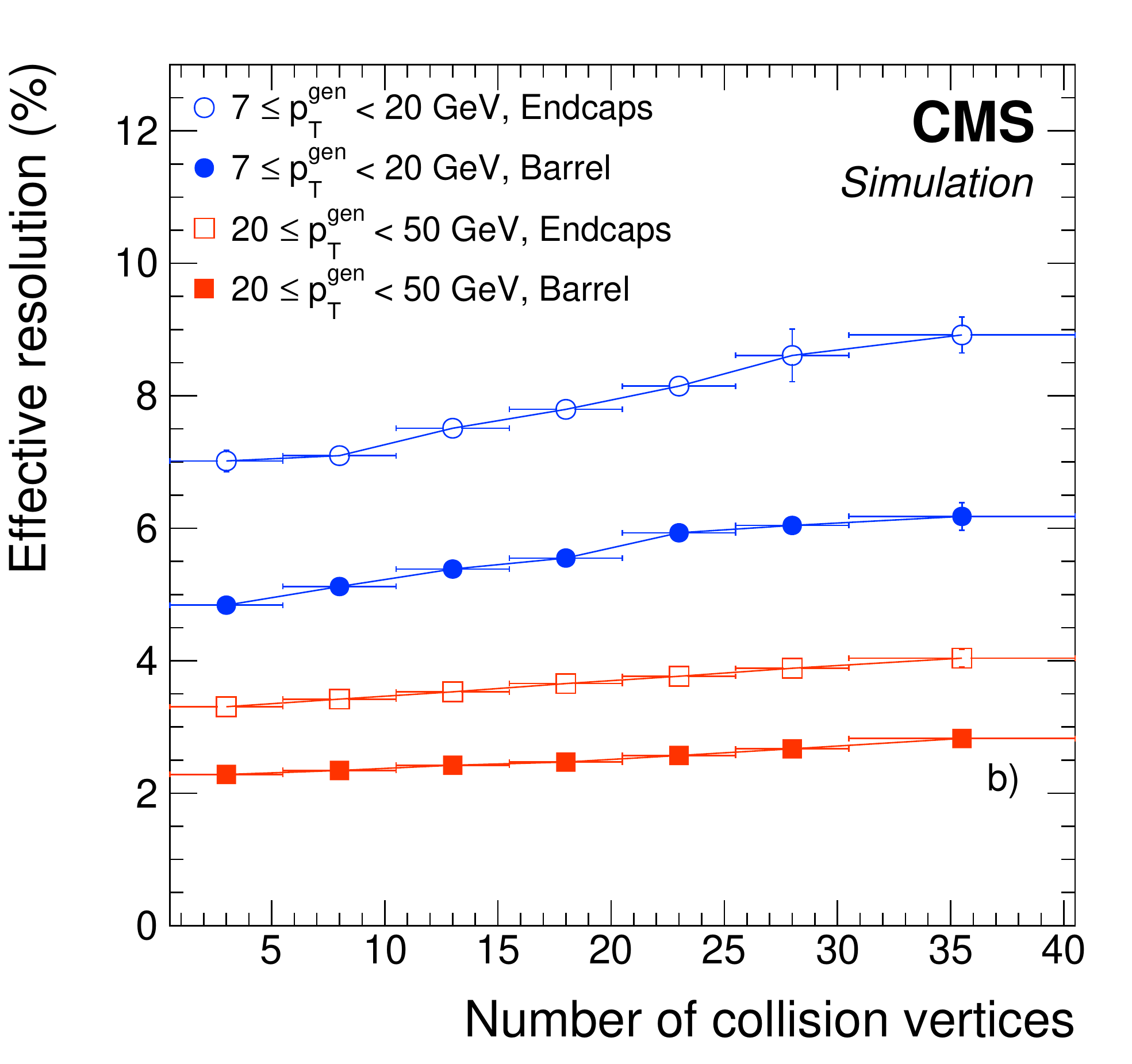}
\caption{\label{fig:regression_Epeak_effrms_nvtx}a) Peak position of $E_{\mathrm{SC}}^{\text{cor}} / E_{\text{gen}}$, and b) effective resolution of $E_{\mathrm{SC}}^{\text{cor}}$, as a function of the number of reconstructed interaction vertices, for electrons in the barrel (solid symbols) and endcaps (open symbols) with $ 7 \leq \pt^{\text{gen}} < 20\GeV$ (circles),
and $ 20 \leq \pt^{\text{gen}} < 50\GeV$ (squares). Electrons are generated with uniform distributions in $\eta$ and \pt.}
\end{figure}

The peak position of $E_{\mathrm{SC}}^{\text{cor}} / E_{\text{gen}}$ and the effective resolution for $E_{\mathrm{SC}}^{\text{cor}}$ are shown in Fig.~\ref{fig:regression_Epeak_effrms_nvtx}, as a function of the number of reconstructed interaction vertices for low-\pt and medium-\pt electrons, in the barrel and in the endcaps. The bias in the peak position is independent of the number of pileup interactions. The effective resolution is in the range of 2--3\% for medium-\pt electrons in the barrel, and in the range of 7--9\% for low-\pt electrons in the endcaps, degrading slowly with increasing number of pileup interactions.

The use of the MVA regression technique compared to a standard parameterization of the  correction for $E_{\mathrm{SC}}$ as a function of the electron $\eta$, category, and \ET, provides significant improvement of $\approx$20\% in the resolution on average and up to $\approx$35\% in the forward regions, while reducing the bias in the peak position for each electron class over the entire range of electron $\eta$ and $\pt$.

Another MVA regression technique, based on the same input variables, is used to estimate the uncertainty in the corrected $E_{\mathrm{SC}}$, separately for electrons in the barrel and in the endcaps, with the absolute difference between $E_{\mathrm{CB}}$ and the corrected $E_{\mathrm{SC}}$ being the target.

\paragraph*{Fine-tuning of calibration and simulated resolution}
\label{par:calib}
The SC energy corrections described above are based on simulation.
Events in data are used to account for any discrepancy between data and simulation in input
variables, as well as to correct for biases.
The applied remnant corrections are quite small. The energy in individual crystals
is already calibrated, and simulation
of showers in the ECAL is rather precise and includes the measured uncertainties in the inter-calibration between crystals.
The main source of discrepancy between the energy estimate in data and in simulation is
the imperfect description of the tracker material in simulation,
which affects differently each category of electrons. The evolution of the transparency of the crystals and of the noise in the ECAL during data taking, if not considered through specific run-dependent simulations, leads to an additional difference between data and simulation.
Another possible source of discrepancy could be the underestimation of uncertainties in
the calibration of individual crystals.
Finally, a difference in the ECAL geometry relative to the nominal one can cause
the corrections discussed in the previous paragraph, which are obtained using simulated events
with the nominal geometry, to be inappropriate for data.
While
it is now understood that at least one of the above effects
contributes to degradation, their relative magnitudes are not as fully clear.
More details on this issue can be found in Ref.~\cite{photonPaper}.

The SC energy scale is corrected in the data to match
that in simulation.
These corrections are assessed using $\Z \to \Pep\Pem$ events, by comparing the dielectron invariant mass in data and in simulation for four $\abs{\eta}$ regions and two categories of electrons, over 50 running periods, following the procedure described in Ref.~\cite{Chatrchyan:2013dga}. The $\eta$ regions are defined from the most central to the most forward values as barrel $\abs{\eta} \leq 1$, barrel $\abs{\eta} > 1$, endcaps $\abs{\eta} \leq 2$, and endcaps $\abs{\eta} > 2$. The $R_9$ variable, defined as the ratio of the energy reconstructed in the $3\times3$ crystals matrix centered on the crystal with most energy and the SC energy, is used to assess the amount of bremsstrahlung emitted by the electron.
 The category of electrons with a low level of \brem is defined by $R_9 \geq 0.94$, and the one with a high level of \brem by $R_9< 0.94$. The \Z boson mass is reconstructed from the SC energies and the opening angles measured from the tracks. The mass distribution in the range between 60 and 120\GeV is fitted using a Breit--Wigner convolved with a Crystal Ball function, both for data and simulation. The scale corrections, obtained from the difference between the peak positions measured in the data and in simulation, are applied to the data, so that the peak position of the \Z boson mass agrees with that in simulation, in each category. Overall, these corrections vary between 0.9880 and 1.0076 and their uncertainties between 0.0002 and 0.0029.

The estimate of the SC energy resolution is also
affected by the sources of discrepancy between data and simulation.
A correction is applied in simulation
to match the resolution observed in data~\cite{Chatrchyan:2013dga}. This correction is independent of time, and evaluated for the above categories of $\eta$ and $R_9$. The SC energy is modified by applying a factor drawn from a Gaussian distribution, centered on the corrected scale value, and with a standard deviation of $\delta \sigma^{\mathrm{e}}$, corresponding to a required additional constant term in the energy resolution. The value of $\delta \sigma^{\mathrm{e}}$ for each electron category is assessed using a maximum-likelihood fit of the data to a resolution-broadened simulated energy. This constant term in the energy resolution ranges from $(0.92 \pm 0.03)\%$ in the $\abs{\eta} < 1$ and $R_9 \geq 0.94$ category, to $(2.90 \pm 0.03)\%$ in the $\abs{\eta} > 2$ and $R_9 < 0.94$ category.
 The uncertainty in the SC energy  is increased accordingly.

\subsubsection{Combination of energy and momentum measurements}

The electron momentum estimate $p_{\text{comb}}$ is improved by combining the ECAL SC energy, after applying the refinements mentioned in the previous sections, with the track momentum. At energies $\lesssim$15\GeV, or for electrons near gaps in detectors, the track momentum is expected to be more precise than the ECAL SC energy.
A regression technique is used to define a weight $w$ that multiplies the track momentum in linear combination with the estimated SC energy as $p_{\text{comb}} = w  p + (1-w)  E_{\mathrm{SC}}$.

The complementarity of the two estimates depends on the amount of emitted bremsstrahlung. The corrected SC energy and its relative uncertainty, and the track momentum and its relative uncertainty are the main input observables. The addition of the $E_{\mathrm{SC}}/p$ ratio and its uncertainty, together with the ratio of the two relative uncertainties, brings a higher-level information that optimizes the performance of the regression.
The electron class and the position in the barrel or endcaps are also included as probes of the quality and amount of emitted \bremns.

\begin{figure}[htbp]
\centering
\includegraphics[width=0.49\textwidth]{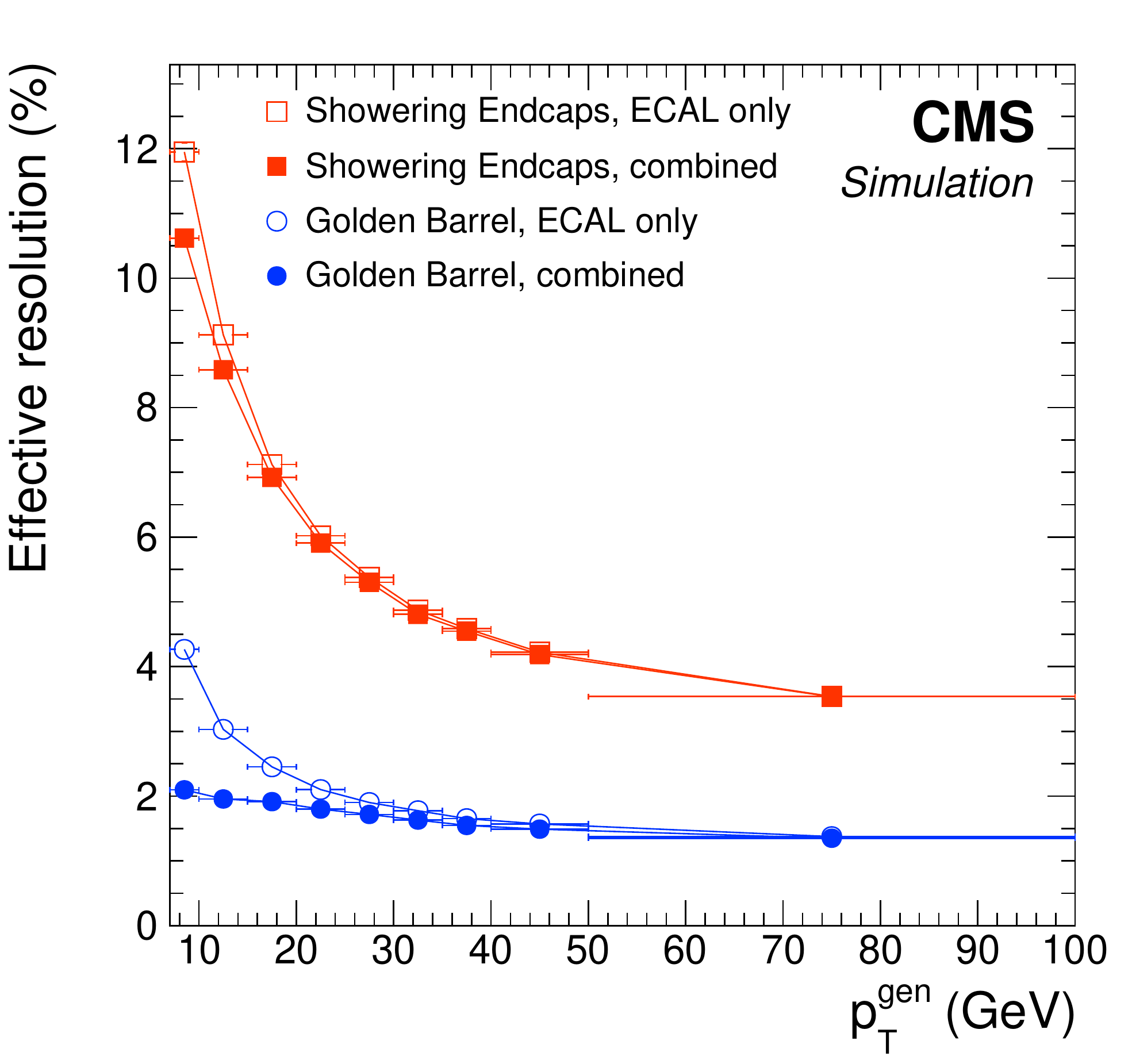}
\caption{\label{fig:Ep_reso}
Effective resolution in electron momentum after combining the $E_{\mathrm{SC}}$ and $p$ estimates
(solid symbols), compared to that of the corrected SC energy (open symbols),
as a function of the generated electron \pt. Golden electrons in the barrel (circles) and showering electrons in the endcaps (squares) are shown as examples.
Electrons are generated with uniform distributions in $\eta$ and \pt, and the resolution is shown after applying the resolution broadening.}

\end{figure}

After combining the two estimates, the bias in the electron momentum is reduced in all regions and all electron classes, except for showering electrons in the endcaps, where the bias becomes slightly worse.
Figure~\ref{fig:Ep_reso} shows the
effective resolution in the electron momentum (in percent), after combining the $E_{\mathrm{SC}}$ and $p$ estimates, as a function of the generated \pt, compared to the effective resolution of the corrected SC energy,
for golden electrons in the barrel and for showering electrons in the endcaps. The improvement is typically 25\% for electrons with
$\pt\approx 15\GeV$ in the barrel and reaches 50\% for golden electrons of $\pt < 10 \GeV$.

\begin{figure}[htbp]
\centering
\includegraphics[width=0.49\textwidth]{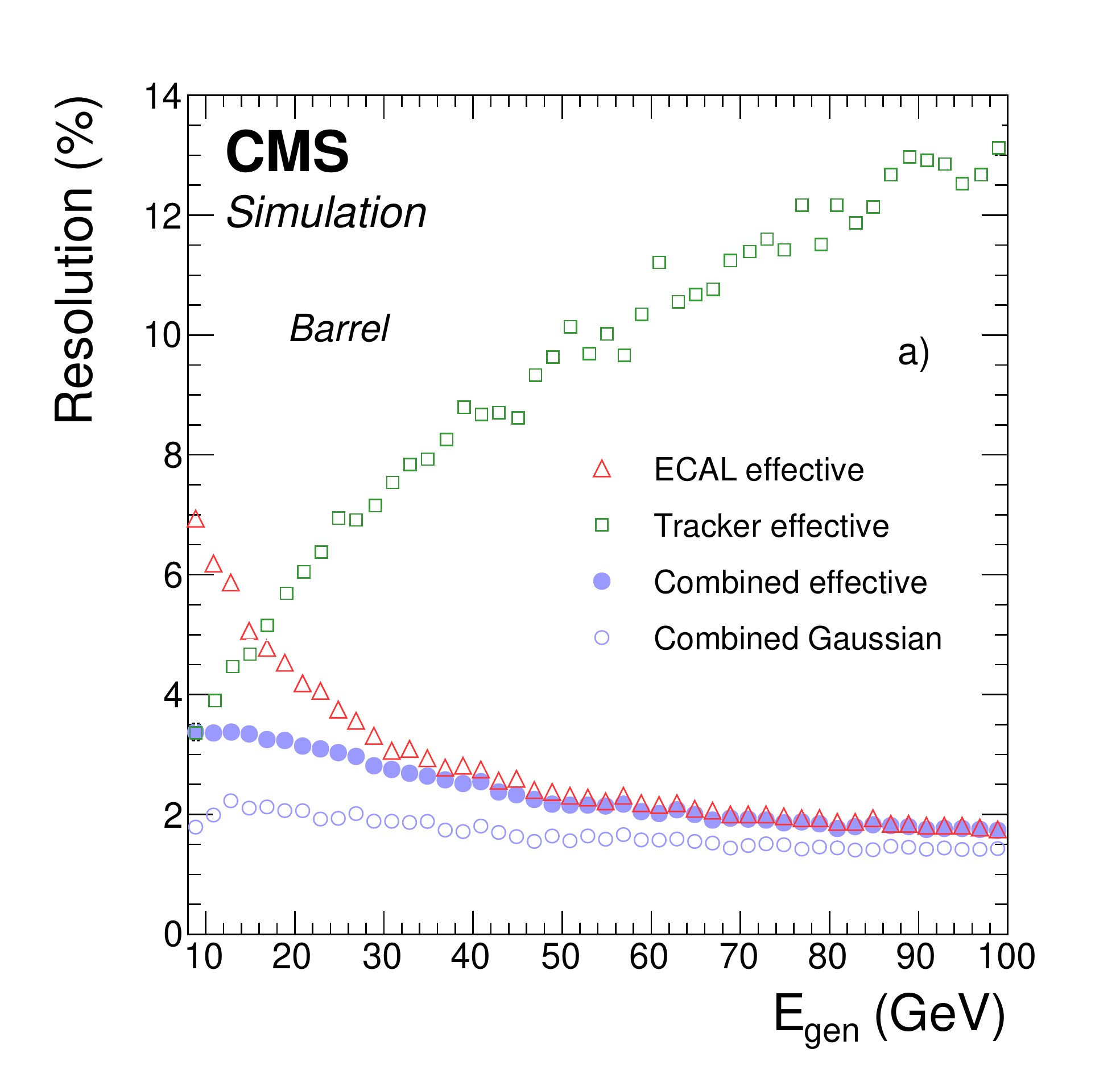}
\includegraphics[width=0.49\textwidth]{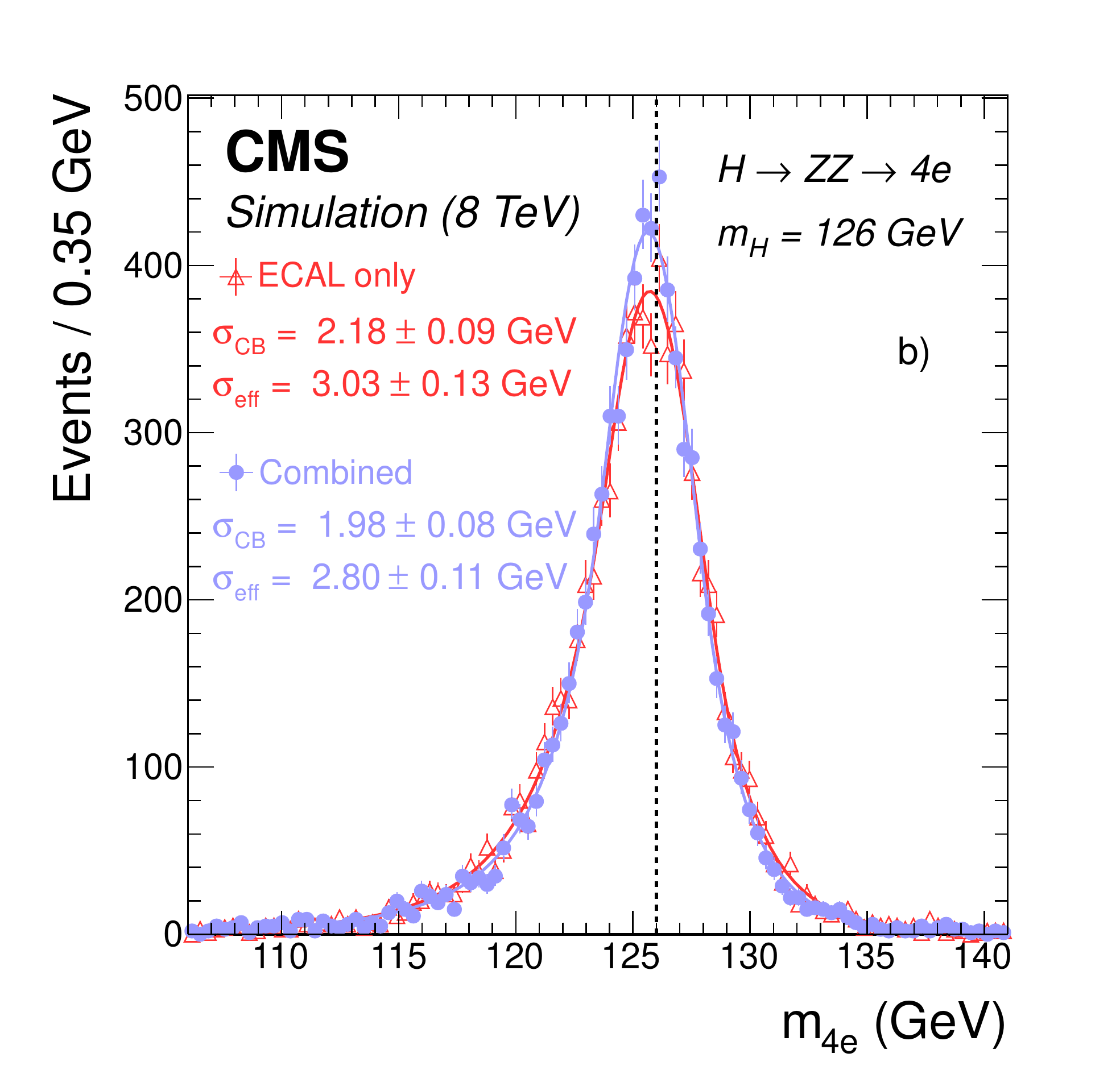}
\caption{\label{Ep_effRMS}
a) Effective resolution in electron momentum after combining $E_{\mathrm{SC}}$ and $p$ estimates (solid circles), compared to those using the corrected SC energy (triangles), and the track momentum (squares), as a function of the generated energy for electrons in the barrel.
Also shown is the resolution in momentum after combining $E_{\mathrm{SC}}$ and $p$ estimates in terms of  $\sigma_{\mathrm{CB}}$ (open circles), to illustrate the contribution of the Gaussian core of the distribution.  Electrons are generated with uniform distributions in $\eta$ and \pt. b) Reconstructed mass of the Higgs boson for
$\PH(126) \to \cPZ\Z^* \to 4\Pe$
simulated events, using either the corrected SC energy (open triangles) or the electron momentum after combining $E_{\mathrm{SC}}$ and $p$ estimates (solid dots)~\cite{legacyHzz}.}

\end{figure}

The improvement in resolution is significant for all electrons in the barrel up to energies of about 35\GeV, as can be seen in Fig.~\ref{Ep_effRMS}~a), which displays
the effective resolution
of the corrected SC energy, of the track momentum, and of the electron momentum after combining $E_{\mathrm{SC}}$ and $p$ estimates,
as a function of the generated electron energy.
Figure~\ref{Ep_effRMS}~b) shows the expected reconstructed mass for a 126\GeV Higgs boson in the
$\PH \to \cPZ\Z^* \to 4\Pe$ decay channel.
The masses reconstructed using the corrected SC energy are compared to those using the electron momentum obtained after combining the $E_{\mathrm{SC}}$ and $p$ estimates.
The improvement in the effective resolution is 7\%.
When considering only the Gaussian core of the distribution, the improvement in the resolution is 9\%.

\subsubsection{Uncertainty in the momentum scale and in the resolution}
\label{sec:scale_perf}
The corrections to the momentum scale
and resolution discussed above are only obtained from correcting the SC energy in $\Z \to \Pep\Pem$  events.
As a consequence, they must be further corrected, first over a large range of \pt, especially for the
$\PH \to \cPZ\Z^*$
analysis which uses electrons with \pt as low as 7\GeV, and second for the $E_{\mathrm{SC}}$ and $p$ combination.
 For this purpose, $\Z \to \Pep\Pem$ events are used together with $\JPsi \to \Pep\Pem$ and $\Upsilon \to \Pep\Pem$ events that provide clean sources of electrons at low \pt. The reconstructed invariant masses of these resonances in data are compared with simulation to probe any remaining differences.

\begin{figure}[htbp]
\centering
\includegraphics[width=0.49\textwidth]{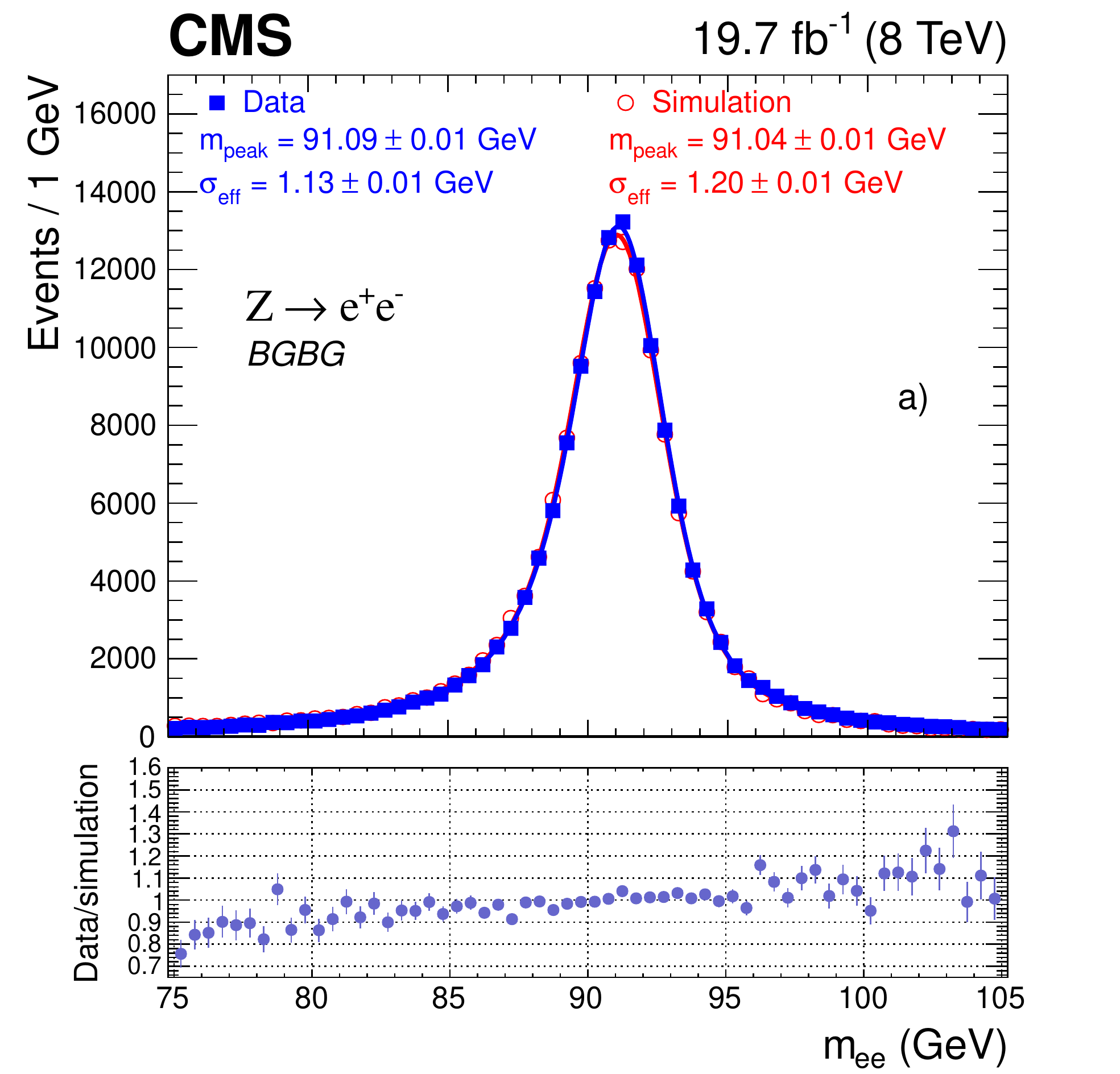}
\includegraphics[width=0.49\textwidth]{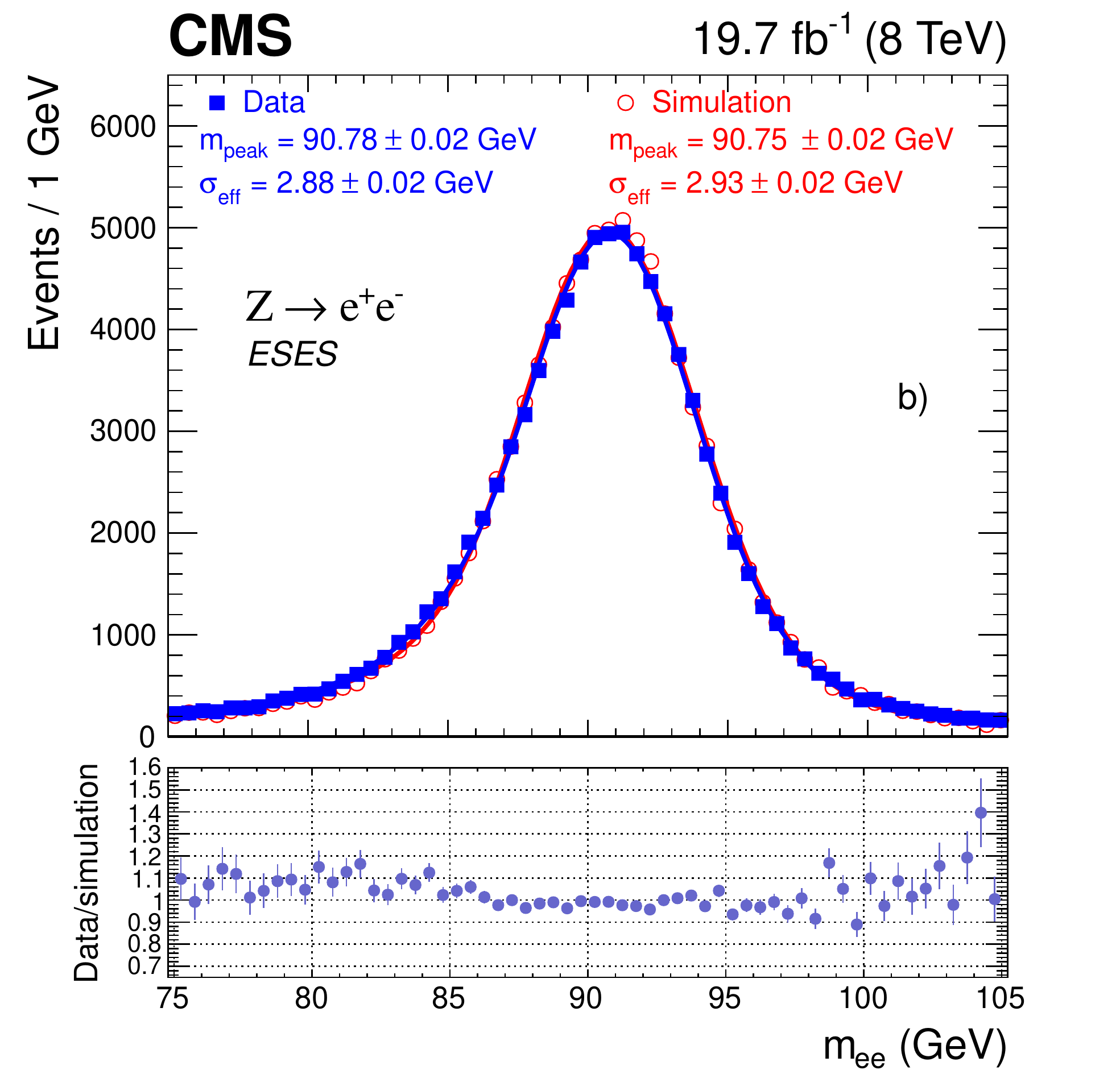}
\caption{\label{Scale_zeemass} Dielectron invariant mass distribution from $\Z \to \Pep\Pem$ events in data (solid squares) compared to simulation (open circles)
fitted with a convolution of a Breit--Wigner function and a Crystal Ball function,
a) for the best-resolved event category with two well-measured single-cluster electrons in the barrel (BGBG), and b) for the worst-resolved category with two more-difficult patterns or multi-cluster electrons in the endcaps (ESES). The masses at which the fitting functions have their maximum values, termed $m_{\text{peak}}$, and the effective standard deviations $\sigma_{\text{eff}}$ are given in the plots.
 The data-to-simulation factors are shown below the main panels.}
\end{figure}

Figure~\ref{Scale_zeemass} shows an example of such comparisons and their degree of agreement for two extreme categories of events: one where each electron is well measured, having a single-cluster SC (golden or big-brem class) in the barrel, and the other one where each electron has a multi-cluster SC, or is poorly-measured (showering, crack, or bad track class) in the endcaps. These two categories represent the breadth of performance in data that enters, for example, in the mass measurement of the benchmark process for Higgs boson decays to four leptons.
The distributions in data and in simulation are fitted with a Breit--Wigner function convolved with a Crystal Ball function,
\begin{equation*}
P(m_{\Pep\Pem}; m_\Z,\Gamma_\Z,\alpha,n,m_{\mathrm{CB}},\sigma_{\mathrm{CB}}) = \mathrm{BW}(m_{\Pep\Pem}; m_\Z,\Gamma_\Z) \otimes f_{\mathrm{CB}}(m_{\Pep\Pem}; \alpha,n,m_{\mathrm{CB}},\sigma_{\mathrm{CB}}),
\end{equation*}
where $m_\Z$ and $\Gamma_\Z$ are fixed to the nominal values of 91.188 and 2.485\GeV~\cite{Olive:2014}.

The effective standard deviation
$\sigma_{\text{eff}}$,
which is indicated in the plots,
is calculated as the effective standard deviation of the function $f_{\mathrm{CB}}$,
which therefore does not include the contribution from the width of the \Z boson.
In both categories of events, the data and simulation show good agreement.
The $\sigma_{\text{eff}}$ in data for the $\Z \to \Pep\Pem$ invariant mass are, respectively for the best and worst categories,
$1.13 \pm 0.01\GeV$ and
$2.88 \pm 0.02\GeV$. Considering only the Gaussian cores of the distribution, the standard deviations ($\sigma_{\mathrm{CB}}$) are
$1.00 \pm 0.01\GeV$ and $2.63 \pm 0.02\GeV$, for the best and the worst categories, respectively.
The effective and Gaussian invariant mass resolutions of dielectron events in the data range, respectively,
from
1.2 and 1.1\%
 for the best category with two well-measured single-cluster electrons in the barrel,
to 3.2 and 2.9\%
 for the worst category with two poorly-measured or multi-cluster electrons in the endcaps.
The effective and Gaussian momentum resolutions for single electrons, approximated by multiplying the dielectron mass resolution by $\sqrt{2}$,
therefore range in data from 1.7 and 1.6\%, to 4.5 and 4.1\%, respectively.

The data-to-simulation comparisons are performed for different categories of events based on $\eta$, $\pt$, and class of electron, and for different instantaneous luminosities. The scale corrections are applied to data, and the resolutions are broadened in the simulated distributions, as discussed in Section~\ref{par:calib}.

\begin{figure}[htbp]
\centering
\includegraphics[width=0.49\textwidth]{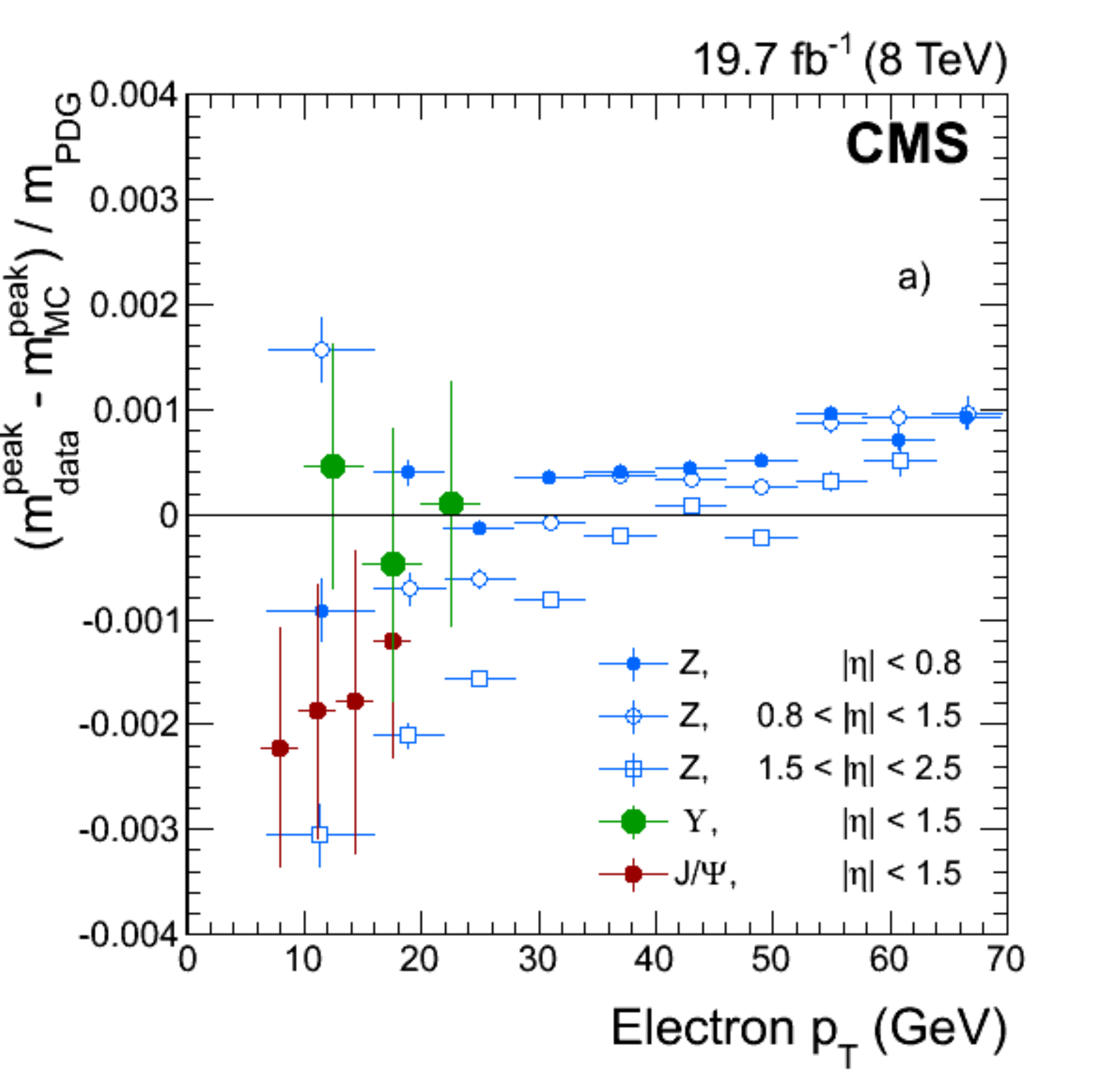}
\includegraphics[width=0.49\textwidth]{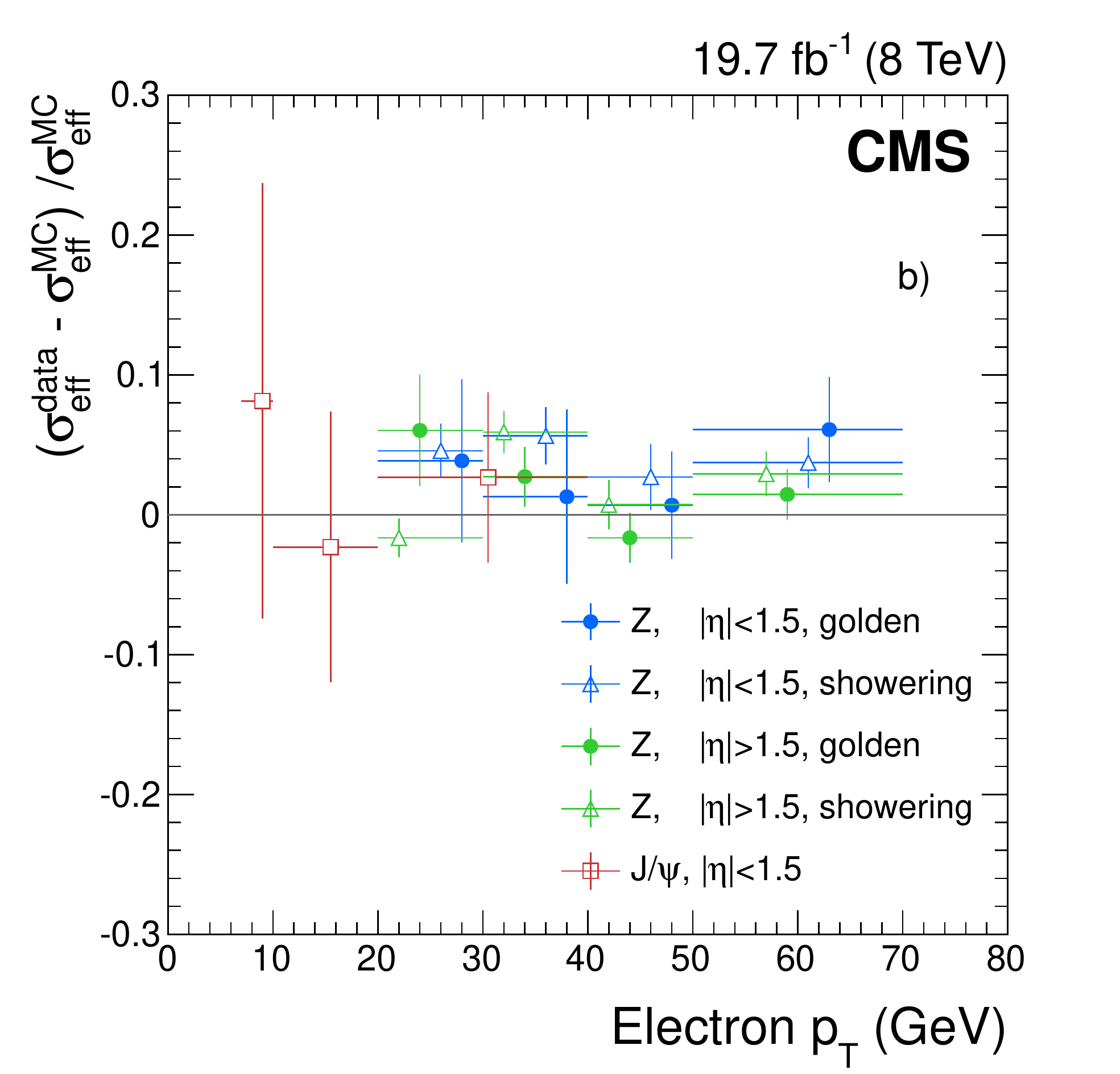}
\caption{\label{Scale_vspt}
Relative differences between data and simulation as a function of electron \pt for different $\abs{\eta}$ regions, a) for the momentum scale measured using $\JPsi \to \Pep\Pem$, $\Upsilon \to \Pep\Pem$, and $\Z \to \Pep\Pem$ events~\cite{legacyHzz}, and b) for the effective momentum resolution of  $\Z \to \Pep\Pem$ and $\JPsi \to \Pep\Pem$ events for different electron categories.}
\end{figure}

For study of the momentum scale,
the \pt and $\eta$ categories are defined according to the \pt and $\eta$ of one of the two electrons,
the other electron is used to tag the \Z event, it satisfies tight identification requirements (as described in Section~\ref{sec:TandP}), and has $\pt>20\GeV$.
The fits are performed using signal templates (obtained from simulation as binned distributions) that are convolved with Gaussians with floating means and standard deviations.
A \pt-dependence of the momentum scale of up to 0.6\% in the barrel and 1.5\% in the endcaps is observed and corrected in the \pt range between 7 and 70\GeV.
The final performance of the momentum scale is shown in Fig.~\ref{Scale_vspt}~a) as the relative difference between data and simulation of the $\JPsi \to \Pep\Pem$, $\Upsilon \to \Pep\Pem$, and the $\Z \to \Pep\Pem$ mass peaks, as a function of the \pt of one electron and for several $\eta$ regions of this electron, integrating over the \pt and $\eta$ of the other electron.
The residual scale difference between data and MC simulation is at most 0.2\% in the barrel and 0.3\% in the endcaps.
These numbers are taken as systematic uncertainties on the momentum scale of electrons in the barrel
and in the endcaps.
For the study of the resolution, the \pt, $\eta$, and class categories are defined for both electrons from the \Z decay.
The fits are performed using a Breit--Wigner function convolved with a Crystal Ball function.
The agreement between data and simulation in effective resolution
is shown in Fig.~\ref{Scale_vspt}~b), in terms of the relative difference between data and simulation for the $\JPsi \to \Pep\Pem$ and $\Z \to \Pep\Pem$ events,
as a function of the \pt of one electron, for different categories of electrons.
Overall the
relative difference in effective resolution between data and simulation is less than 10\% for all the categories in this comparison.

\subsubsection{High-energy electrons}

For high-energy electrons, the $E_{\mathrm{SC}}$ and $p$ combination is dominated entirely by the energy measurement in the ECAL.
Because of this and for reasons of simplicity, analyses exploiting high-energy electrons, with typical energies above 250\GeV,
estimate the electron momentum using only the SC information.
Moreover, energy deposition from very high-energy electrons (from about 1500\GeV in the barrel and from about 3000\GeV in the endcaps) lead to a saturation of the front-end electronics~\cite{Chatrchyan:2008aa}.

Both the calibration of high-energy electrons and the energy correction for saturated crystals are tuned with
$\Z \to \Pep\Pem$ events through a method
that estimates the energy contained in the central (highest energy) crystal of a $5\times5$ matrix, using the 24 lower-energy surrounding crystals. The energy fraction contained in the central crystal relative to the $5\times5$ matrix ($E_1/E_{5\times5}$) is parameterized as a function of the electron $\eta$, $E_{5\times5}$, as well as other SC shower-shape variables, using simulated high-mass DY events. The parameterization is validated with data through a comparison of the central crystal energy with the energy estimated from the parameterization.
The energy scale is validated at the 1--2\% level using electrons with energy larger
than 500\GeV in data.
The dominant uncertainty is mainly from the limited number of high-energy electrons available for this study.

\section{Electron selection}
\label{sec:selection}
\subsection{Identification}
\label{sec:eleID}
Several strategies are used in CMS to identify prompt isolated electrons (signal),
and to separate them from background sources, mainly originating from photon conversions,
jets misidentified as electrons,
or electrons
from semileptonic decays of~\cPqb~and~\cPqc~quarks.
Simple and robust algorithms have been developed to apply sequential selections on a set of discriminants.
More complex algorithms combine variables in an MVA analysis to achieve better discrimination.
In addition, dedicated selections are used for highly energetic electrons.

Variables that provide discriminating power are grouped into three main categories:
\begin{itemize}
\item
Observables that compare measurements obtained from the ECAL and the tracker (track--cluster matching,
including both geometrical as well as SC energy--track momentum matching).
\item
Purely calorimetric observables used to separate genuine electrons (signal electrons or electrons from photon
conversions) from misidentified electrons (e.g., jets with large electromagnetic components), based on the
transverse shape of electromagnetic showers in the ECAL and exploiting the fact that electromagnetic showers are narrower
than hadronic showers. Also utilized are the energy fractions deposited in the HCAL (expected to be small,
as electromagnetic showers are essentially fully contained in the ECAL), as well the
energy deposited in the preshower in the endcaps.
\item
Tracking observables employed to improve the separation between electrons and charged hadrons,
exploiting the information obtained from the GSF-fitted track, and the difference between the information from the KF and GSF-fitted tracks.
\end{itemize}

\begin{figure*}[hbtp]\centering
\includegraphics[width=0.45\textwidth]{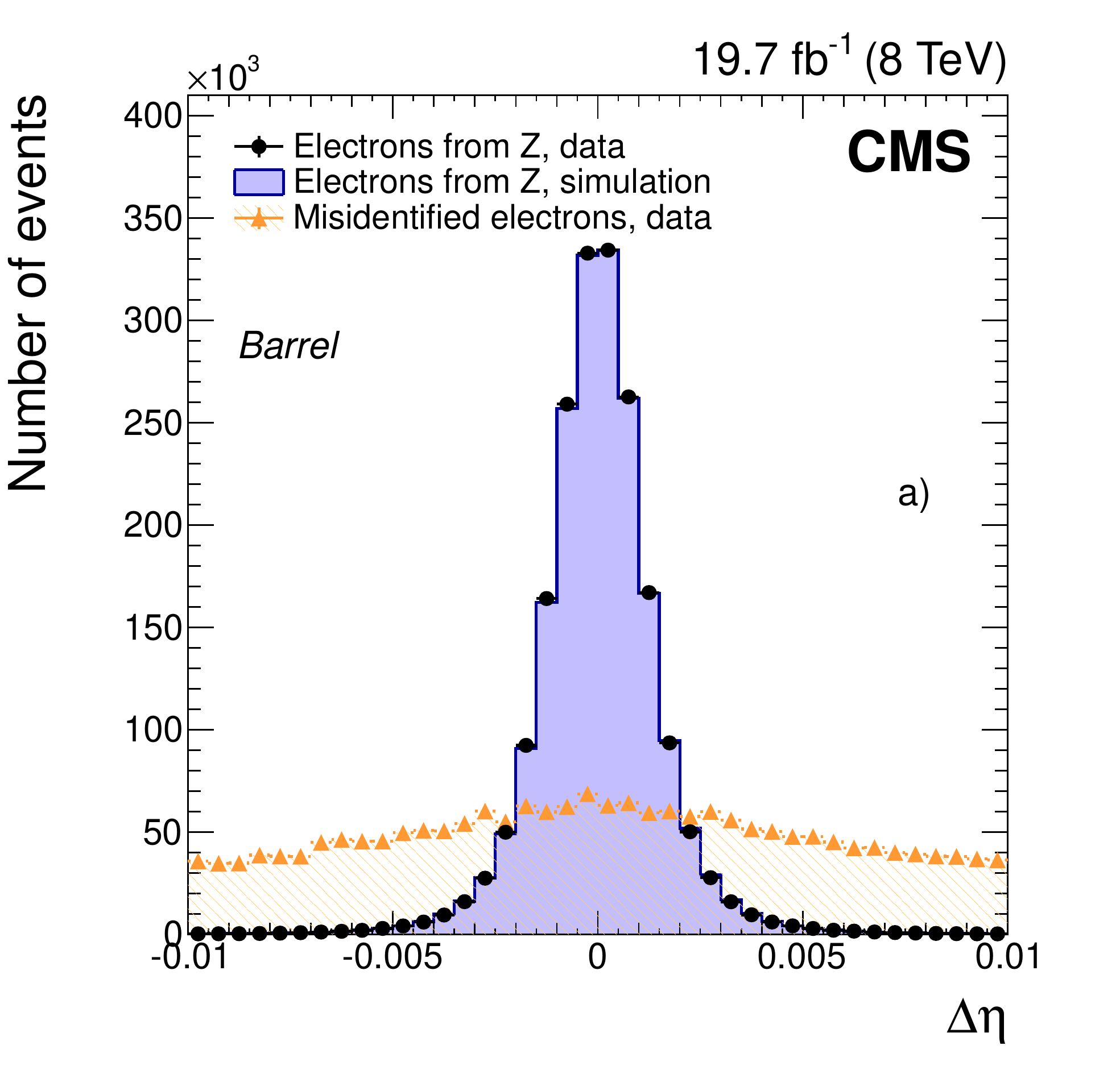}
\includegraphics[width=0.45\textwidth]{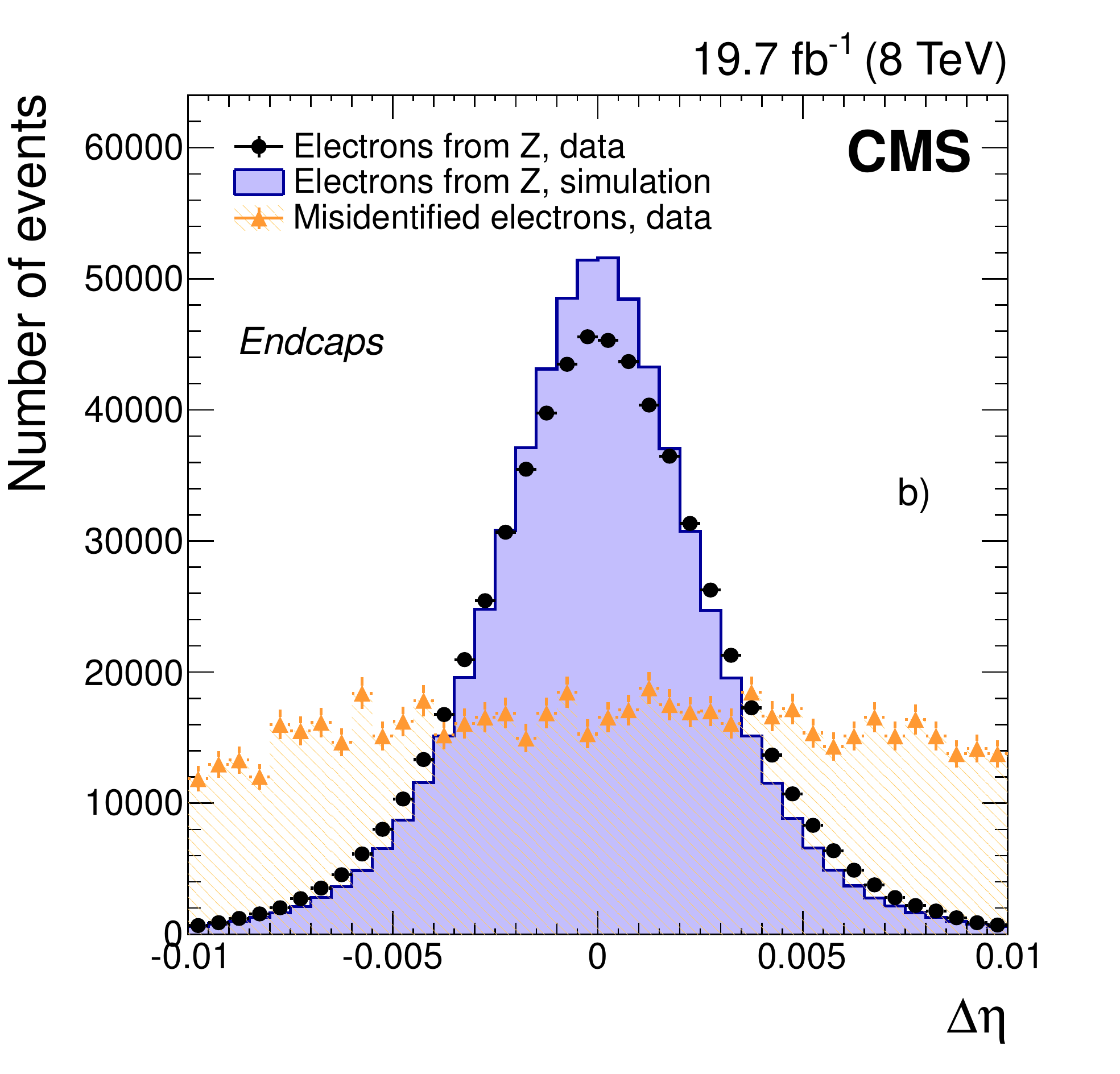}\\
\includegraphics[width=0.45\textwidth]{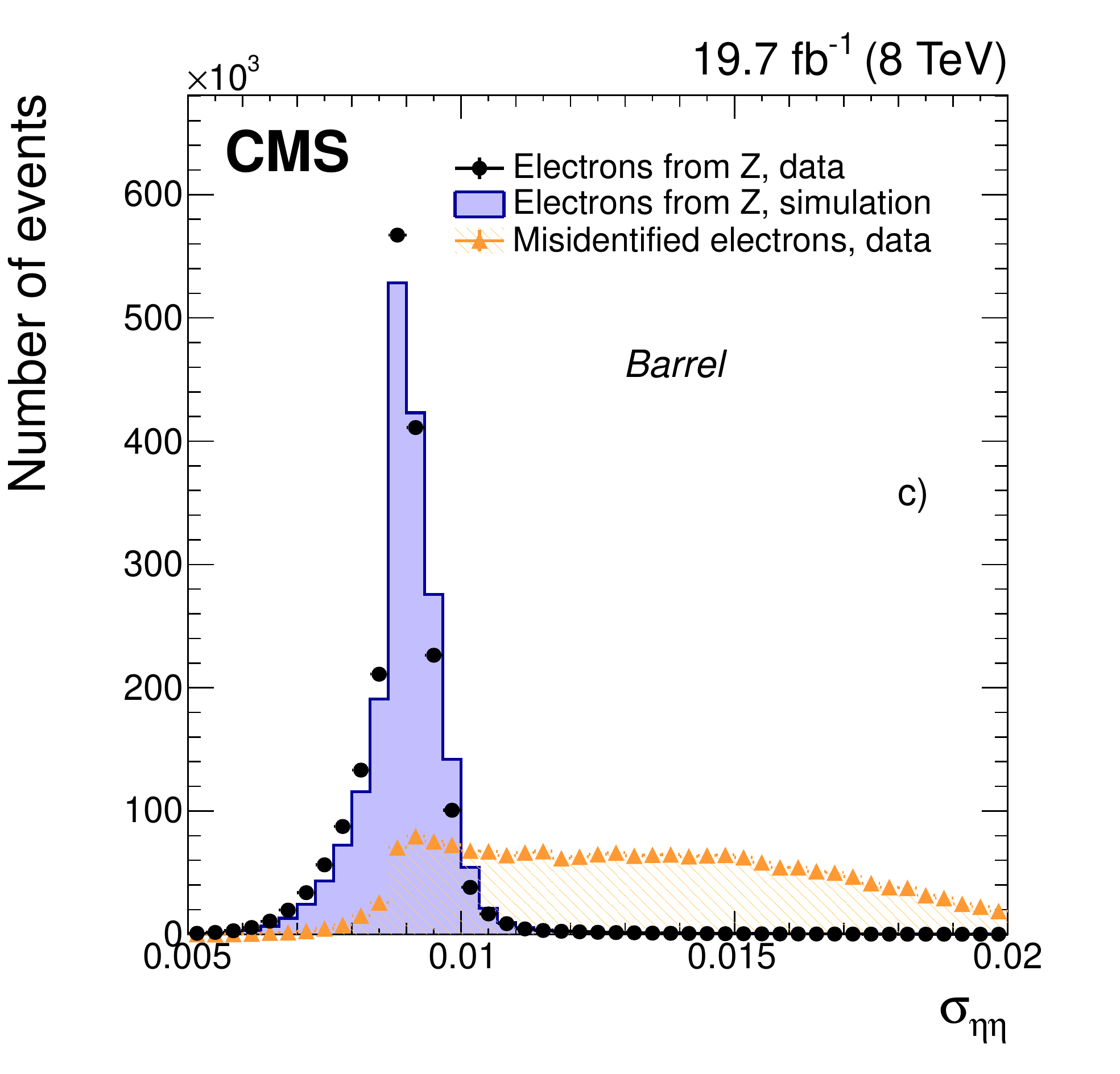}
\includegraphics[width=0.45\textwidth]{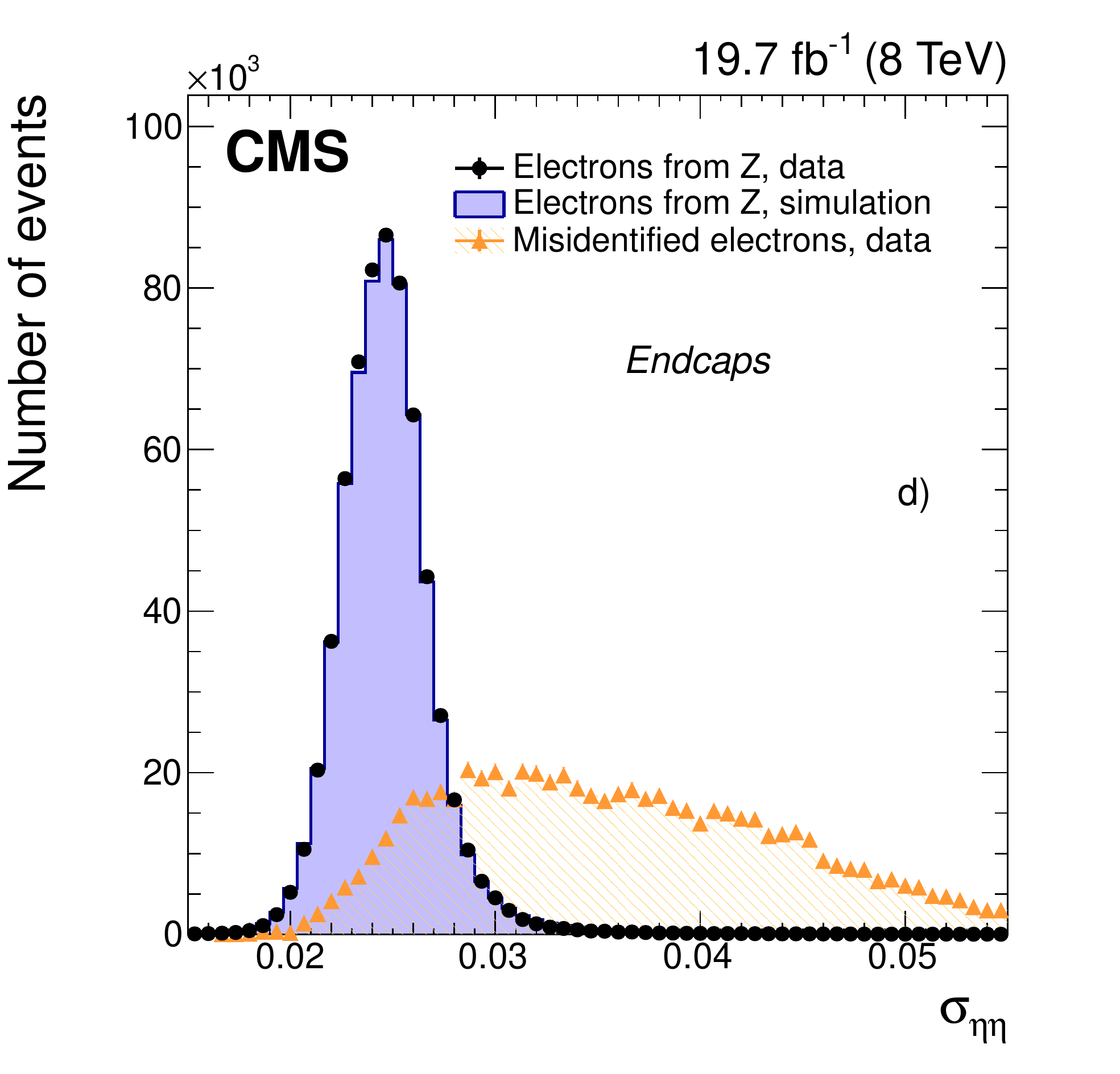}\\
\includegraphics[width=0.45\textwidth]{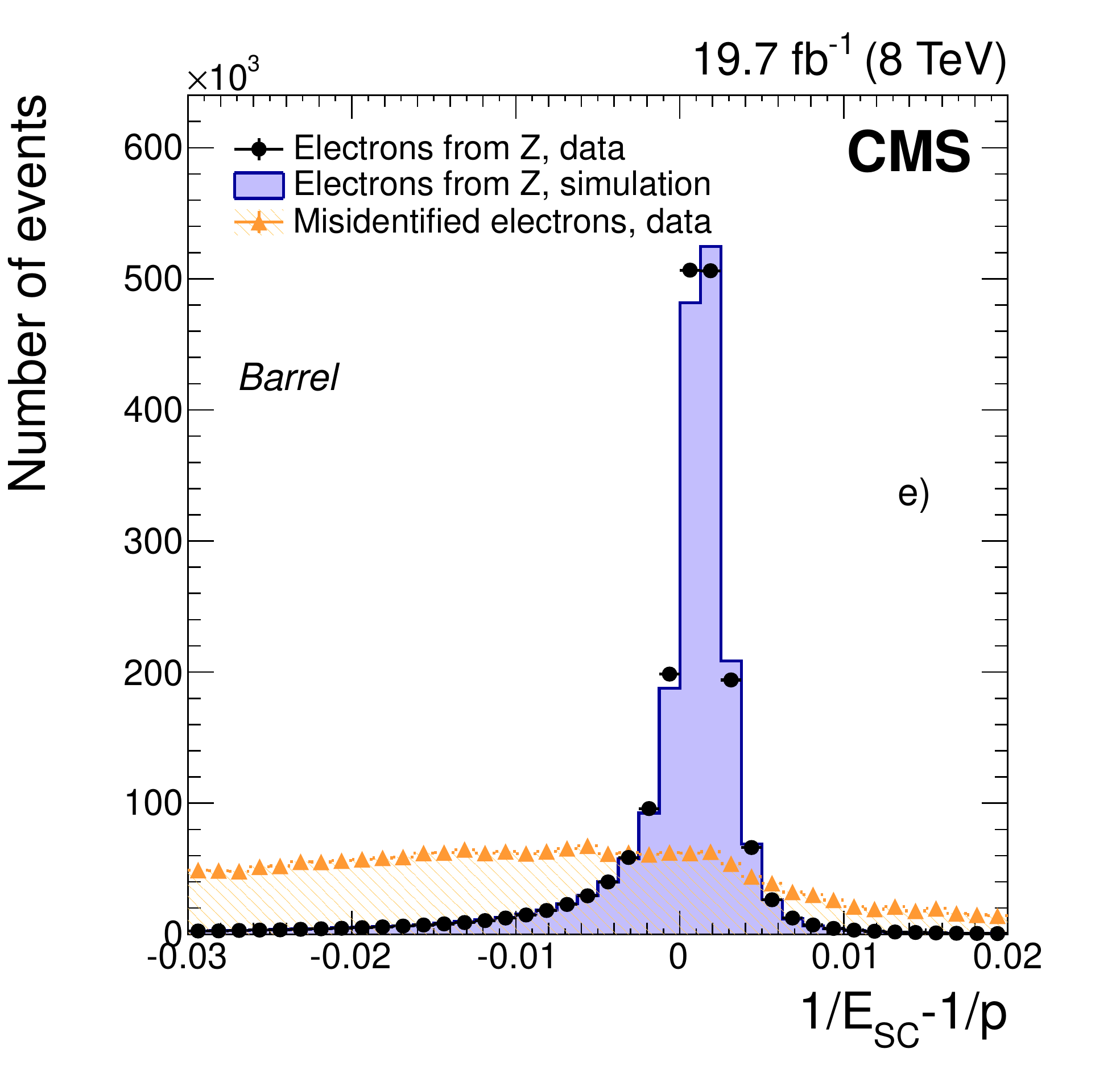}
\includegraphics[width=0.45\textwidth]{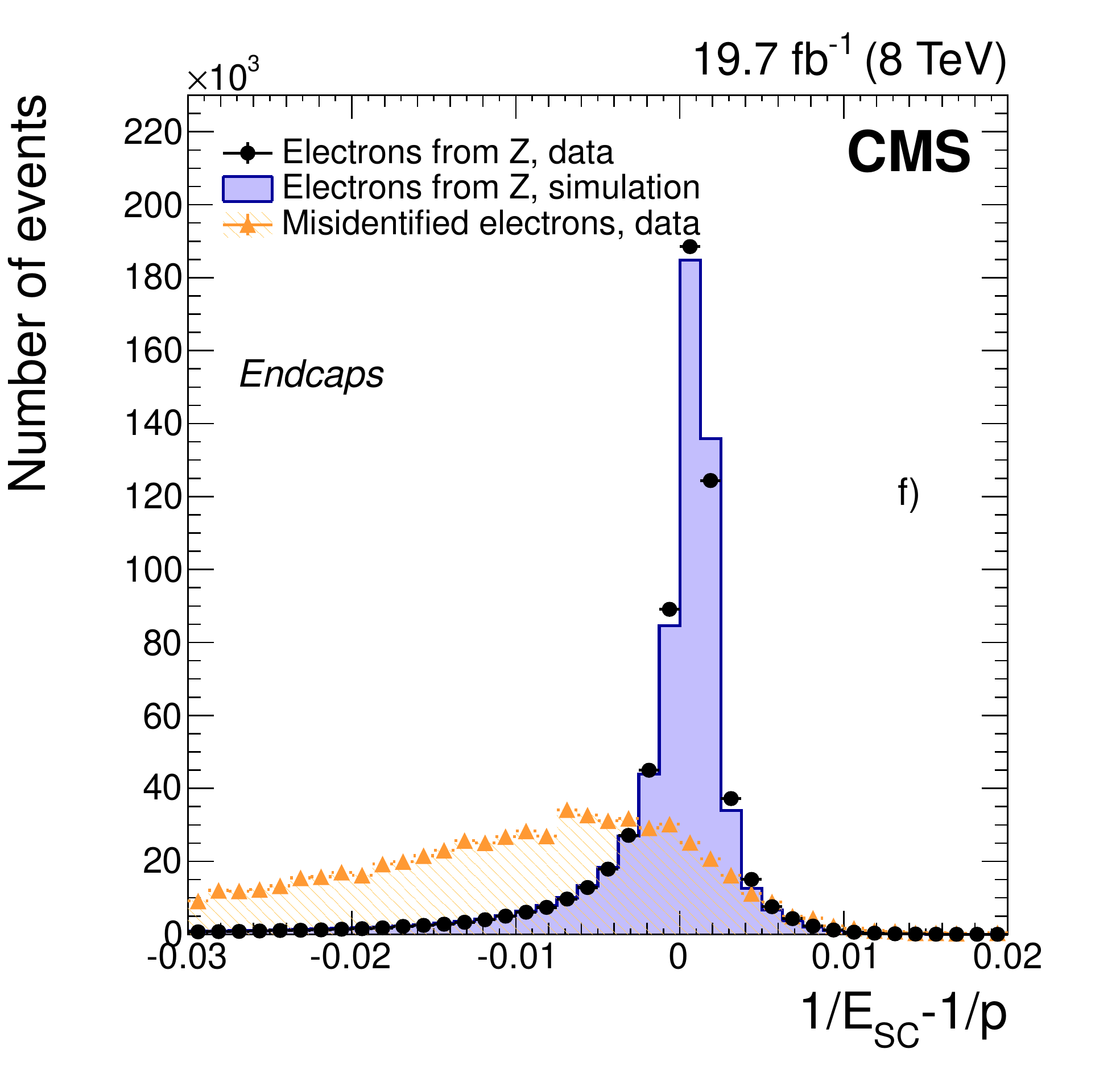}\
\caption{Distributions in the distance $\Delta \eta$ between the position of the SC and the track extrapolated to the point of closest approach to the SC are shown for a) the barrel and b) the endcaps. Distributions in the shower-shape variable $\sigma_{\eta \eta}$, defined in the text, are shown in c) and d). Distributions in energy-momentum matching $1/E_{\mathrm{SC}}- 1/p$, as defined in the text, are shown in e) and f).
Distributions are shown for electrons from $\Z\to \Pep\Pem$ data (dots) and simulated (solid histograms) events, and from background-enriched events in data (triangles).
All distributions are normalized to their respective areas of the $\Z\to \Pep\Pem$ data.
(See text for details on the samples composition.)
}
\label{fig:variables_SoB_dataMC}\end{figure*}

An example of the purely-tracking variable $f_{\text{brem}}$ was given in Fig.~\ref{fig:fbrem}.
Figure~\ref{fig:variables_SoB_dataMC} shows examples of ECAL-only and track--cluster matching variables.
The simulated signal consists of reconstructed electrons compatible with those generated from
$\Z\to \Pep\Pem$ decays, using a run-dependent version of the simulation.
The data are electrons reconstructed in a sample dominated by $\Z\to \Pep\Pem$ events.
To achieve sufficient purity in data, a stringent requirement of $\abs{m_{\Pep\Pem}-m_\Z}<7.5\GeV$
is made again in data and in simulation, on
the invariant mass of the two electrons.
Both electrons are required to be isolated: for each electron, the scalar sum of the transverse
momenta of the PF candidates in a cone around its direction (excluding the electron) is required to be
$<$10\% of the electron \pt.
The background sample consists of misidentified electrons from jets in Z+jets data.
This sample is selected by requiring a pair of identified leptons (electrons or muons) with an invariant mass
compatible with that of the \Z boson. To suppress the contribution from
events with associated production of W and \Z bosons,
the imbalance in the transverse momentum of the event is required to be smaller than 25\GeV (which also suppresses \ttbar events).
One additional electron candidate must be present in the event, which is required
not to be isolated by inverting the selection used for signal.
In the $\Pep\Pem$+jets events, the invariant mass of the dielectron pair with one misidentified-electron candidate
and an electron of opposite sign from the $\Z\to \Pep\Pem$ decay must be greater than 4\GeV,
in order to reject contributions from lower-mass resonances.
As a consequence of these requirements, the control sample consists largely of events
with one \Z boson and one jet that is misidentified as the additional electron.
All signal and background electrons are also required to have $\pt>20\GeV$
and satisfy some simple criteria to reject electrons from photon conversions.

The distance $\Delta\eta$, previously defined in Section~\ref{sec:association}, is shown in
Figs.~\ref{fig:variables_SoB_dataMC}~a) and b).
The agreement between data and simulation is very good for electrons in the barrel. Disagreement is observed in
the endcaps,
which is related to the mismodelled material in simulation. The $\Delta\eta$ indeed
increases with the amount of \bremns, which for the endcaps is somewhat larger in data than in simulation.

The lateral extension of the shower along the $\eta$ direction
is expressed in terms of the variable $\sigma_{\eta \eta}$, which is defined as
$(\sigma_{\eta \eta})^2 = [\sum (\eta_i - \overline{\eta})^2 w_i]/\sum w_i$.
The sum runs over the 5$\times$5 matrix of crystals around the highest \ET crystal of the SC,
and $w_i$ is a weight that depends logarithmically on the contained energy.
The positions $\eta_i$ are expressed in units of crystals,
which has the advantage that the variable-size gaps between ECAL crystals
(in particular at modules boundary) can be ignored.
The variable $\sigma_{\eta \eta}$ is shown in Figs.~\ref{fig:variables_SoB_dataMC}~c) and d).
The discrimination power of $\sigma_{\eta \eta}$ is greater than the analogous variable in $\phi$,
because
\brem strongly affects the pattern of energy deposition in the ECAL along the $\phi$ direction.
A small disagreement between data and simulation is visible in the barrel,
and is mainly due to the limited tuning of electromagnetic showers in simulation
(improved in \GEANTfour Release 10.0~\cite{Geant4Release10}).
For electrons in the endcaps, the main factor determining the resolution of the shower-shape
variables is the pileup. Since this is well described in the run-dependent version
of simulation, the agreement between data and simulation in these plots is regarded as quite good.

Finally, Figs.~\ref{fig:variables_SoB_dataMC}~e) and f) show the distributions in
$1/E_{\mathrm{SC}}- 1/p$, where $E_{\mathrm{SC}}$ is the SC energy
and $p$ the track momentum at the point of closest approach to the vertex.
Good agreement is observed between data and simulation
both in the barrel and in the endcaps.
In all cases, the distributions for signal and background electrons are well separated.

\begin{figure*}[hbtp]
\centering
\includegraphics[width=0.49\textwidth]{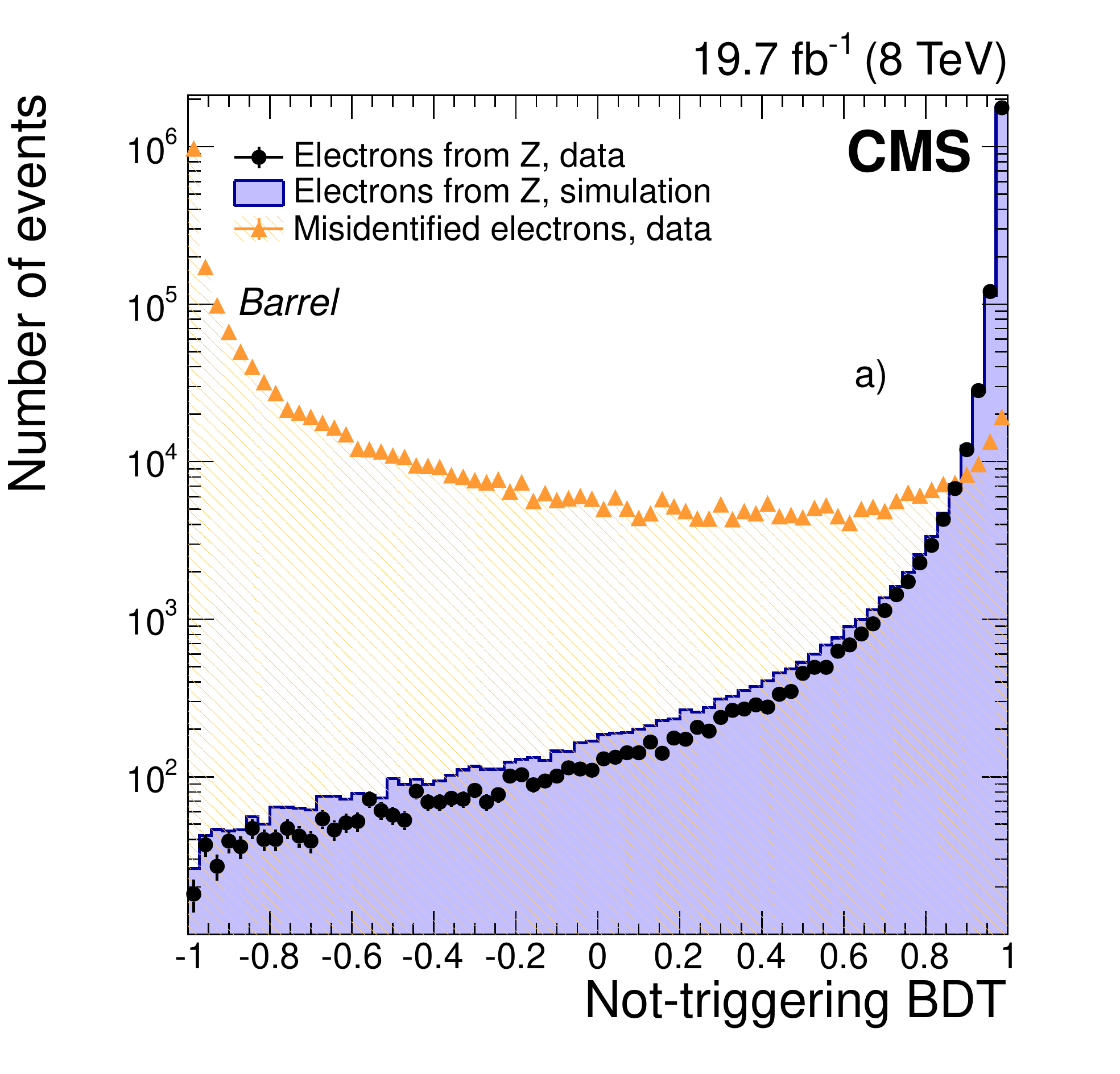}
\includegraphics[width=0.49\textwidth]{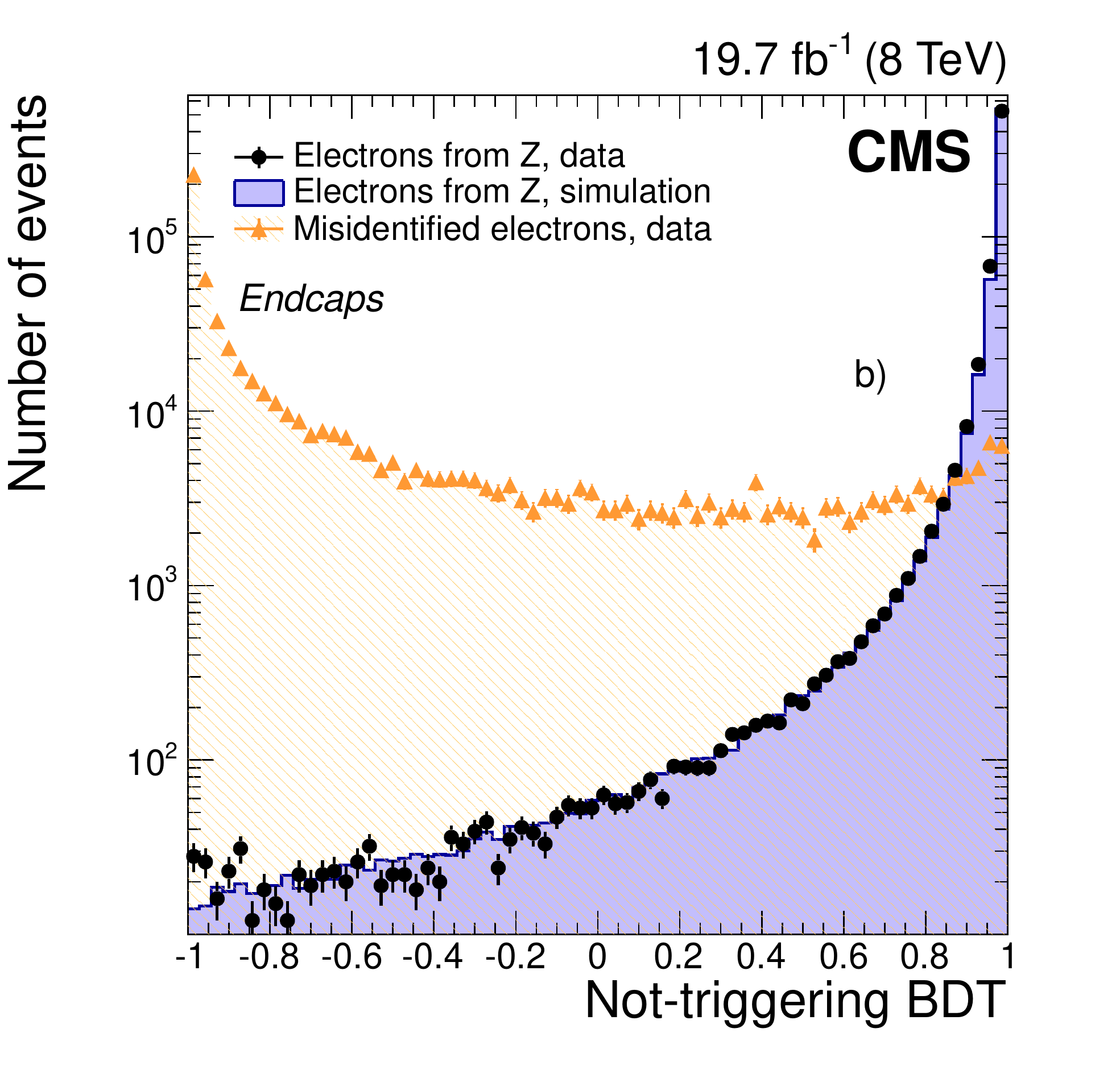}
\caption{
Output of the electron-identification BDT for electrons from $\Z\to \Pep\Pem$ data (dots) and simulated (solid histograms) events, and from background-enriched events in data (triangles),
in the ECAL a) barrel, and b) endcaps.
All the distributions are normalized to the area of the respective $\Z\to \Pep\Pem$ data.
(See text for details on the samples composition.)
}
\label{fig:MVAoutputSB_dataMC}\end{figure*}

To maximize the sensitivity of electron identification, several variables are
combined using
the ``boosted decision tree'' (BDT) algorithm~\cite{TMVA}.
The set of observables in each category is extended relative to the simpler sequential selection
as follows:
the track--cluster matching observables are computed both at the ECAL surface and at the vertex,
the SC substructure is exploited, more information related to the cluster shape is used, as well as
the $f_{\text{brem}}$ fraction. Similar sets of variables are used for electrons
in the barrel and in the endcaps.
Two types of BDT are defined that depend on whether the electron passes HLT identification
requirements (``\textit{triggering} electron'') or does not (``\textit{not-triggering} electron'').
For triggering electrons, loose identification and isolation requirements
are applied as a preselection, to mimic the requirements applied at the HLT.
Dedicated training then can exploit the variables discriminating
power at best in the remaining phase space.
In the following, results are presented just for not-triggering electrons,
since the training and performance of the two algorithms are similar.
The BDT is trained in several bins of \pt and $\eta$. To model
the signal,
reconstructed electrons are used when they match electrons
with \pt in the range between 5 and 100\GeV
in generated events.
The background is modelled using misidentified electrons reconstructed in \PW+jets events in data.
The distribution of variables in these training samples is found
to be in agreement with the one observed in the samples used in the analyses.
The signal and background BDT output distributions are compared in Fig.~\ref{fig:MVAoutputSB_dataMC},
where there is also a comparison given between data and simulation for signal electrons.
The same selections are used as in Fig.~\ref{fig:variables_SoB_dataMC},
and the same signal and background samples.
The discriminating power of the BDT algorithm is evident, and
the agreement between data and simulation is good.
The small difference observed is due to the differences in input variables,
which were described in the previous paragraphs.

\begin{figure*}[hbtp]
\centering
\includegraphics[width=0.49\textwidth]{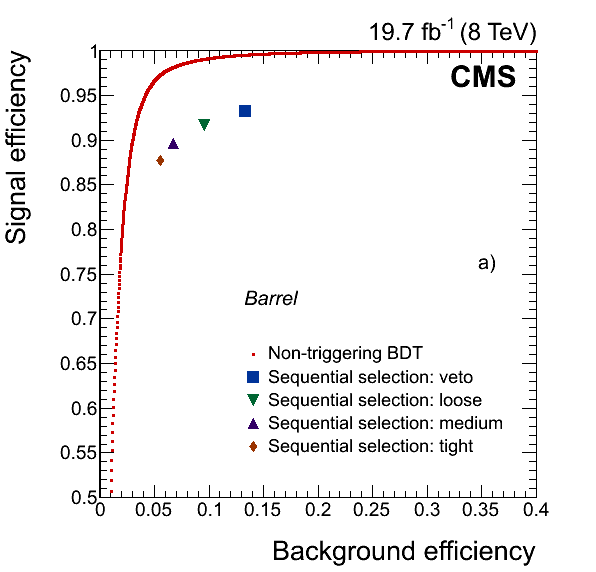}
\includegraphics[width=0.49\textwidth]{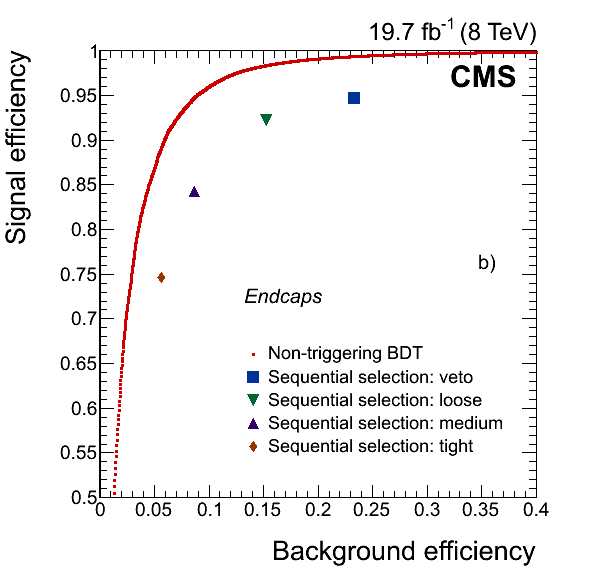}
\caption{Performance of the electron BDT-based identification algorithm (red dots) compared
with results from working points of the sequential selection (only the identification part)
for electron candidates in the ECAL a) barrel, and b) endcaps.
(See text for details on the samples composition.)
}
\label{fig:mvaIdRocs}\end{figure*}

The results on the performance of the BDT-based and the sequential electron-identification algorithms for four selected working points are compared in Fig.~\ref{fig:mvaIdRocs} for electrons
with $\pt>20$\GeV.
Signal electrons from $\Z\to \Pep\Pem$ events in a simulated sample
are compared with misidentified electrons from jets reconstructed in data.
The same selections and samples are used as in Fig.~\ref{fig:variables_SoB_dataMC}.
As expected, better performance is obtained when the variables are combined in an MVA
discriminant such as the BDT.
In the ECAL barrel and endcaps, a working point of the sequential selection with respective efficiency for signal electrons
of about 90\% and 84\%, has an efficiency of about 7\% and 9\% on background electrons.
For the same signal efficiency, the misidentification probability using the BDT algorithm is reduced
by about a factor of two.

\begin{figure*}[hbtp]
\centering
\includegraphics[width=0.49\textwidth]{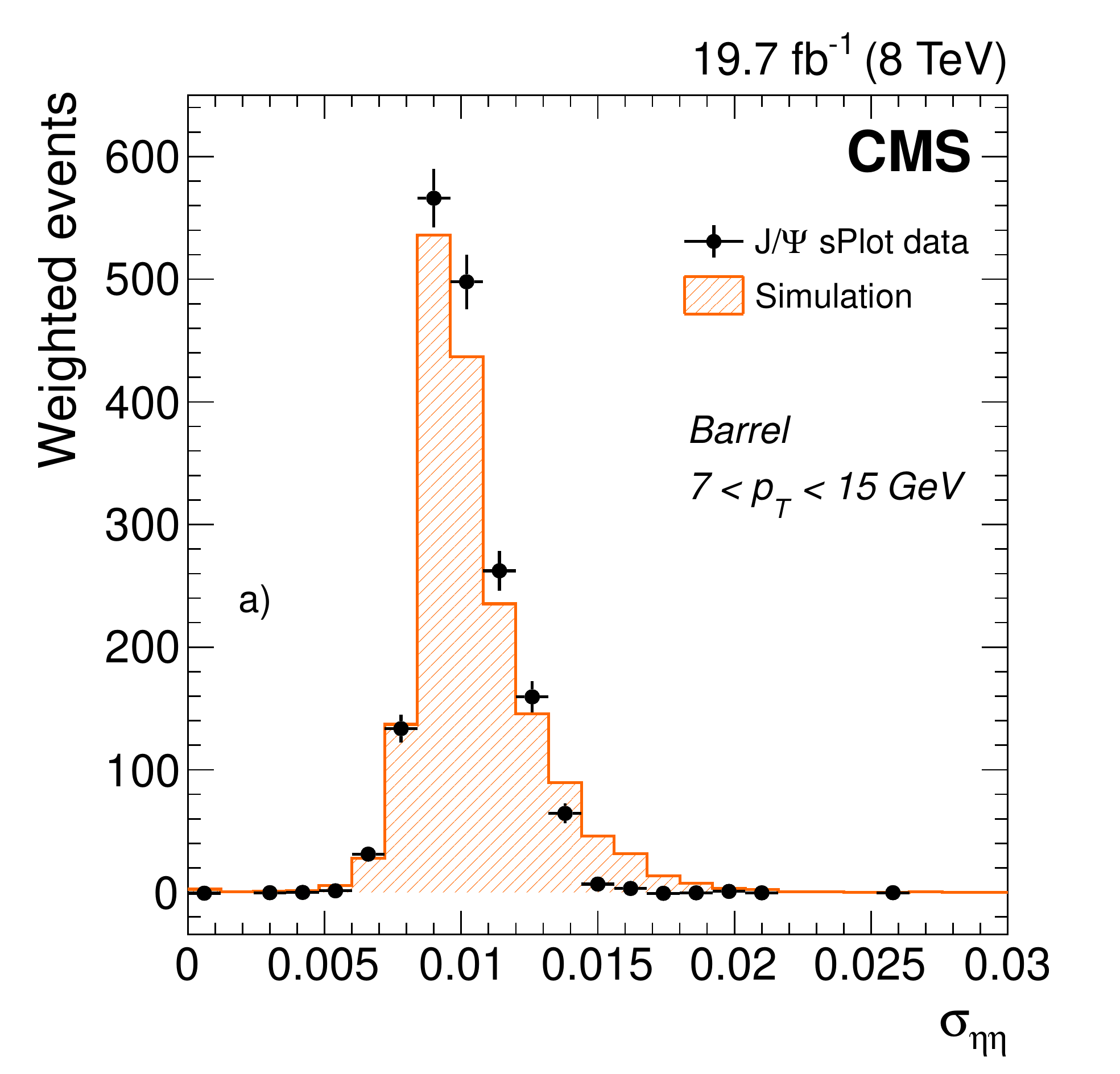}
\includegraphics[width=0.49\textwidth]{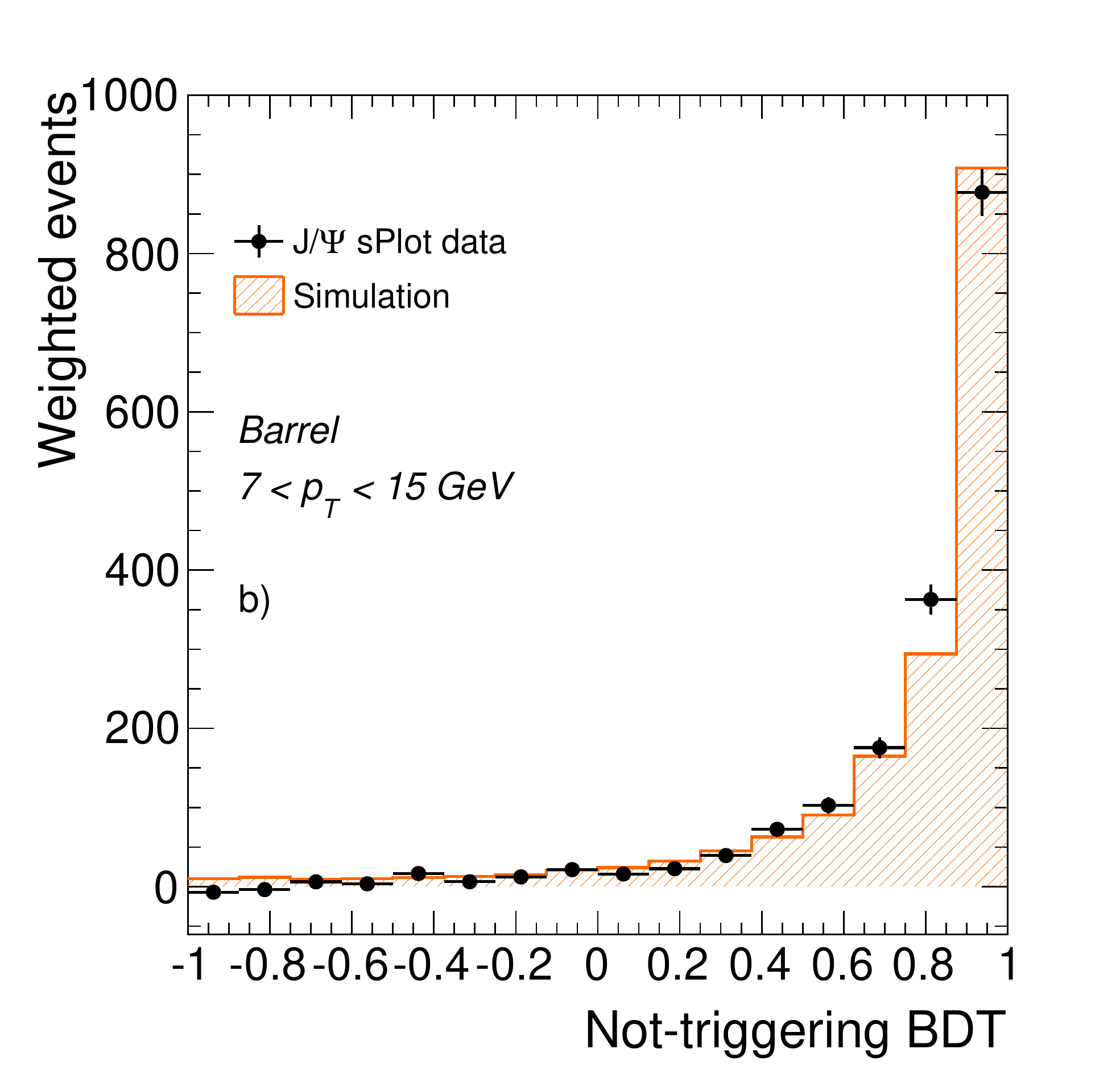}\\
\caption{Distribution of a) the shower-shape variable $\sigma_{\eta \eta}$, defined in the text, and b) the output of the BDT electron identification algorithm for electron candidates in the ECAL barrel, in data (symbols) and simulation (histograms).
A statistical subtraction of the background is applied using the
\sPlot
technique. (See text for details.)}
\label{fig:jpsi}\end{figure*}

Although the focus of the analysis thus far has been on electrons with $\pt > 20$\GeV,
this identification strategy is also adopted at smaller \pt. The
agreement between data and simulation in the $\pt$ range between 7
and 15\GeV was studied using electrons from $\JPsi$ meson decays. As
an illustration, Fig.~\ref{fig:jpsi} shows a comparison between data and simulation
for two variables, using events with both electrons in the barrel, and the
run-dependent version of simulation. The remnant background
is subtracted statistically, using the
\sPlot
technique~\cite{splots}, through a
fit to the dielectron invariant mass. The agreement between data and
simulation is very good both for variables such as $\sigma_{\eta \eta}$
in Fig.~\ref{fig:jpsi} a), but also for more complex ones, such as the BDT output
shown in Fig.~\ref{fig:jpsi} b).

\subsection{Isolation requirements}
\label{sec:eleIso}
A significant fraction of background to isolated primary electrons
is due to misidentified jets or to genuine electrons within a jet resulting from
semileptonic decays of b or c quarks.
In both cases, the electron candidates have
significant energy flow near their trajectories, and requiring electrons
to be isolated from such nearby activity greatly reduces these sources of background.
The isolation requirements are separated from electron identification, as the interplay between them
tends to be analysis-dependent. Moreover, the inversion of isolation requirements,
independent of those used for identification, provides control of different sources of
such backgrounds in data.

Two isolation techniques are used at CMS.
The simplest one is referred to as detector-based isolation,
and relies on the sum of energy depositions either in the ECAL or in the HCAL around each electron trajectory,
or on the scalar sum of the \pt of all tracks reconstructed
from the collision vertex.
These sums are usually computed within cone radii of $\Delta R=0.3$ or 0.4
around the electron direction, and remove contributions from the electron through smaller exclusion cones.
This procedure, which has good performance in rejecting jets misidentified as electrons, is used by the HLT,
and in certain analyses in which just mild background rejection suffices.

Most of the offline analyses, however, benefit from the PF
technique for defining isolation quantities. Rather than using energy measurements in independent
subdetectors, the isolation is defined using the PF candidates reconstructed
with a momentum direction within some chosen cone of isolation. In this way,
the correct calibration can be used,
and a possible double-counting of energy assigned to particle candidates is avoided.
When an electron candidate is misidentified by the PF as another particle,
it enters the isolation sum, and artificially
increases the size of the isolation observable.
This effect increases when the identification efficiency of the PF decreases.
Electron-candidate identification using PF performs very well for electrons
in the ECAL barrel, where no additional
corrections
for removing electron contributions to the
isolation sum are needed.
However, in the endcaps, and in the version of the reconstruction used for the results discussed in this paper, the electron identification applied through the PF is not fully efficient. Therefore, in line with what is done in the detector-based approach, veto cones are applied for charged hadrons and photons when the isolation sums are computed.

\begin{figure*}[hbtp]
\centering
\includegraphics[width=0.45\textwidth]{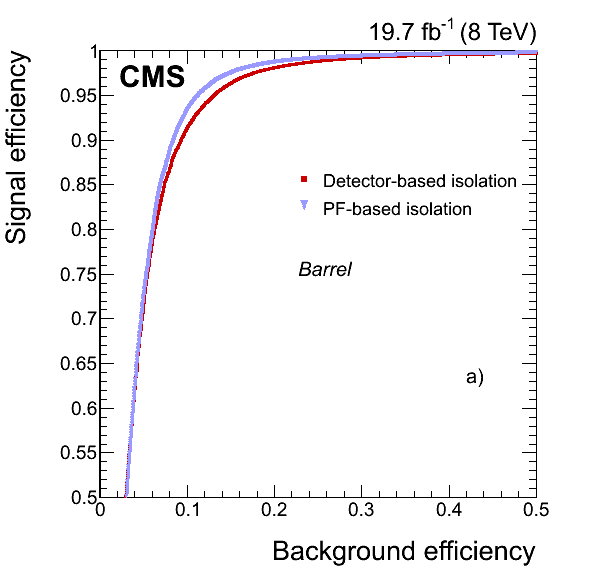}
\includegraphics[width=0.45\textwidth]{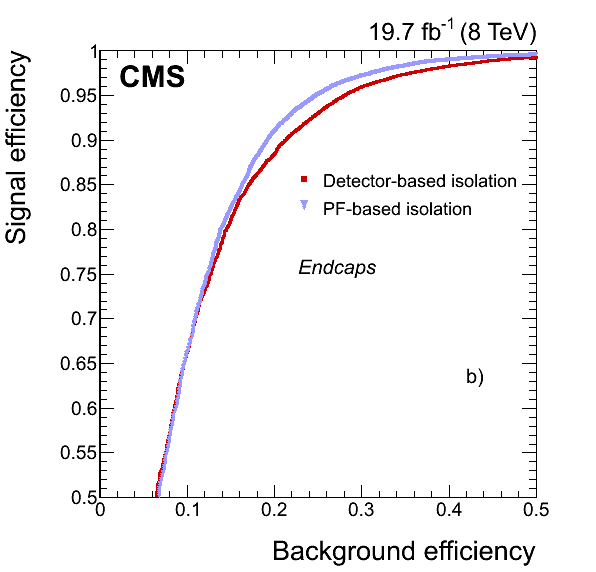}
\caption{Performance of the detector-based isolation algorithm (red squares) compared with that using PF (blue triangles) in the ECAL a) barrel, and b) endcaps.
(See text for the definition of the samples.)
}
\label{fig:detVsPFiso}\end{figure*}

A comparison between the performance of the two techniques is given in Fig.~\ref{fig:detVsPFiso}
for electrons with $\pt>20\GeV$ (with no pileup correction applied). Signal electrons from
$\Z\to \Pep\Pem$ events in a simulated sample
are compared with misidentified electrons from jets
reconstructed in Z+jets data.
The run-dependent version of the simulation is used.
A loose identification is applied in reconstructing PF electrons,
and only the electron candidates that pass this selection are considered in performing a meaningful comparison.
Better performance is obtained when the information
from all detectors is combined using the PF technique, especially in the endcaps.

\begin{figure*}[hbtp]
\centering
\includegraphics[width=0.49\textwidth]{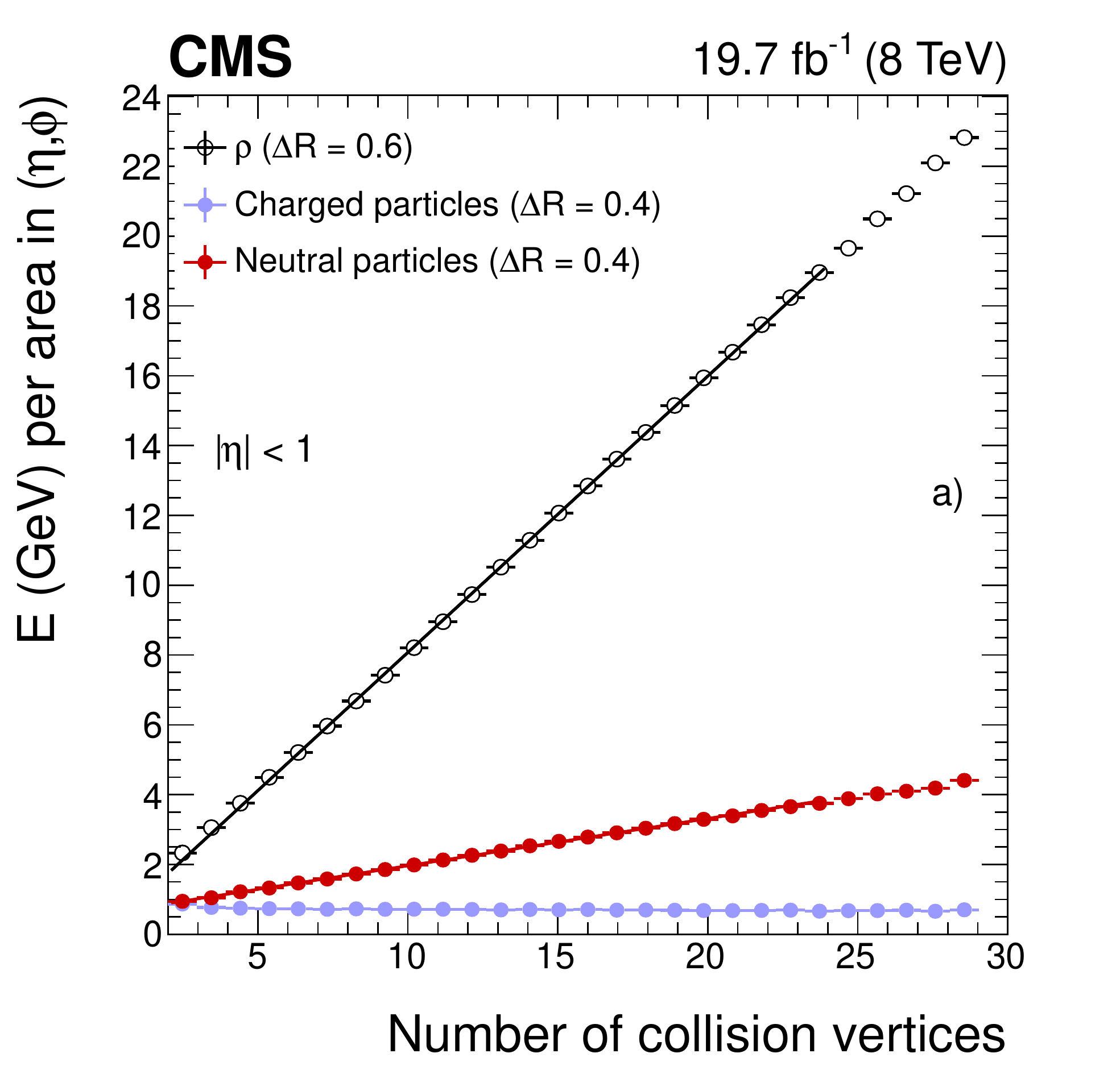}
\includegraphics[width=0.49\textwidth]{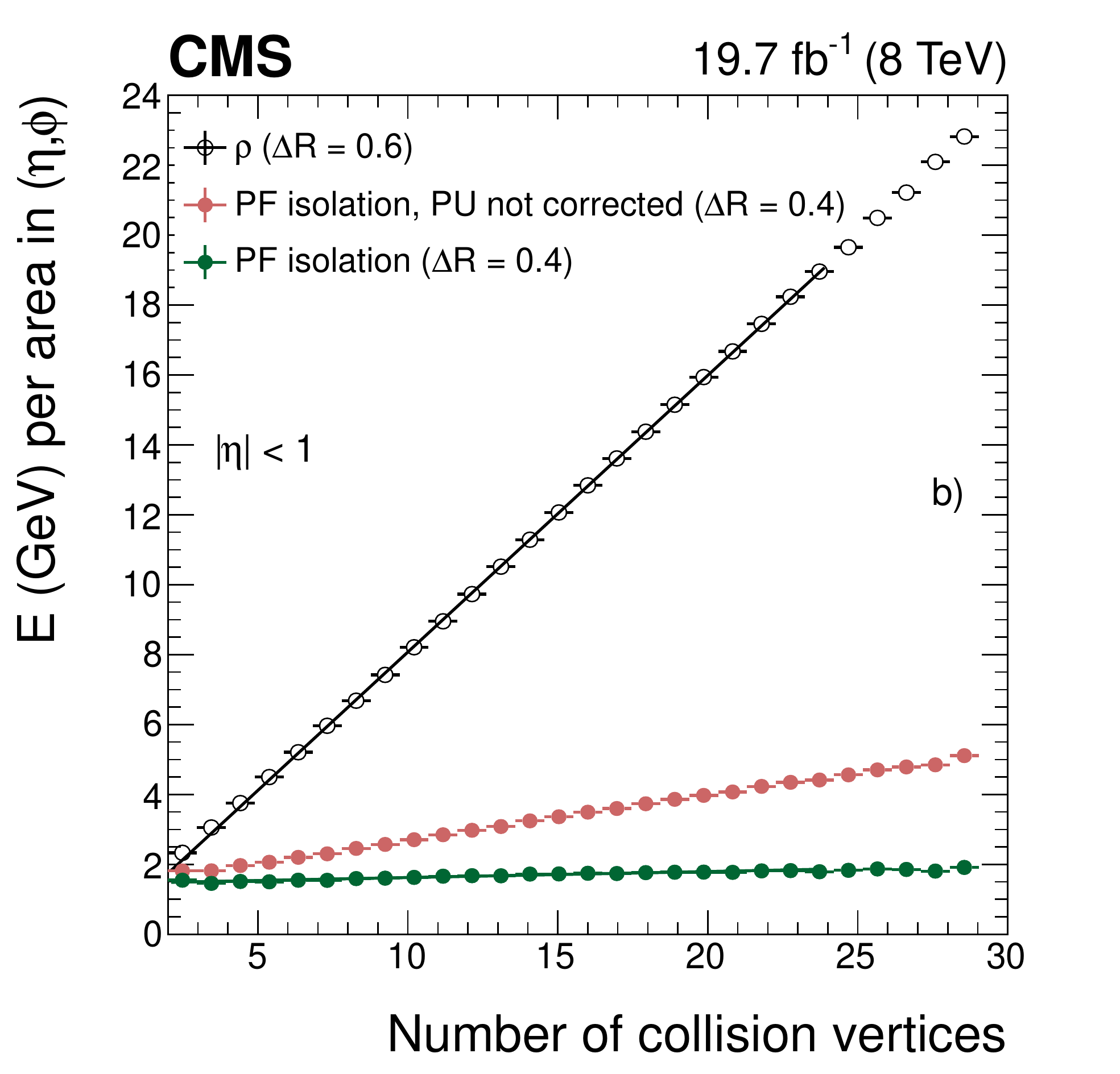}\\
\caption{Average energy density
as a function of the number of reconstructed proton-proton collision vertices,
for electron candidates with $\pt> 20\GeV$ and $\abs{\eta}<1$ from data dominated
by $\Z \to \Pep\Pem$ events.
The energy density $\rho$ (open dots) is shown, along with each component of the particle isolation: a) neutral particles (red dots) and charged particles associated with the vertex (blue dots), and b) before (pink dots) and after (green dots) the correction for pileup on PF isolation.
}
\label{fig:pucorrForIso}\end{figure*}

The PF isolation is defined as
\begin{equation}
\label {eq:iso}
\mathrm{Iso_{\mathrm{PF}}} = \sum \pt^{\text{charged}} + \max\Bigl[0,\; \sum \pt^{\text{neutral had}} + \sum \pt^{\gamma} - \pt^{\mathrm{PU}}\Bigr],
\end{equation}
where the sums run over the charged PF candidates, neutral hadrons and photons,
within a chosen $\Delta R$ cone around the electron direction. The charged candidates
are required to originate from the vertex of the event of interest,
and $\pt^{\mathrm{PU}}$ is a correction related to event pileup.
The isolation-related quantities are among the observables most sensitive to the extra energy
from pileup interactions (either occurring in the same or earlier bunch crossings), which spoils the isolation efficiency
when there are many interactions per bunch crossing.
The contribution from pileup in the isolation cone,
which must be subtracted, is computed using the \textsc{FastJet} technique~\cite{fastjet1,fastjet2,fastjet3},
assuming $\pt^{\mathrm{PU}} = \rho A_{\text{eff}}$
(the variable $\rho$ is defined in Section~\ref{sec:scEneCorr}).
The dependence of $\rho$ on pileup is shown in Fig.~\ref{fig:pucorrForIso}~a),
and refers to electrons selected in a data sample dominated by $\Z \to \Pep\Pem$ events.
The dependence of both the charged and neutral components of the PF-based
isolation is also shown as a function of the number of reconstructed proton-proton collision vertices.
The charged component of the isolation becomes independent of pileup
once only candidates compatible with the vertex of interest are considered. For both $\rho$
and the neutral component of the isolation, the dependence is almost linear.
The effective area $A_{\text{eff}}$ in ($\eta$, $\phi$)
is defined, for each component of the isolation, by $(\Delta R)^2$, scaled by
the ratio of the slopes for $\rho$ and for the considered component shown in Fig.~\ref{fig:pucorrForIso}~a).
Once the correction is applied to the neutral components, the dependence on the number of vertices is much reduced,
as shown in Fig.~\ref{fig:pucorrForIso}~b). The plots refer to electrons with
 $\abs{\eta}<1$, but similar conclusions hold in any range of $\eta$.

\begin{figure*}[hbtp]
\centering
\includegraphics[width=0.49\textwidth]{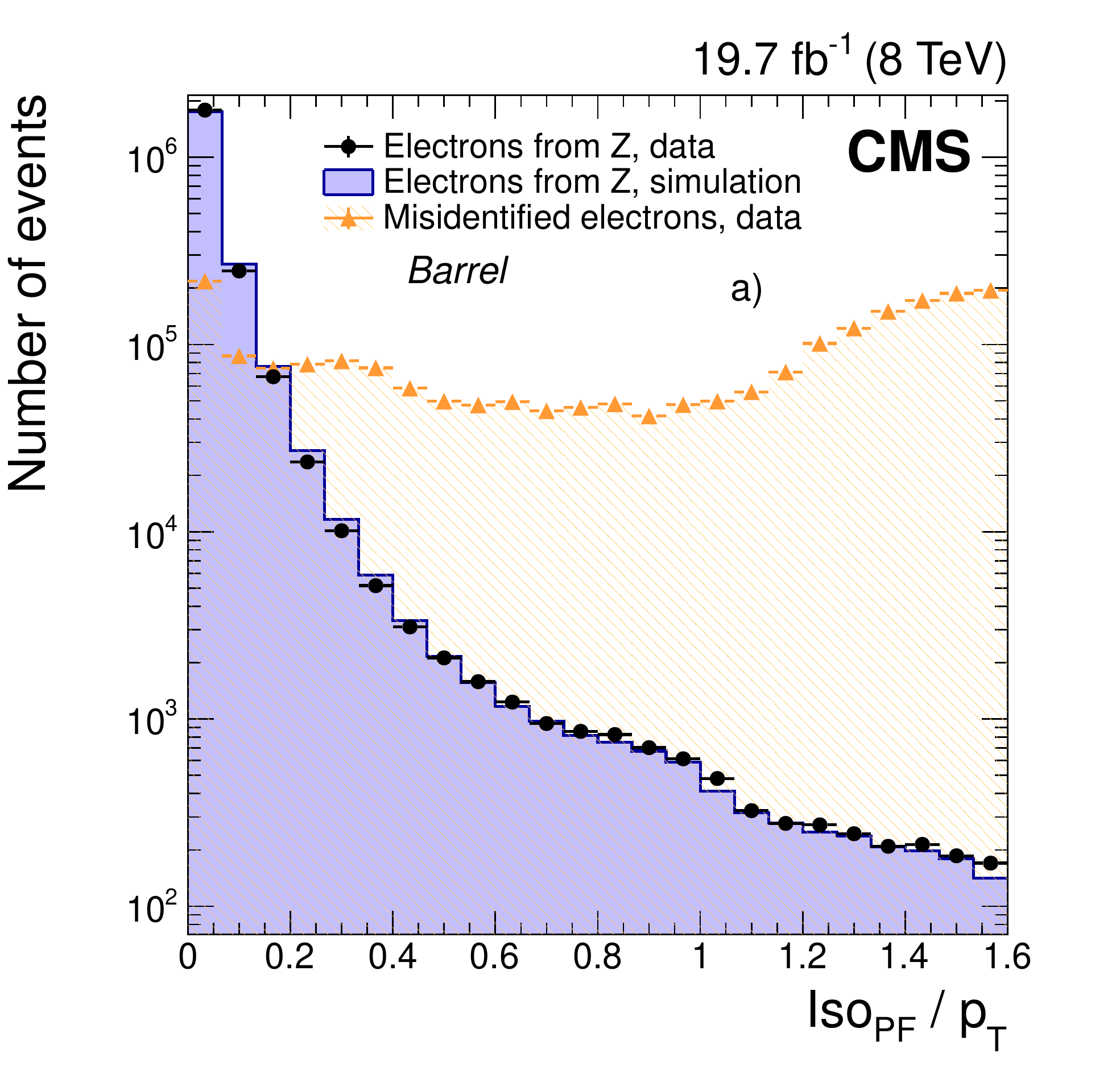}
\includegraphics[width=0.49\textwidth]{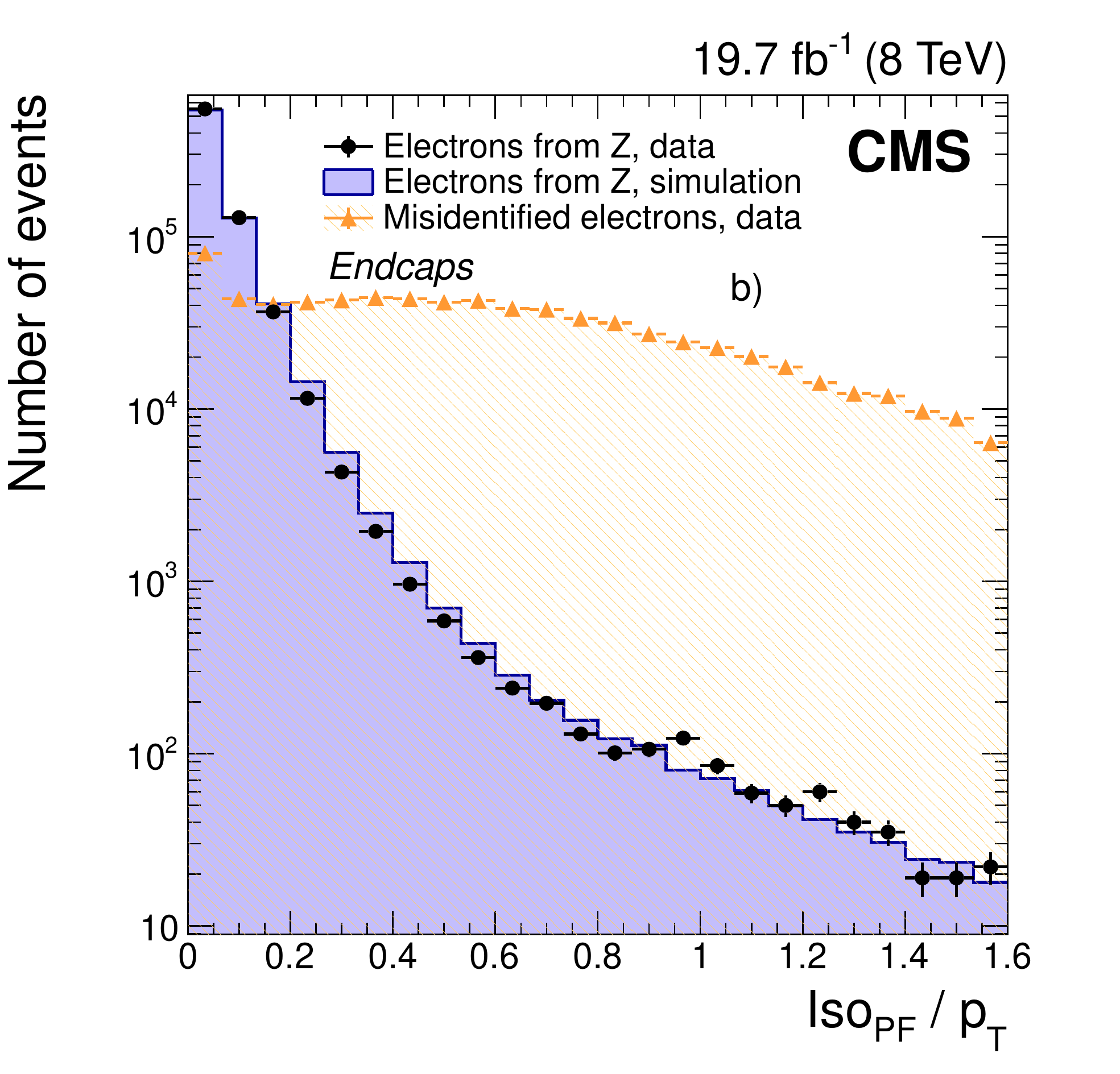}
 \caption{Distributions of PF isolation divided by electron \pt, after applying the pileup
correction discussed in the text, for electrons from $\Z \to \Pep\Pem$ data (dots) and simulated (solid histograms) events, and from background-enriched events in data (triangles), in the ECAL
a) barrel, and b) endcaps.
(See text for more details on the compositions of the samples.)}
\label{fig:iso_SoB_dataMC}\end{figure*}

Figure~\ref{fig:iso_SoB_dataMC} shows the distributions of the $\mathrm{Iso_{\mathrm{PF}}}$
variable divided by the electron \pt, for signal and background electrons, after the correction for pileup
contributions.
For signal electrons, both data and simulation are shown.
The samples and selection criteria presented in Section~\ref{sec:eleID}
are used without the isolation requirement which is replaced by a loose selection on the BDT identification discriminant.
Excellent discrimination is observed between signal and background, and there is also good
agreement between data and simulation.
The remnant discrepancy in the endcaps is mostly due to the difference of the PF
electron identification efficiency in data and in simulation,
which is reflected in different
contributions from misidentified particles to the
isolation sums as discussed above. This difference is not completely recovered through the use of the additional exclusion
cones.

\subsection{Rejection of converted photons}
\label{sec:convRej}
An important source of background to prompt electrons arises from secondary electrons produced in conversions of photons in the tracker material.

To reject this background, CMS algorithms exploit the pattern of track hits.
When photon conversions take place inside the volume of the tracker, the first hit on electron tracks from the converted photons
is often not located in the innermost layer of the tracker, and missing hits are therefore present
in that region.
For prompt electrons, whose trajectories
start from the beamline, no missing hits are expected in the inner layers.
In addition to the missing hits, photon conversion candidates can also be
rejected using a fit to the reconstructed electron tracks.
Since the photon is massless, and momentum transfer is in general small, the conversions
have a well defined topology, with tracks that have essentially the same
tangent at the conversion vertex in the $(r,\phi)$ and $(r,z)$ planes.
The strategy for rejecting these candidates consists of fitting the track pairs
to a common vertex,
incorporating this topological constraint, and then rejecting the
converted photon candidates according to the $\chi^2$ probability of the fit.
Also, the impact parameters ($\mathrm{ip}$) of the electron, such as the transverse ($d_0$)
and longitudinal ($d_z$) distance to the vertex at the point of closest approach
in the transverse plane, or the ratio of the uncertainties in the three-dimensional impact parameter
relative to its value
$(\sigma_\mathrm{ip}/\mathrm{ip})$
are used to reject secondary electrons.

Overall, when the requirement of no missing hits together with a selection on the
$\chi^2$ probability of the described fit to a common vertex are applied, the inefficiency for
prompt electrons in a simulated $\Z \to \Pep\Pem$ sample is of the order of a percent.
The rejection factor computed for the background data described in the previous
paragraphs is about 45\%.
These performance figures depend strongly on the selections applied to define the
electron candidates, since that affects the background composition,
and therefore the fraction of photon conversions.
The quoted numbers refer to electron candidates passing the ``MVA selection''
detailed in Section~\ref{sec:refSel}, without using the
selection based on the number of missing hits.

The algorithms described above are used in combination with other selection variables
discussed in the next section to select prompt electrons.

\subsection{Reference selections}
\label{sec:refSel}

Scientific analyses must balance efficiency and purity, depending on the levels of signal and background,
by defining their own electron selections through a combination of different algorithms.
This subsection summarizes some of the basic selections used widely at CMS.
The efficiency and misidentification rates, along with a discussion of a
tag-and-probe method used to check the performance, are given in Section~\ref{sec:TandP}.

The sequential selection applies requirements on five identification variables
among those discussed previously:
$\Delta \eta$, $\Delta \phi$, $H/E_{\mathrm{SC}}$, $\sigma_{\eta \eta}$, and
$1/E_{\mathrm{SC}}- 1/p_{\text{in}}$.
In addition, a selection is also applied on the combined PF isolation relative to the electron \pt,
and on the variables used to reject converted photons.
Finally, the impact parameters of the electron, $d_0$ and $d_z$, are required to be small for the electron
to originate from the vertex of interest.
The sequential selection, initially developed for measuring the W boson and \Z boson cross sections,
is still used in standard model analyses,
where the yield of signal is not too small so that the efficiency is not the most important issue.
Three working points were originally designed to have average efficiencies of about
90, 80, and 70\% for electrons
from $\Z \to \Pep\Pem$ events, and were optimized separately for electrons in the ECAL
barrel and endcaps.
For the analysis of 8\TeV data, four working points are defined: loose, medium, tight,
and a very loose point for analyses aiming at vetoing electrons.
The selections corresponding to the medium working point
are given in Table~\ref{tab:cbcuts}.
\begin{table}[htb]
\centering
\topcaption{\label{tab:cbcuts} Requirements corresponding to the medium working point of the sequential selection for electrons in the ECAL barrel and endcaps. At most one missing hit is allowed.}
\begin{tabular}{*{3}{l}}
\hline
\multicolumn{1}{c}{Variable} & \multicolumn{1}{c}{Upper value, barrel} & \multicolumn{1}{c}{Upper value, endcaps} \\
\hline
$\abs{\Delta\eta}$       & 0.004 & 0.007 \\
$\abs{\Delta \phi}$       & 0.06 rad  & 0.03 rad \\
$H/E_{\mathrm{SC}}$                 & 0.12  & 0.10  \\
$\sigma_{\eta \eta} $ & 0.01  & 0.03  \\
$\abs{1/E_{\mathrm{SC}} - 1/p}$         & 0.05 $\GeV^{-1}$ & 0.05 $\GeV^{-1}$ \\
$\mathrm{Iso_{\mathrm{PF}}}$ ($\Delta$R=0.3) / $\pt$  & 0.15  & 0.15  \\
$\abs{d_0}$                & 0.02\unit{cm}  & 0.02\unit{cm} \\
$\abs{d_z}$                & 0.1\unit{cm}& 0.1\unit{cm}  \\
Missing hits           & 1     & 1     \\
Conversion-fit probability   & 10$^{-6}$ & 10$^{-6}$ \\
\hline
\end{tabular}
\end{table}

The MVA selection combines requirements on the output of the identification BDT
described in Section~\ref{sec:eleID}, on the combined PF isolation,
and on rejection variables for photon conversion.
The example discussed in this paper is the selection used in the search for the $\PH \to \cPZ\Z^* \to 4 \ell$
process~\cite{legacyHzz}, which exploits the BDT optimized to identify electrons that are
not required to pass the trigger selection.
In the training, the BDT for these not-triggering electrons does not use any variables related to electron
impact parameters, or variables used to suppress conversions.
Therefore such variables can be exploited in scientific analyses.
For the $\PH \to \cPZ\Z^* \to 4 \ell$ analysis,
a requirement on the significance of the three-dimensional impact
parameter
$\abs{\sigma_\mathrm{ip}/\mathrm{ip}} < 4$
is applied, and the number of missing hits is required to be at most 1.
The combined
$\mathrm{Iso_{\mathrm{PF}}}/\pt$
is required to be less than 0.4 in a cone of $\Delta R = 0.4$.
The selection is optimized in six categories of electron $\pt$ and $\eta$
to maximize the expected sensitivity, using two $\pt$ ranges ($7<\pt<10\GeV$,
and $\pt>10\GeV$), and three $\abs{\eta}$ regions
($\abs{\eta} <0.80$, $0.80 <\abs{\eta}<1.48$, and $1.48 <\abs{\eta}< 2.50$),
corresponding to two regions in the barrel with different amounts of material in front of the ECAL,
and one region in the endcaps.
The MVA selection is used mainly in analyses
that require high efficiency down to low $\pt$, as well as sufficient background
rejection. Examples of such analyses are the Higgs boson searches in leptonic final states.

In addition, CMS has developed a specialized algorithm for the selection of
high-$\pt$ electrons (HEEP, i.e. High Energy Electron Pairs).
Variables similar to those in the sequential selection are used to select large-\pt
electrons, starting at 35\GeV and up to about 1\TeV.
The main difference is the usage of the detector-based isolation instead of
PF isolation
(the two algorithms offer similar performance).
Also, in the barrel, the ratio of the energy collected in
$n \times m$ arrays of crystals
(either $E_{1\times5}/E_{5\times5}$ or $E_{2\times5}/E_{5\times5}$)
is used, since this is found to be more effective at high
\pt than using $\sigma_{\eta \eta}$.
This selection was adopted in many of the searches for exotic particles
published by the CMS experiment, e.g.~Ref.~\cite{zprime}.

\section{Electron efficiencies and misidentification probabilities}
\label{sec:TandP}

A method based on the tag-and-probe (T\&P) technique~\cite{WZxsec}
exploits $\Z/\gamma^{*} \to \Pep\Pem$ events in data to estimate the
reconstruction and selection efficiencies for signal electrons.
The method requires one electron candidate, called the ``tag'', to satisfy tight selection requirements.
Different criteria are tried to define the tag electron, and it is found
that the estimated efficiencies are
almost insensitive to any specific definition of the tag.
For the results in this paper, tag electrons are required to satisfy
$\pt>25\GeV$ and
the tight working point of the sequential selection or, for analyses involving very high-\pt electrons, to satisfy
$\pt>35\GeV$ and
 the
HEEP selection.
A second electron candidate, called the ``probe'', is required to pass specific criteria that depend on the
efficiency under study.
The invariant mass of the two electrons is required to be within a window around the \Z boson mass of $60<m_{\Pep\Pem}<120\GeV$,
which extends sufficiently far from the peak region to enable the background component to be extracted in the fit,
and which is matched to the window used by the analyses that rely on this method.
A requirement for having leptons of opposite charge can also be enforced.
When more than two tag--probe matches are found, they are all used in the procedure
to minimize possible biases produced by some specific choice.

The number of probes passing any chosen selection is determined from fits to the invariant
mass distribution that include contributions from signal and background.
Different models can be used in the fit to disentangle the two components.
In absence of a kinematic selection on the tag-and-probe candidates, the background component in the mass spectrum is well described by a falling exponential. However, the kinematic restrictions on the \Z candidates in each \pt and $\eta$ range of the probe candidate distort the mass spectrum in a way which is well described by an error function. Consequently, the background component of the mass spectrum is described by a falling exponential multiplied by an error function.
In the fits, all parameters of the exponential and of the error function are allowed to float.
The fit to the signal component can use analytic expressions, or be based on templates from simulation.
When using analytical functions, a Breit--Wigner function with the \Z boson mass
and natural width taken from Ref.~\cite{Olive:2014} is convolved with a Crystal Ball function
that acts as the resolution function, and multiplied by a falling exponential function, to model the signal in the mass region between 60 and 70\GeV.
If a template from simulation is used, the signal part of the distribution is modelled through a sample
of simulated electrons from $\Z\to \Pep\Pem$ decays,
convolved with a resolution function to account for any remnant
differences in resolution between data and simulation.
In all cases, a simultaneous fit is performed for events where probes pass or fail the requirements, to account for their correlation.
An alternative to fitting is the subtraction of the background contribution using
predictions from simulation or techniques based on control samples in data. This is the case of the
HEEP selection efficiency, as detailed in the following.

The same T\&P technique is applied to data and simulated events to compare
efficiencies, and to evaluate the data-to-simulation ratios.
In many analyses, these scaling factors are applied as corrections to the
simulation, or are used in computing systematic uncertainties.
The efficiency in simulation is estimated from $\Z\to \Pep\Pem$ signal samples
that contain no background. A geometrical match with generated electrons is often requested
to resolve ambiguities that may arise, mainly at low $\pt$.
In data, the events used in the T\&P procedure are required to satisfy HLT paths that do not bias the efficiency under study.
For the reconstruction efficiency, only triggers requiring one electron and one SC are used,
where the tag is matched to the trigger-electron candidate and the probe is matched to the trigger SC.
For selection efficiencies, triggers requiring two electrons with requirements that are
less restrictive than those under study can also be used.
In such cases, the offline tag and probe are requested to match a trigger-electron candidate.

\begin{figure*}[hbtp]
\centering
 \includegraphics[width=0.49\textwidth]{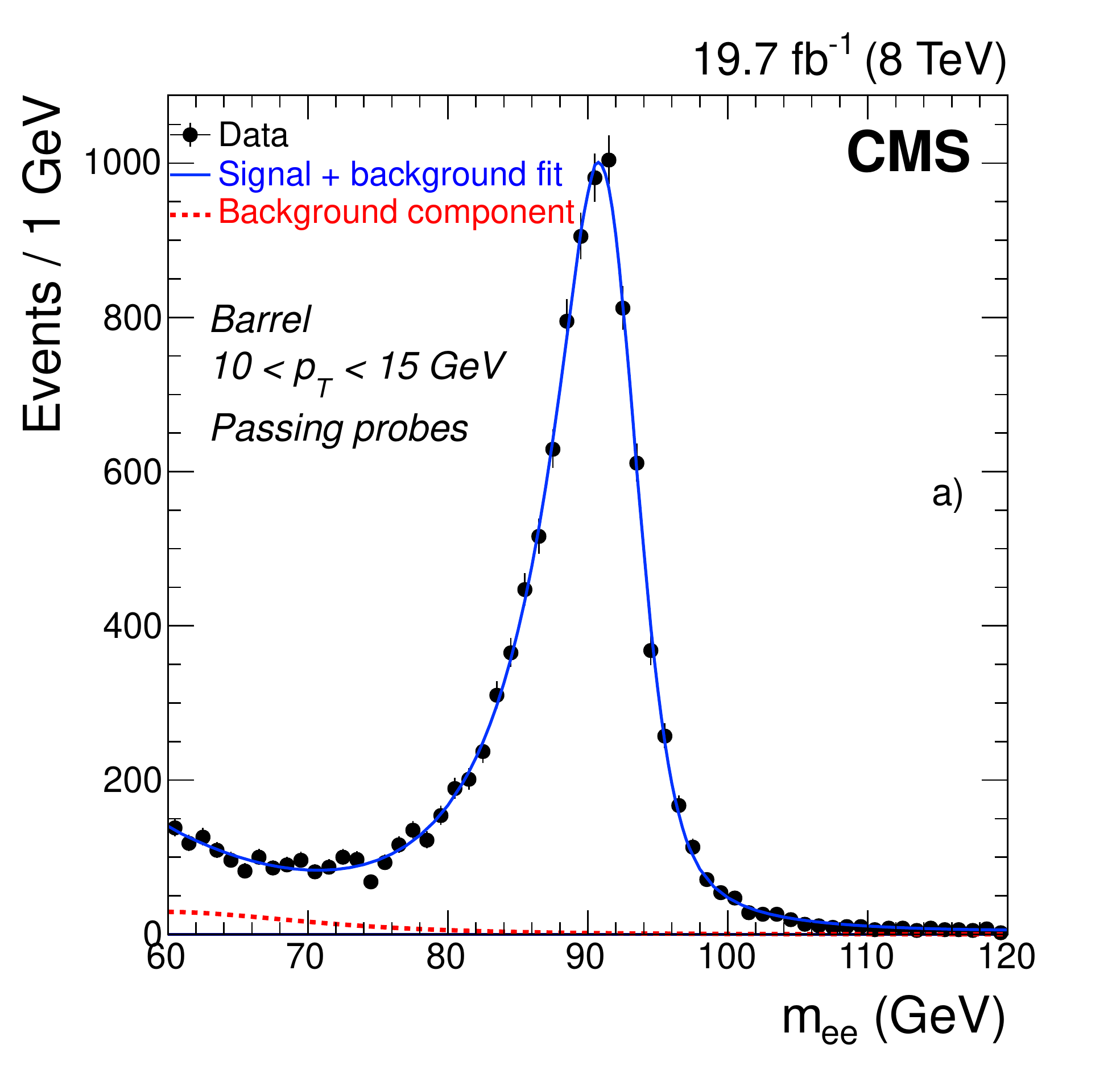}
 \includegraphics[width=0.49\textwidth]{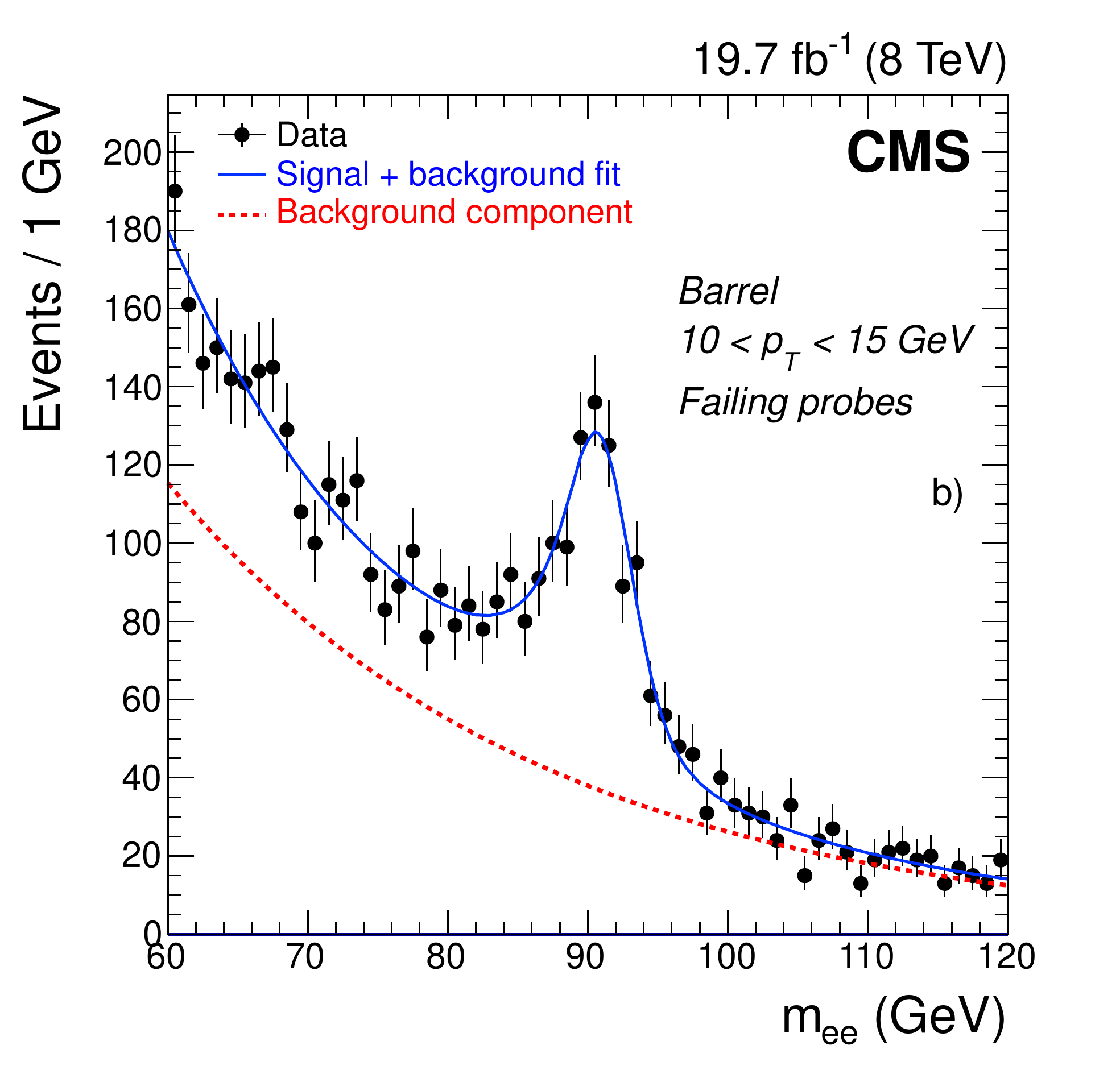}
 \caption{Example of fits to dielectron invariant mass distributions for probe electrons with $10<\pt<15\GeV$ in the ECAL barrel that a) pass or b) fail the selections on isolation and impact parameter of the MVA selection used in Ref.~\cite{legacyHzz}.
Fits are shown for the signal+background hypothesis (full line), and for the background
component alone (dashed line).
}
\label{fig:TaPFitExample}\end{figure*}

The fits are performed in $\eta$ and $\pt$ bins, and an example of a fit to data is shown
in Fig.~\ref{fig:TaPFitExample}. The fits to probe electrons that pass or fail the
selections are shown, respectively in a) and b). The signal in the mass region between 60 and 70\GeV corresponds to contributions from $\gamma^*$ events, from final state radiation, and from poorly measured electrons, essentially located in the ECAL cracks.

Several sources of systematic uncertainty are considered in the fits.
The main uncertainty is related to the model used in the fit,
and is estimated by comparing
alternative distributions for signal and background,
in addition to comparing analytic functions with templates
from simulation.
Only a small dependence is found on the number of bins used in the fits
and on the definition of the tag, such
as on the
reweighting of the simulation to match the pileup in data.
Different event generators are also compared in the analyses,
and the differences among them are found to be negligible.

The results discussed in the next paragraphs illustrate the method applied to several reference selections,
and the performance that is reached.

\subsection{Reconstruction efficiency}
\label{recoEff}
\begin{figure*}[hbtp]
\centering
\includegraphics[width=0.49\textwidth]{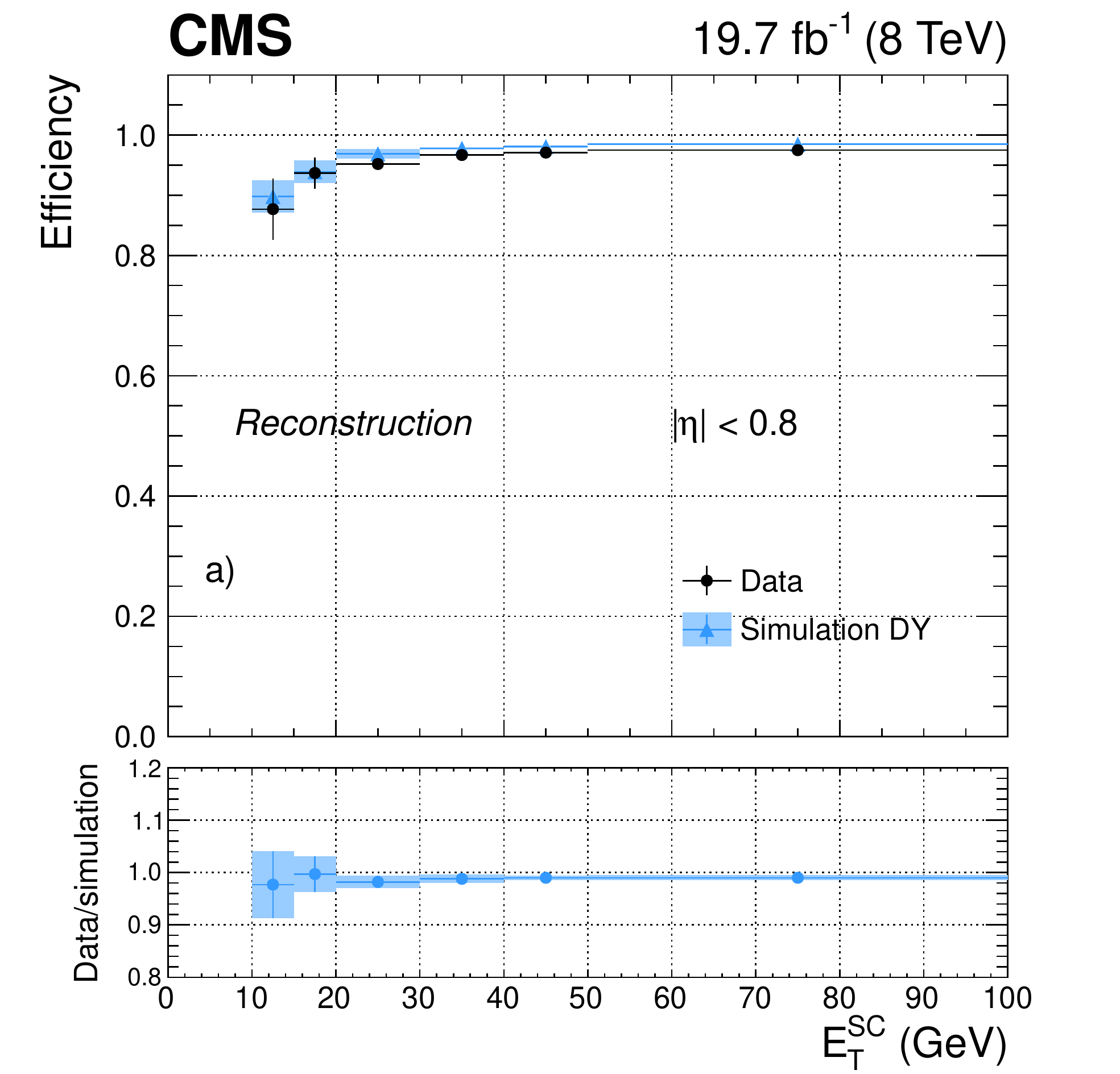}
\includegraphics[width=0.49\textwidth]{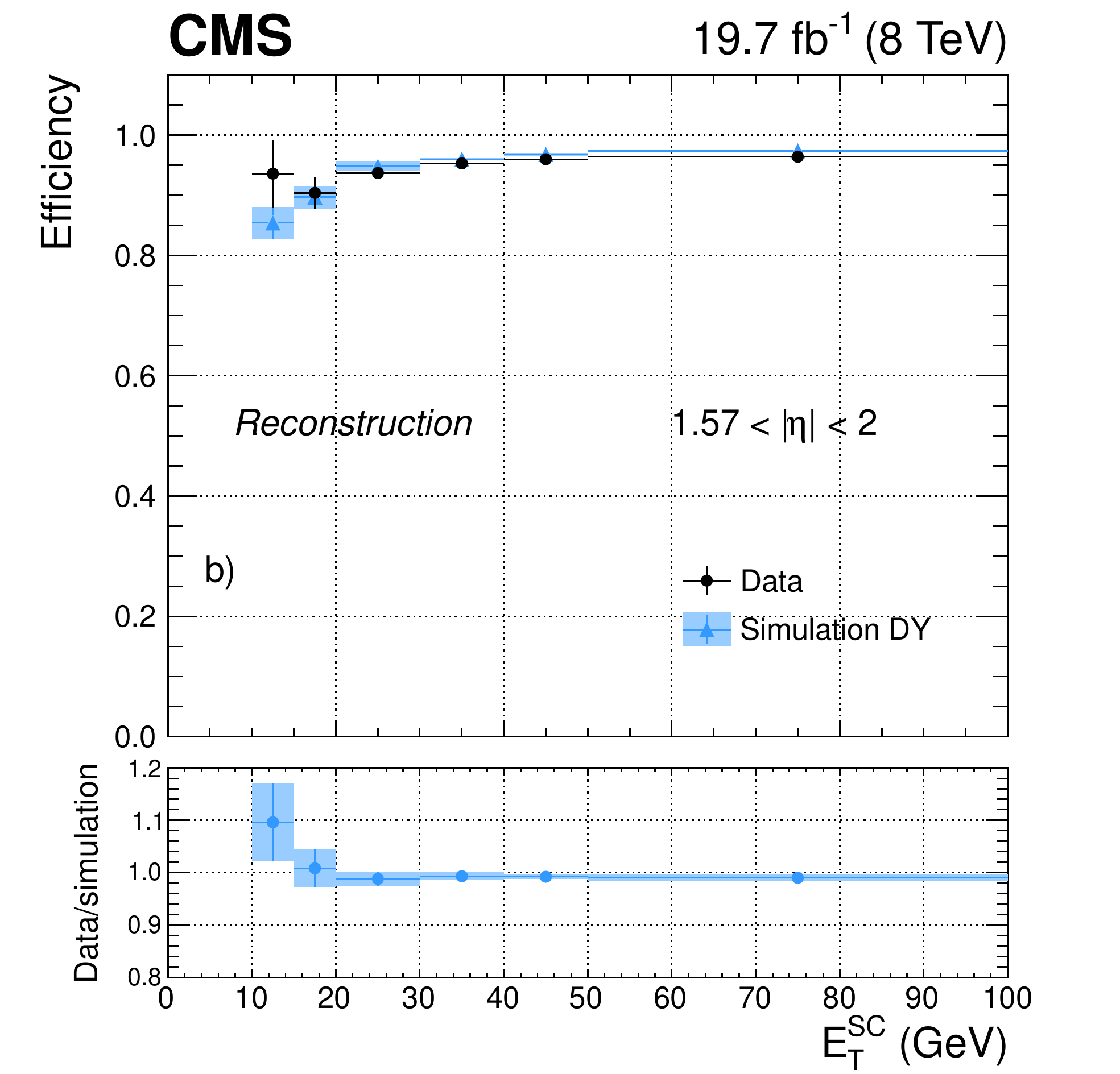}
 \caption{
Electron reconstruction efficiency measured in dielectron events in data (dots) and DY simulation (triangles),
as a function of the electron $\ET^{\mathrm{SC}}$ for
a) $\abs{\eta}<0.8$, and b) $1.57<\abs{\eta}<2$.
The bottom panels show the corresponding data-to-simulation scale factors. The uncertainties shown in the plots correspond to the quadratic sum of the statistical and systematic contributions.
}
\label{fig:recoEffVsPt}\end{figure*}

The reconstruction efficiency is computed as a function of the
$\ET^{\mathrm{SC}}$ and $\eta$ of the SC, and covers all reconstruction effects.
The SC reconstruction efficiency for $\ET^{\mathrm{SC}}>5\GeV$ is close to 100\%.
To illustrate the nature of the results, the electron reconstruction efficiencies measured in data and
in DY simulated samples
are shown in Fig.~\ref{fig:recoEffVsPt},
together with the data-to-simulation scale factors, as a function of $\ET^{\mathrm{SC}}$, for
a)~$\abs{\eta}<0.8$, and b)~$1.57<\abs{\eta}<2$.

The efficiencies are found to be $>$85\% for $\ET^{SC}>10\GeV$, for all $\eta$.
They are compatible in data and simulation, giving scale factors consistent
with unity almost in the entire range.
The uncertainties shown on the plots correspond to the quadratic sum of the statistical and systematic contributions,
dominated by the systematic components, at the level of a few percent for
$\ET^{\mathrm{SC}}<20\GeV$ and decreasing to $<$1\% as $\ET^{\mathrm{SC}}$ increases.
The main uncertainty is related to the
fitting function.
The background contamination is large in the estimation of reconstruction efficiency,
and additional requirements are therefore applied,
such as requiring the imbalance in \pt in the event to be $<$20\GeV.
Also, the probe must be isolated, which requires the scalar \pt sum
of all tracks from the vertex of interest that fall into the isolation cone
to be $<$15\% of the probe $\ET^{\mathrm{SC}}$.
The impact of changing the definitions of these extra requirements corresponds to the second-highest source of systematic uncertainty in this measurement.

\subsection{Selection efficiency}
\label{selEff}
\begin{figure*}[hbtp]
\centering
\includegraphics[width=0.49\textwidth]{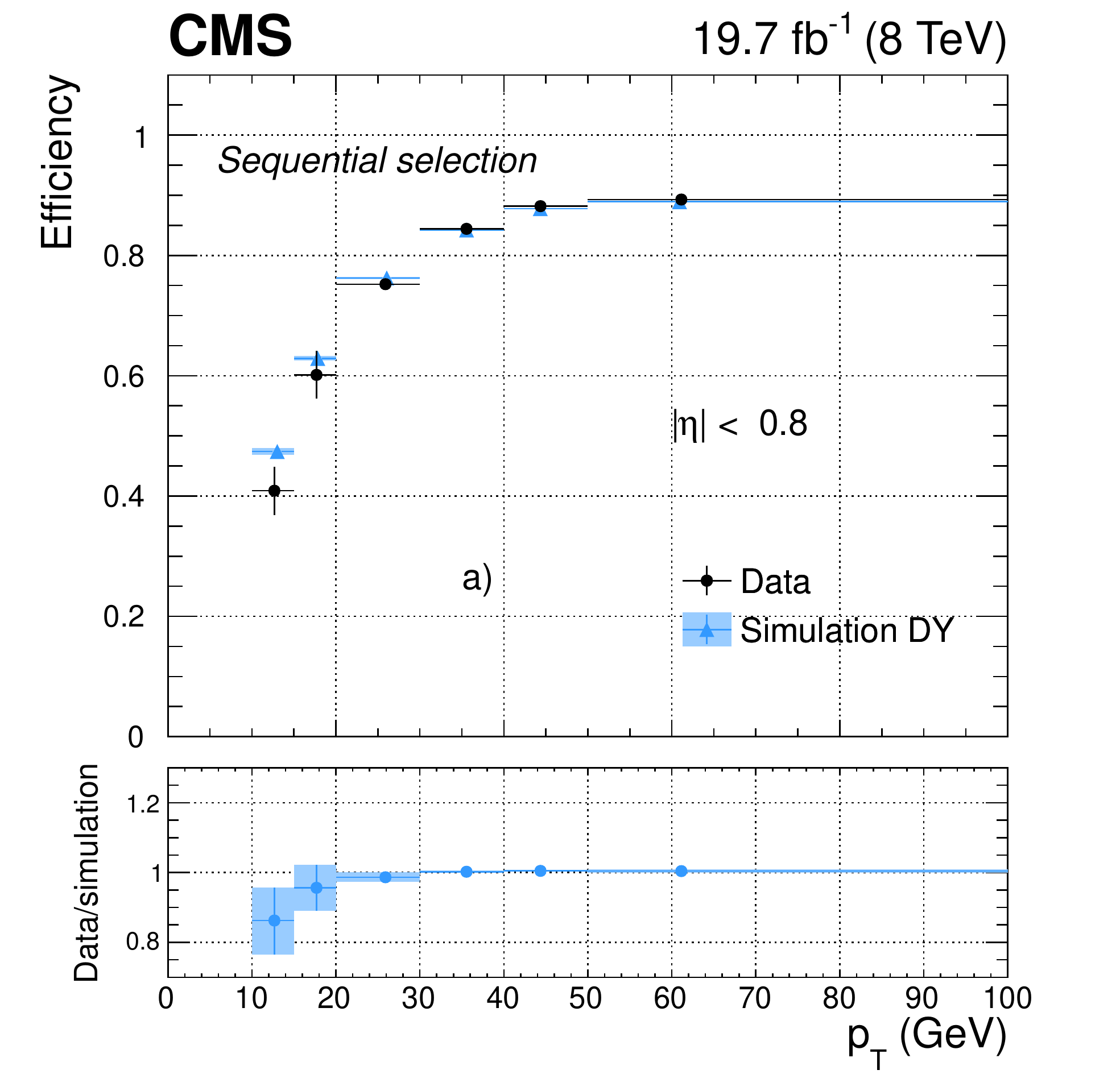}
\includegraphics[width=0.49\textwidth]{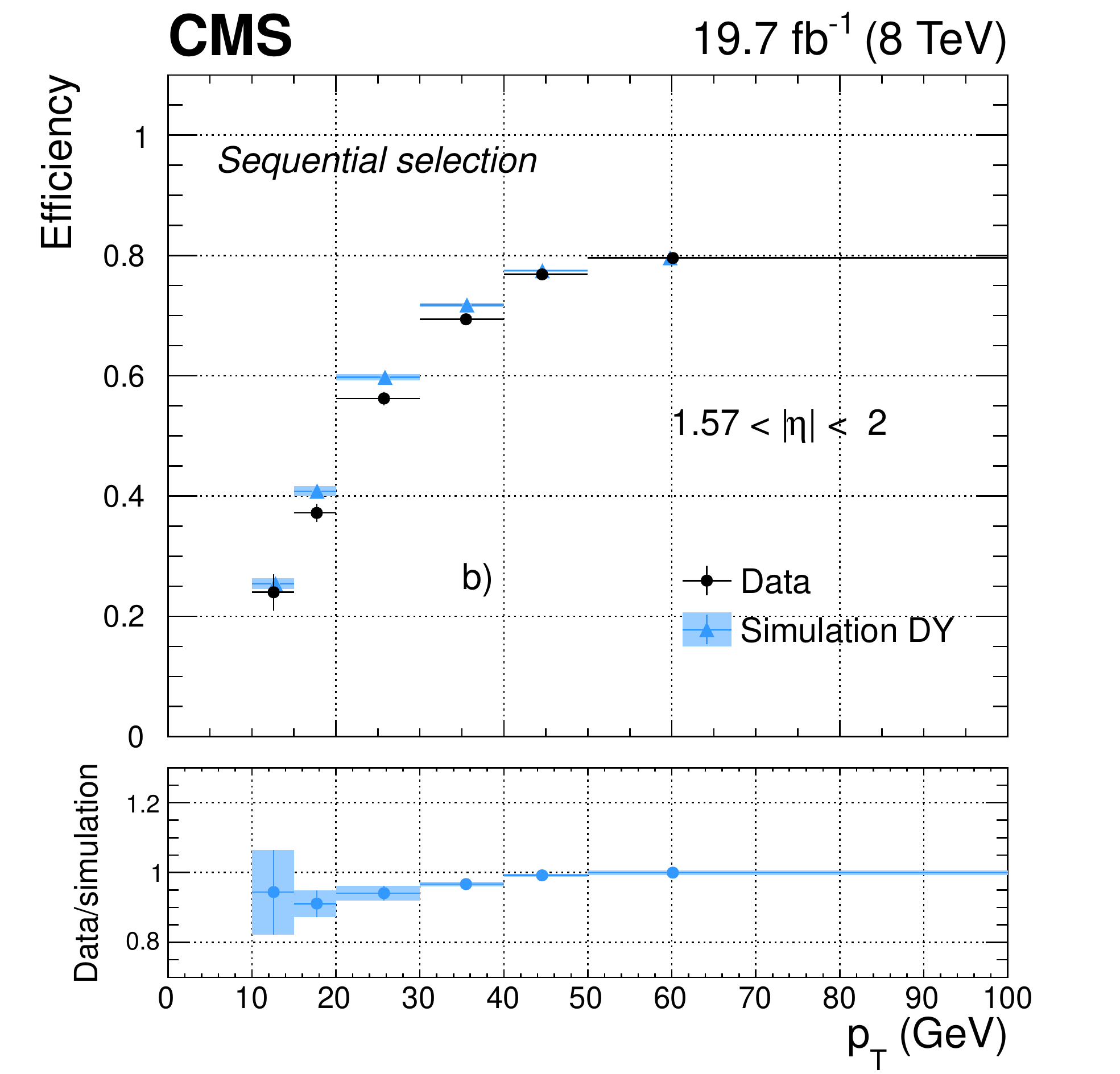}\\
\includegraphics[width=0.49\textwidth]{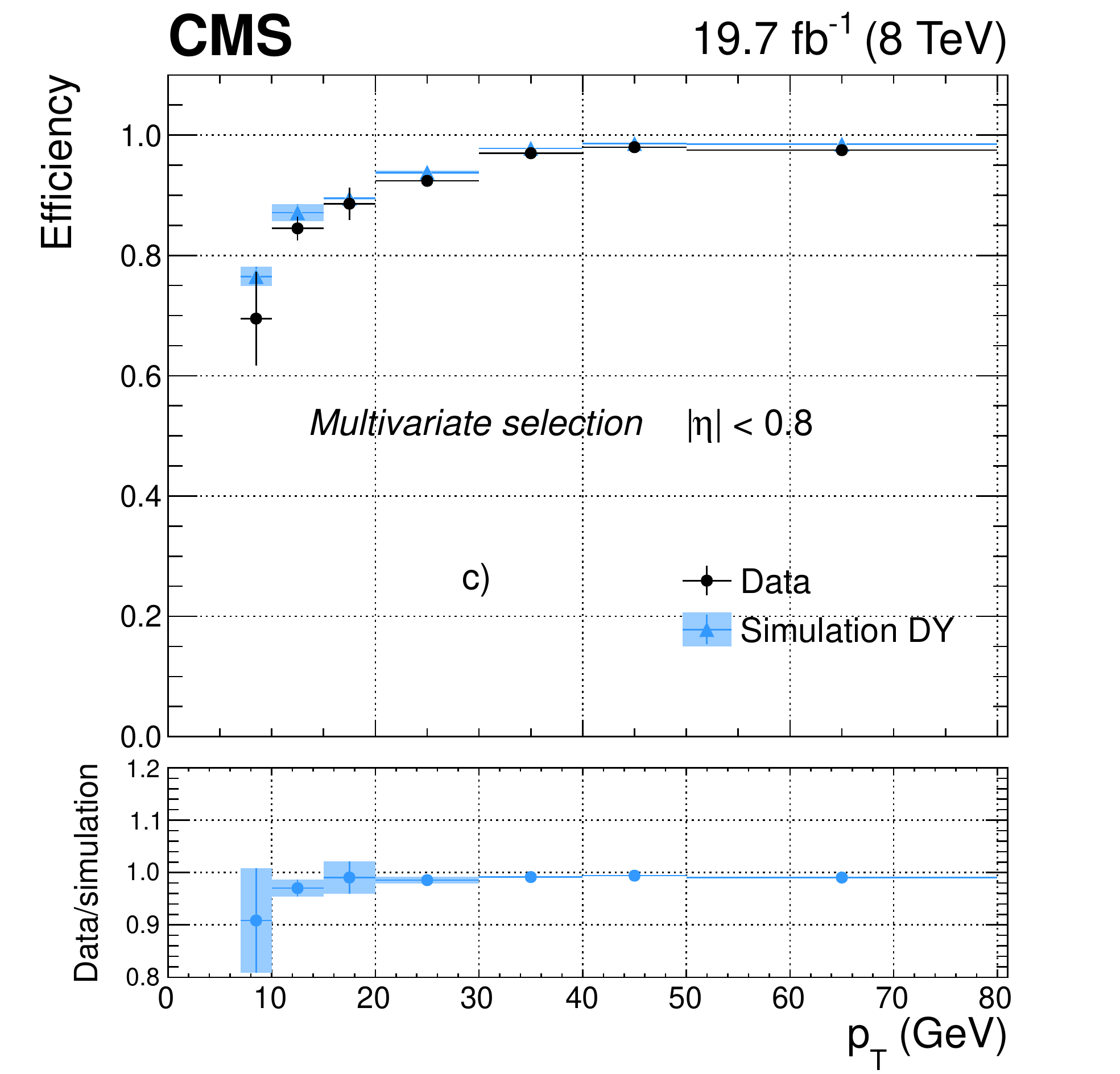}
\includegraphics[width=0.49\textwidth]{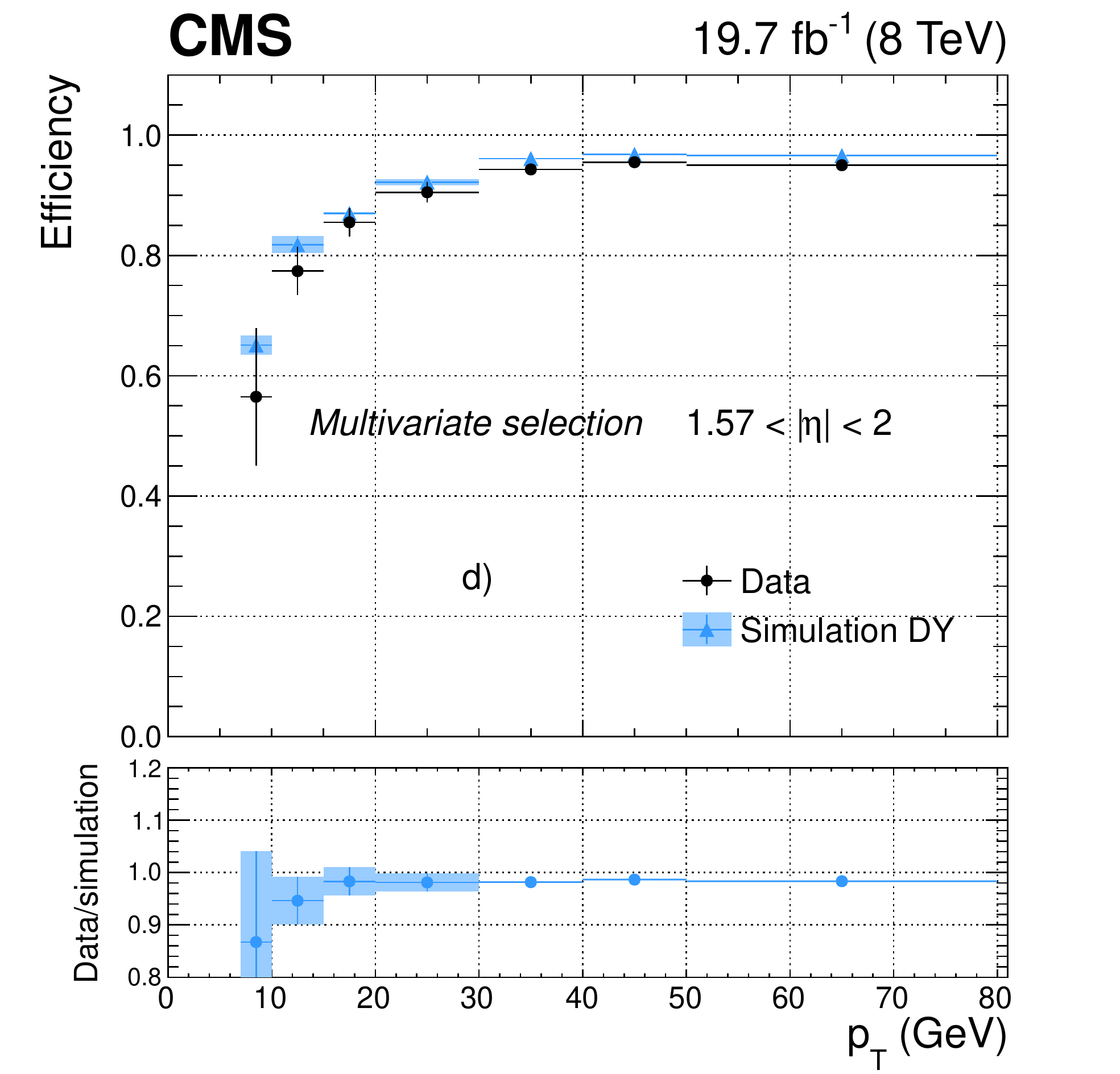}
 \caption{Efficiency as a function of electron $\pt$ for dielectron events
in data (dots) and DY simulation (triangles), for
the medium working point of the sequential selection in a) $\abs{\eta}<0.8$, and
b) $1.57<\abs{\eta}<2$;
and for the MVA selection used in Ref.~\cite{legacyHzz} in c) $\abs{\eta}<0.8$,
and d) $1.57<\abs{\eta}<2$.
The corresponding data-to-simulation scale factors are shown in the bottom panels of each plot. The uncertainties shown in the plots correspond to the quadratic sum of the statistical and systematic contributions. }
\label{fig:effCutMediumVsPt}\end{figure*}

The selection efficiency is computed for reconstructed electrons
in bins of the electron $\pt$ and of the $\eta$ of the SC.
For the sequential selection, the efficiencies of the medium working point in data and in simulation
are presented as a function of electron $\pt$ in Fig.~\ref{fig:effCutMediumVsPt}
for a)~$\abs{\eta}<0.8$, and b)~$1.57<\abs{\eta}<2$. The corresponding
data-to-simulation scale factors are shown in the bottom panels.
Similarly, Figs.~\ref{fig:effCutMediumVsPt}~c) and d) show
the efficiencies as a function of $\pt$
for the BDT selection, discussed in the previous section.
The selections are optimized respectively for $\pt>10\GeV$ and $\pt>7\GeV$, which are the ranges shown in the plots.
In general, data and simulation agree well. The scale factors are compatible with unity,
with the exception of the low-\pt region ($7<\pt<15\GeV$), where they can be as low
as 0.85--0.90 depending on the selections.
The uncertainties shown include contributions from both the statistical and systematic sources.
They are again dominated by systematic contributions, which are at the level of several percent for
$\pt<20\GeV$, and decrease below 2\% when $\pt$ increases, with the exception of the transition region between the barrel and the endcap.
As for reconstruction efficiencies, the main uncertainty originates from the choice of the fitting function.
It is verified that efficiencies are almost uniform as a function of the number of reconstructed interaction
vertices. As expected, the less restrictive the selection, the smaller is the remnant dependence on pileup.
For the working points illustrated in Fig.~\ref{fig:effCutMediumVsPt},
the efficiencies decrease only by about 5\% and 2\% for up to 50 primary vertices,
meaning that the proposed selections are almost independent of pileup.
The average number of proton-proton interactions per bunch crossing is about 21 in the 8\TeV data.

\begin{figure*}[hbtp]
\centering
\includegraphics[width=0.49\textwidth]{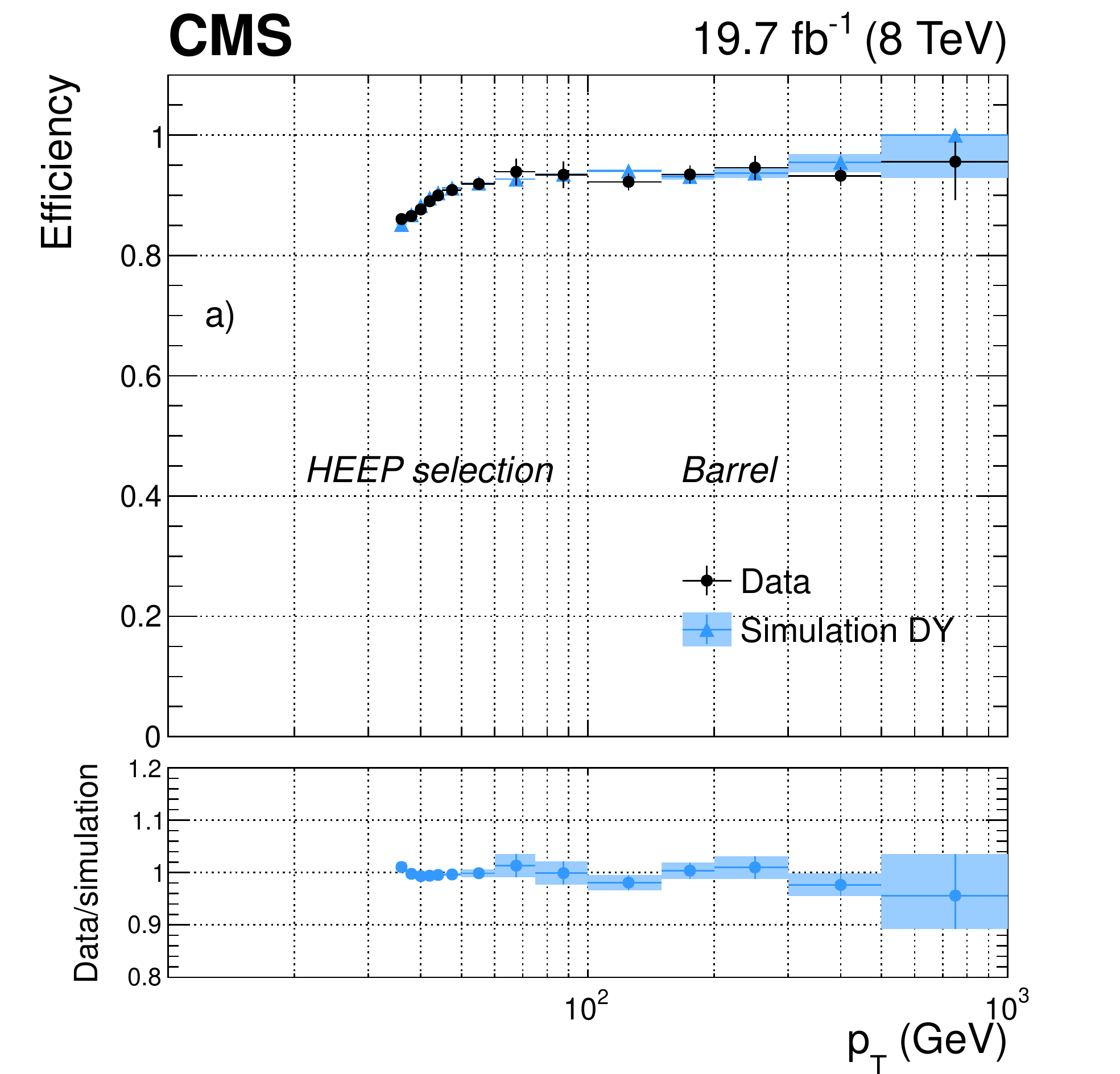}
\includegraphics[width=0.49\textwidth]{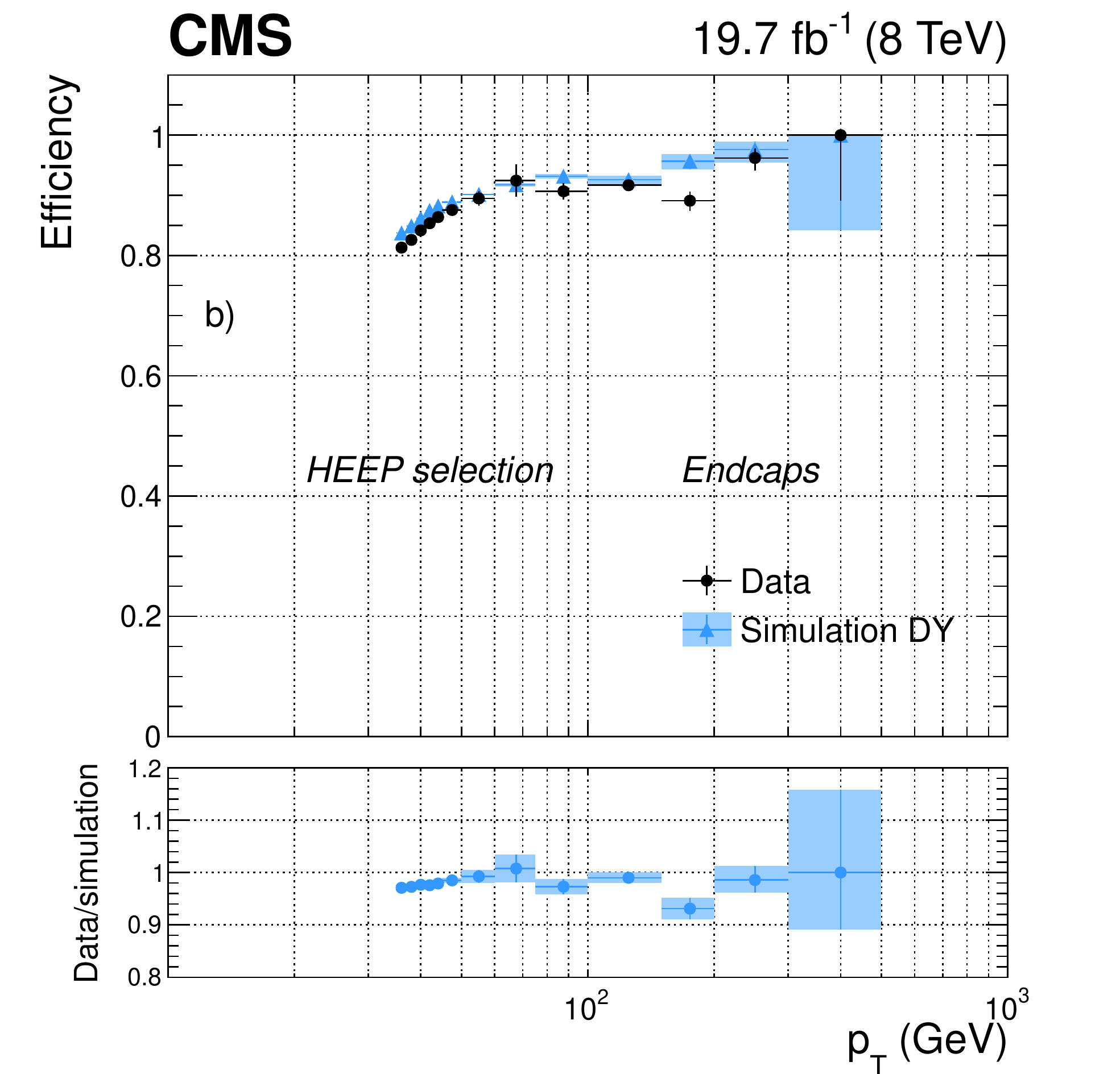}\\
 \caption{Efficiency of the HEEP selection as a function of electron $\pt$ for dielectron events in data (dots) and DY simulation (triangles) in the ECAL a) barrel, and b) endcaps. The uncertainties shown in the plots correspond to the quadratic sum of the statistical and systematic contributions.}
\label{fig:effHEEPVsPt}\end{figure*}

For the HEEP selection, the efficiency is computed by subtracting the background contribution
estimated from simulation, instead of using a fit. This is done especially because of the small number of events
at large \pt in data.
Multijet production, which is among the dominant contributions to the backgrounds to Z+jets,
is estimated directly from data using the jet-to-electron misidentification probabilities measured
in a dedicated control sample.
The measured uncertainty of about 40\% in the estimated background
is the main source of systematic uncertainty.
The efficiency of the HEEP selection in data and in simulation
is shown as a function of electron $\pt$ in Fig.~\ref{fig:effHEEPVsPt},
together with the data-to-simulation scale factors.
Because of the limited number of events, only two $\eta$ bins are considered, corresponding to the ECAL
barrel and endcaps.
The $\pt$ region is restricted to $\pt>35\GeV$, and a wider $\pt$ range is covered in the barrel
because of the presence of more events there than in the endcaps.
In the barrel, the efficiency ranges from 85 to 95\%, and the data-to-simulation scale factors are
compatible with unity. In the endcaps, the fluctuations are larger, with efficiencies ranging from
about 80 to close to 100\%.
The uncertainties shown in the plots correspond to the quadratic sum of the statistical and systematic contributions.
For electrons with \pt~$<$~100\GeV, the uncertainty is dominated by
systematic sources, since this is the
region where the background is more important, while above about 100\GeV the statistical uncertainty dominates.

\subsection{Misidentification probability}
To each efficiency corresponds a misidentification probability,
defined as the fraction of background candidates reconstructed as electrons
that pass some set of selection criteria.
The results have their misidentification probability computed using
data enriched in \Z bosons that also contain an additional electron, as
described in Section~\ref{sec:eleID}.

\begin{figure*}[h!tbp]
\centering
\includegraphics[width=0.45\textwidth]{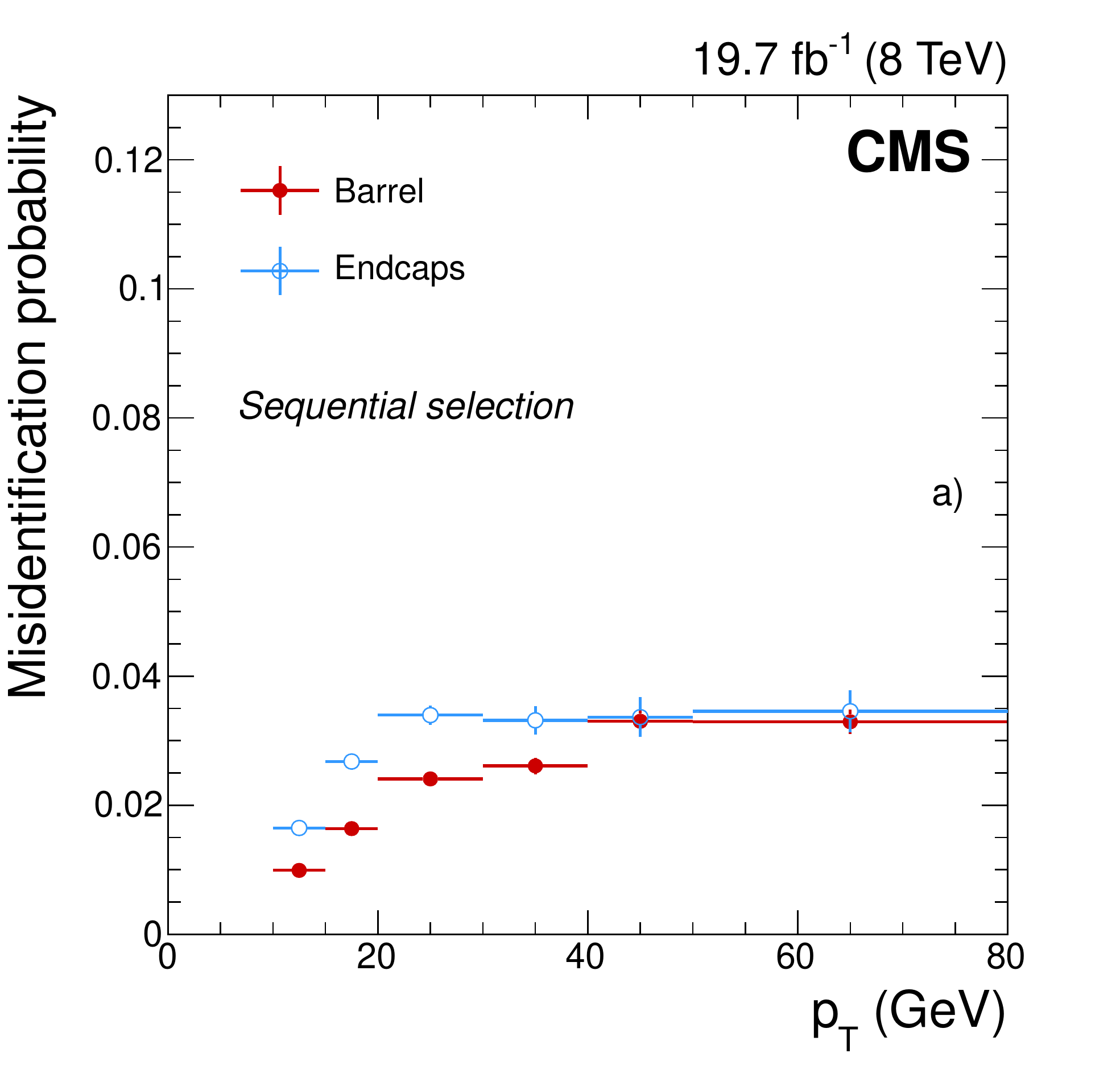}
\includegraphics[width=0.45\textwidth]{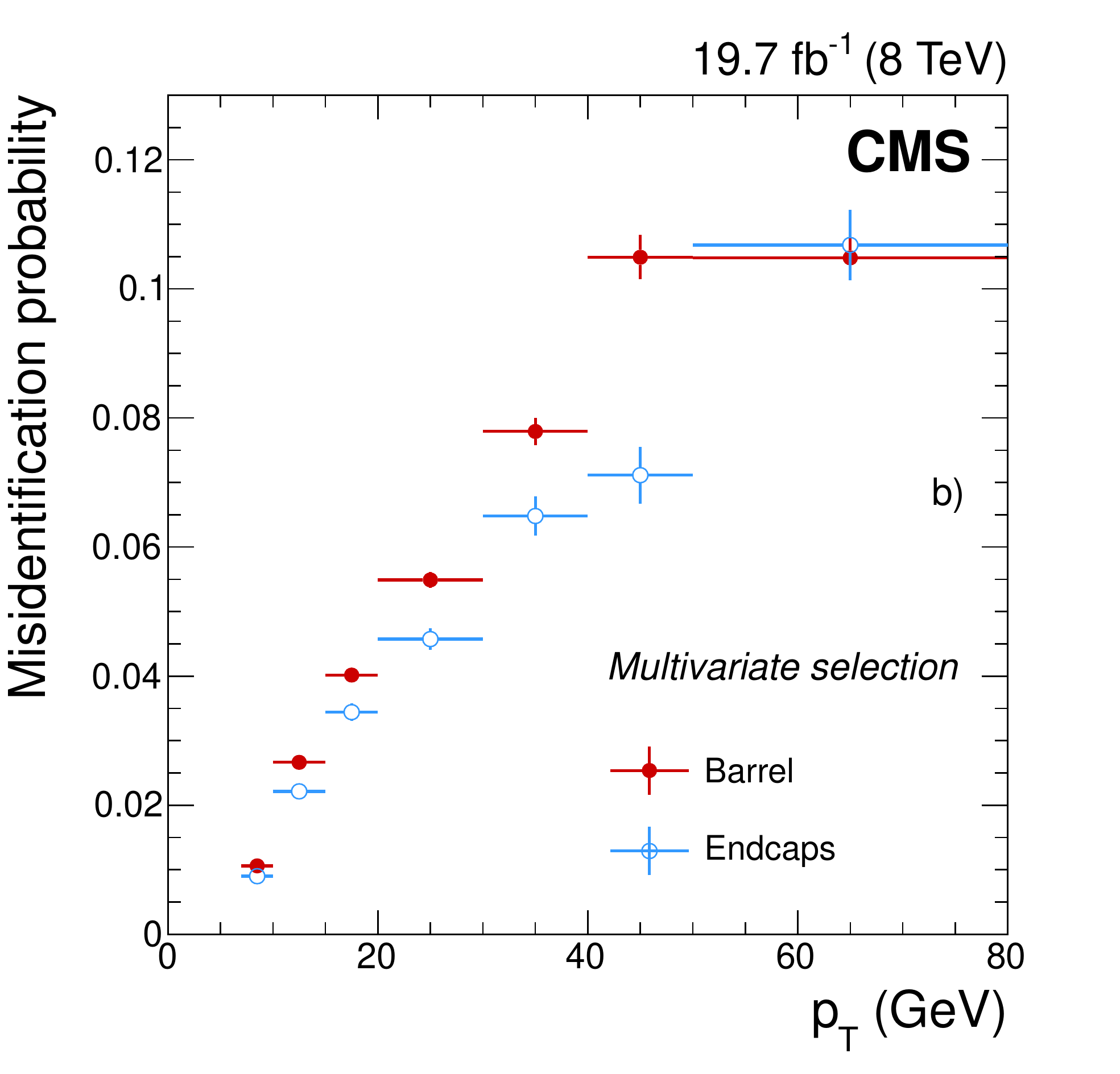}
\caption{Misidentification probability, measured in data as described in the text,
as a function of the electron $\pt$ in the barrel (red dots) and endcaps (blue dots)
for candidates passing a) the medium working point of the sequential selection,
and b) the working point of the MVA selection used in Ref.~\cite{legacyHzz}. The uncertainties shown in the plots correspond to just the statistical contributions.
 }
\label{Fig:FRVsPt}\end{figure*}

The fraction of events in which additional reconstructed electron candidates
from background contributions
pass the medium
working point of the sequential selection is shown in Fig.~\ref{Fig:FRVsPt}~a) as a
function of the candidate $\pt$. The same fraction is shown in Fig.~\ref{Fig:FRVsPt}~b) for the MVA selection.
The uncertainties shown in the plots correspond to just the statistical contributions.
In both cases, the misidentification probability increases with the \pt of the candidate.
For the working point of the sequential selection,
it ranges from 1 to 3.5\%, depending on $\pt$ and on the region of the detector.
For the MVA selection, the chosen working point~\cite{legacyHzz} is less restrictive and the misidentification probability is therefore
larger (from 1 to 10.5\%).

The main source of systematic uncertainty in the misidentification probability is related to the
composition of the sample used to extract its value. For this particular choice, it is mainly related to
the contamination from processes with genuine electrons, such as the associated production of
W and \Z bosons, and \ttbar events.
The selection on the imbalance in transverse momentum
strongly reduces such contamination, and therefore the systematic uncertainty, with the consequence that the
main uncertainty in the analyses comes from the difference
between the samples used to extract the misidentification probability and the one to which the result is applied.
This is strongly analysis-dependent and therefore not discussed further.

\section{Summary and conclusions}
\label{sec:conclusion}

The performance of electron reconstruction and selection in CMS has been studied using
data collected in proton-proton collisions at $\sqrt{s}=8\TeV$ corresponding to an integrated
luminosity of 19.7\fbinv.

Algorithms used to reconstruct electron
trajectories and energy deposits
in the tracker and ECAL respectively,
have been presented.
A Gaussian sum filter algorithm used for track reconstruction provides
a way to follow the track curvature and to account for \brem loss up to the entrance into the ECAL.
The strategies for finding seeds for electron tracks, constructing trajectories, and fitting
 track parameters
are optimized to reconstruct the electrons down to small \pt values with high efficiency and accuracy.
Moreover, the clustering of energy in the ECAL and its optimization to recover bremsstrahlung photons are discussed.
Dedicated algorithms are used to correct the energy measured in the ECAL as well as to estimate the
electron momentum by combining independent measurements in the ECAL and in the tracker.

The overall
momentum scale is calibrated with an uncertainty smaller than 0.3\% in the \pt range from 7 to
70\GeV.
For electrons from \Z boson decays, the effective momentum resolution varies from 1.7\%,
for well-measured electrons with a single-cluster supercluster in the barrel,
to 4.5\%
for electrons with a multi-cluster supercluster, or poorly measured, in the endcaps.
The electron momentum resolution is modelled in simulation with a precision better than 10\% up to a
\pt of 70\GeV.

The performance of the reconstruction algorithms in data is studied
together with those of several benchmark selections designed
to cover the needs of the physics programme of the CMS experiment.
Good agreement is observed between data and predictions from simulation for most of the variables relevant
to electron reconstruction and selection.
 The origin of small remaining discrepancies is understood and
corrections will be implemented in the future.

The reconstruction efficiency as well as the efficiency of all the selections
are measured using $\Z \to \Pep\Pem$ samples in data and in simulation.
The reconstruction efficiency in the data ranges from 88\% to 98\% in the barrel and
from 90\% to 96\% in the endcaps in the \pt range from 10 to 100\GeV.
The ratios of efficiencies of data to simulation, both for reconstruction
and for the different proposed selections, are found to be in general compatible with unity within
the respective uncertainties, over the full \pt range, down to a \pt as low as 7\GeV. Differences of up to 5\% between data and simulation are observed
in most cases, while differences of up to~15\% are measured for
a few points at small \pt values.

The analysis of electron performance with data has shown that, despite the challenging conditions of pileup at the LHC and the significant level of bremsstrahlung in the tracker, using
dedicated algorithms and a large number of recorded $\Z\to \Pep\Pem$ decays
provided successful means of reconstructing and identifying electrons in CMS. The quality of simulation at the beginning of the experiment
was sufficiently good to require only a few adjustments to the originally conceived reconstruction algorithms, and also enabled quick deployment of sophisticated developments, such as PF reconstruction and the use of MVA methods for electron identification and, later, for momentum correction. The reconstruction and selection of electrons at low \pt have been achieved with a performance level close to that anticipated at the time the detector was designed. These achievements, especially for low-\pt electrons, played an essential role in the discovery of the Higgs boson at CMS~\cite{discoveryPaper,Chatrchyan:2013lba},
and in the measurement of its properties~\cite{Khachatryan:2014jba} in the
$\PH \to \cPZ\Z^* \to 4 \ell$ channel.

\begin{acknowledgments}
\label{sec:acknowledgements}
\hyphenation{Bundes-ministerium Forschungs-gemeinschaft Forschungs-zentren} We congratulate our colleagues in the CERN accelerator departments for the excellent performance of the LHC and thank the technical and administrative staffs at CERN and at other CMS institutes for their contributions to the success of the CMS effort. In addition, we gratefully acknowledge the computing centres and personnel of the Worldwide LHC Computing Grid for delivering so effectively the computing infrastructure essential to our analyses. Finally, we acknowledge the enduring support for the construction and operation of the LHC and the CMS detector provided by the following funding agencies: the Austrian Federal Ministry of Science, Research and Economy and the Austrian Science Fund; the Belgian Fonds de la Recherche Scientifique, and Fonds voor Wetenschappelijk Onderzoek; the Brazilian Funding Agencies (CNPq, CAPES, FAPERJ, and FAPESP); the Bulgarian Ministry of Education and Science; CERN; the Chinese Academy of Sciences, Ministry of Science and Technology, and National Natural Science Foundation of China; the Colombian Funding Agency (COLCIENCIAS); the Croatian Ministry of Science, Education and Sport, and the Croatian Science Foundation; the Research Promotion Foundation, Cyprus; the Ministry of Education and Research, Estonian Research Council via IUT23-4 and IUT23-6 and European Regional Development Fund, Estonia; the Academy of Finland, Finnish Ministry of Education and Culture, and Helsinki Institute of Physics; the Institut National de Physique Nucl\'eaire et de Physique des Particules~/~CNRS, and Commissariat \`a l'\'Energie Atomique et aux \'Energies Alternatives~/~CEA, France; the Bundesministerium f\"ur Bildung und Forschung, Deutsche Forschungsgemeinschaft, and Helmholtz-Gemeinschaft Deutscher Forschungszentren, Germany; the General Secretariat for Research and Technology, Greece; the National Scientific Research Foundation, and National Innovation Office, Hungary; the Department of Atomic Energy and the Department of Science and Technology, India; the Institute for Studies in Theoretical Physics and Mathematics, Iran; the Science Foundation, Ireland; the Istituto Nazionale di Fisica Nucleare, Italy; the Ministry of Science, ICT and Future Planning, and National Research Foundation (NRF), Republic of Korea; the Lithuanian Academy of Sciences; the Ministry of Education, and University of Malaya (Malaysia); the Mexican Funding Agencies (CINVESTAV, CONACYT, SEP, and UASLP-FAI); the Ministry of Business, Innovation and Employment, New Zealand; the Pakistan Atomic Energy Commission; the Ministry of Science and Higher Education and the National Science Centre, Poland; the Funda\c{c}\~ao para a Ci\^encia e a Tecnologia, Portugal; JINR, Dubna; the Ministry of Education and Science of the Russian Federation, the Federal Agency of Atomic Energy of the Russian Federation, Russian Academy of Sciences, and the Russian Foundation for Basic Research; the Ministry of Education, Science and Technological Development of Serbia; the Secretar\'{\i}a de Estado de Investigaci\'on, Desarrollo e Innovaci\'on and Programa Consolider-Ingenio 2010, Spain; the Swiss Funding Agencies (ETH Board, ETH Zurich, PSI, SNF, UniZH, Canton Zurich, and SER); the Ministry of Science and Technology, Taipei; the Thailand Center of Excellence in Physics, the Institute for the Promotion of Teaching Science and Technology of Thailand, Special Task Force for Activating Research and the National Science and Technology Development Agency of Thailand; the Scientific and Technical Research Council of Turkey, and Turkish Atomic Energy Authority; the National Academy of Sciences of Ukraine, and State Fund for Fundamental Researches, Ukraine; the Science and Technology Facilities Council, UK; the US Department of Energy, and the US National Science Foundation.

Individuals have received support from the Marie-Curie programme and the European Research Council and EPLANET (European Union); the Leventis Foundation; the A. P. Sloan Foundation; the Alexander von Humboldt Foundation; the Belgian Federal Science Policy Office; the Fonds pour la Formation \`a la Recherche dans l'Industrie et dans l'Agriculture (FRIA-Belgium); the Agentschap voor Innovatie door Wetenschap en Technologie (IWT-Belgium); the Ministry of Education, Youth and Sports (MEYS) of the Czech Republic; the Council of Science and Industrial Research, India; the HOMING PLUS programme of Foundation for Polish Science, cofinanced from European Union, Regional Development Fund; the Compagnia di San Paolo (Torino); the Consorzio per la Fisica (Trieste); MIUR project 20108T4XTM (Italy); the Thalis and Aristeia programmes cofinanced by EU-ESF and the Greek NSRF; and the National Priorities Research Program by Qatar National Research Fund.

\end{acknowledgments}
\bibliography{auto_generated}

\cleardoublepage \appendix\section{The CMS Collaboration \label{app:collab}}\begin{sloppypar}\hyphenpenalty=5000\widowpenalty=500\clubpenalty=5000\textbf{Yerevan Physics Institute,  Yerevan,  Armenia}\\*[0pt]
V.~Khachatryan, A.M.~Sirunyan, A.~Tumasyan
\vskip\cmsinstskip
\textbf{Institut f\"{u}r Hochenergiephysik der OeAW,  Wien,  Austria}\\*[0pt]
W.~Adam, T.~Bergauer, M.~Dragicevic, J.~Er\"{o}, M.~Friedl, R.~Fr\"{u}hwirth\cmsAuthorMark{1}, V.M.~Ghete, C.~Hartl, N.~H\"{o}rmann, J.~Hrubec, M.~Jeitler\cmsAuthorMark{1}, W.~Kiesenhofer, V.~Kn\"{u}nz, M.~Krammer\cmsAuthorMark{1}, I.~Kr\"{a}tschmer, D.~Liko, I.~Mikulec, D.~Rabady\cmsAuthorMark{2}, B.~Rahbaran, H.~Rohringer, R.~Sch\"{o}fbeck, J.~Strauss, W.~Treberer-Treberspurg, W.~Waltenberger, C.-E.~Wulz\cmsAuthorMark{1}
% ...
\vskip\cmsinstskip
% ...
1:~~Also at Vienna University of Technology, Vienna, Austria\\
2:~~Also at CERN, European Organization for Nuclear Research, Geneva, Switzerland\\
% ...

\end{sloppypar}
\end{document}